\documentclass[aps,prc,reprint,superscriptaddress]{revtex4-1}

\usepackage{amsmath}
\usepackage{amsfonts}
\usepackage[hidelinks]{hyperref}
\usepackage{graphicx,rotating}
\usepackage{dcolumn}
\usepackage{multirow}
\usepackage{verbatim}
\usepackage{paralist}

\begin{document}


\title{Measurement of the $\beta$-asymmetry parameter of $^{67}$Cu in search for tensor type currents in the weak interaction}

\author{G. Soti}
\email[Corresponding author: ]{gergelj.soti@fys.kuleuven.be}
\affiliation{KU Leuven, Instituut voor Kern- en Stralingsfysica, Celestijnenlaan 200D, 3001 Leuven, Belgium}
\author{F. Wauters}
\altaffiliation[Current address: ]{Center for Experimental Nuclear Physics and Astrophysics, University of Washington, Seattle, Washington 98195, USA}
\author{M. Breitenfeldt}
\affiliation{KU Leuven, Instituut voor Kern- en Stralingsfysica, Celestijnenlaan 200D, 3001 Leuven, Belgium}
\author{P. Finlay}
\affiliation{KU Leuven, Instituut voor Kern- en Stralingsfysica, Celestijnenlaan 200D, 3001 Leuven, Belgium}
\author{P. Herzog}
\affiliation{Helmholtz-Institut fuer Strahlen- und Kernphysik, Universitaet Bonn, D-53115 Bonn, Germany}
\author{A. Knecht}
\affiliation{KU Leuven, Instituut voor Kern- en Stralingsfysica, Celestijnenlaan 200D, 3001 Leuven, Belgium}
\affiliation{PH Department, CERN, CH-1211 Geneva 23, Switzerland}
\author{U. K\"{o}ster}
\affiliation{Institut Laue Langevin, 6 rue Jules Horowitz, F-38042 Grenoble Cedex, France}
\author{I. S. Kraev}
\affiliation{KU Leuven, Instituut voor Kern- en Stralingsfysica, Celestijnenlaan 200D, 3001 Leuven, Belgium}
\author{T. Porobic}
\affiliation{KU Leuven, Instituut voor Kern- en Stralingsfysica, Celestijnenlaan 200D, 3001 Leuven, Belgium}
\author{P. N. Prashanth}
\affiliation{KU Leuven, Instituut voor Kern- en Stralingsfysica, Celestijnenlaan 200D, 3001 Leuven, Belgium}
\author{I. S. Towner}
\affiliation{Cyclotron Institute, Texas A\&{M} University, College Station, Texas 77843, USA }
\author{C. Tramm}
\affiliation{Helmholtz-Institut fuer Strahlen- und Kernphysik, Universitaet Bonn, D-53115 Bonn, Germany}
\author{D. Z\'akouck\'y}
\affiliation{Nuclear Physics Institute, ASCR, 250 68 \v{R}e\v{z}, Czech Republic}
\author{N. Severijns}
\affiliation{KU Leuven, Instituut voor Kern- en Stralingsfysica, Celestijnenlaan 200D, 3001 Leuven, Belgium}

\date{\today}

\begin{abstract}
\begin{description}
 \item[Background] Precision measurements at low energy search for physics beyond the Standard Model in a way complementary to searches for new particles at colliders. In the weak sector the most general $\beta$ decay Hamiltonian contains, besides vector and axial-vector terms, also scalar, tensor and pseudoscalar terms. Current limits on the scalar and tensor coupling constants from neutron and nuclear $\beta$ decay are on the level of several percent.
 \item[Purpose] Extracting new information on tensor coupling constants by measuring the $\beta$-asymmetry parameter in the pure Gamow-Teller decay of $^{67}$Cu, thereby testing the V-A structure of the weak interaction.
 \item[Method] An iron sample foil into which the radioactive nuclei were implanted was cooled down to milliKelvin temperatures in a $^3$He-$^4$He dilution refrigerator. An external magnetic field of 0.1\,T, in combination with the internal hyperfine magnetic field, oriented the nuclei. The anisotropic $\beta$ radiation was observed with planar high purity germanium detectors operating at a temperature of about 10\,K. An on-line measurement of the $\beta$ asymmetry of $^{68}$Cu was performed as well for normalization purposes. Systematic effects were investigated using Geant4 simulations.
 \item[Results] The experimental value, $\tilde{A}$~=~0.587(14), is in agreement with the Standard Model value of 0.5991(2) and is interpreted in terms of physics beyond the Standard Model. The limits obtained on possible tensor type charged currents in the weak interaction hamiltonian are -0.045 $< (C_T+C'_T)/C_A <$ 0.159 (90\% C.L.).
 \item[Conclusions] The obtained limits are comparable to limits from other correlation measurements in nuclear $\beta$~decay and contribute to further constraining tensor coupling constants.
\end{description}
\end{abstract}


\maketitle

\section{Introduction}

In their original paper Lee and Yang \cite{Lee1956} formulated the most general $\beta$ decay Hamiltonian supposing only Lorentz invariance. Apart from the vector and axial-vector weak interaction components observed in nature it also contains scalar, tensor and pseudoscalar terms. The latter does not contribute to experimental observables in nuclear $\beta$-decay or neutron decay at the present level of precision as the pseudoscalar hadronic current vanishes in the non-relativistic treatment of nucleons. While scalar and tensor terms are components that can in principle occur, experimental limits from nuclear $\beta$ decay and neutron decay restrict their potential contribution to 7\% and 8\%, respectively (95.5\% C.L.) \cite{Severijns2006} (see also Ref.~\cite{Wauters2014}). Many recent experimental efforts try to improve the sensitivity to these non-standard model weak currents in searches for differences of experimental observables from their standard model predictions in both neutron decay \cite{Abele2008, Konrad2010, Dubbers2011, Mumm2011, Kozela2009, Kozela2012, Mund2013} and nuclear $\beta$ decay \cite{Severijns2006, Severijns2011, Pitcairn2009, Flechard2011, Vetter2008, Wauters2009, Wauters2010, Beck2011, Li2013}.

Measurements of the angular distribution of $\beta$ particles emitted in the decay of polarized nuclei are potentially very sensitive to deviations from the Standard Model \cite{Wauters2009,Wauters2010}. The angular distribution is given by \cite{Jackson1957}:
\begin{eqnarray}
\label{eqn:jtw}  W(\theta) \propto \left[ 1 + b \frac{m}{E_e} +
\frac{\bf{p}}{E_e} \cdot A \bf{J}  \right] ,
\end{eqnarray}
with $E_e$ and $\bf{p}$ the total energy and momentum
of the $\beta$~particle, $m$ the mass of the electron, $\textbf{J}$ the nuclear vector polarization and $b$ the Fierz interference term.
The actual quantity that is determined experimentally is not $A$ but
\begin{equation}
\label{actual-observable}
\tilde{A} = \frac{A} { 1 + \langle b^\prime \rangle }
\end{equation}
with $b^\prime \equiv (m / E_e) b$ and where $\langle \, \rangle$ stands for the weighted average over the observed part of the $\beta$ spectrum.

For an allowed pure Gamow-Teller (GT) $\beta$ transition this can be written as (assuming maximum parity violation and time-reversal invariance for the axial-vector part of the interaction) \cite{Jackson1957}:
\begin{equation}\label{eq:asym}
\begin{split}
 \widetilde{A}_{GT}^{\beta^\mp} & = A_{SM}\left[ 1 \mp \Re\left(\frac{C_TC'_T}{C^2_A}\right) - \frac{ \alpha Z m } {p_e}\Im\left( \frac{ C_T + C^{\prime}_T}{C_A} \right) \right] \\
 &~~~ \times \left[ 1 + \frac{1}{2}\frac{|C_T|^2 +|C'_T|^2}{C_A^2} \pm \frac{\gamma m}{E_e}\Re\left(\frac{ C_T + C^{\prime}_T}{C_A}\right)\right]^{-1} 
\end{split}
\end{equation}
with $C_T$, $C'_T$ and $C_A$ the coupling constants of the tensor and axial-vector parts of the weak interaction and with primed (unprimed) coupling constants for the parity conserving (violating) parts of the interactions, respectively. Further, the upper (lower) sign refers to $\beta^{-}$($\beta^{+}$) decay,  and $\gamma = [ 1 - ( {\alpha Z} )^2 ]^{1/2}$ with $\alpha$ the fine structure constant and $Z$ the atomic number of the daughter isotope. The Standard Model value of the asymmetry parameter for pure GT transitions is $A_{SM}=-1$ for $J \to J-1$, $A_{SM}=-1/(J+1)$ for $J \to J$ and $A_{SM}=J/(J+1)$ for $J \to J+1$ transitions.
As existing limits on the possible time-reversal violating tensor coupling constants are already at the 1\% level \cite{Huber2003}, the results of this paper are interpreted assuming time-reversal invariance. Eq.~(\ref{eq:asym}) can be further simplified by neglecting all second-order terms such as $|C_T|^2/|C_A|^2$, as they are known to be sufficiently small \cite{Severijns2006}. One then obtains for the first-order approximation
\begin{equation}\label{eq:asym_firstorder}
 \widetilde{A}_{GT}^{\beta^\mp} \simeq  \frac{A_{SM}} {1 \pm \frac{\gamma m}{E_e}\frac{ C_T + C^{\prime}_T}{C_A}} \equiv A_{SM} T ~\text{.}
\end{equation}
\noindent Any departure in the measured value of $\widetilde{A}$ from the Standard-Model value $A_{SM}$ is then sensitive to the tensor couplings $(C_T + C^{\prime}_T)$.
Further, with the factor $\gamma$ being of order unity, the sensitivity to these tensor couplings is inversely proportional to the $\beta$ energy and therefore is enhanced for low endpoint energy $\beta$ transitions.

As precisions of the order of 1\% or better can be obtained for the $\beta$-asymmetry parameter $\tilde{A}$, higher-order corrections to the value predicted by the Standard Model become significant (see e.g.\ Ref.~\cite{Wauters2010}). These recoil corrections, induced by the strong interaction between the quarks, and radiative corrections are discussed in Sec.~\ref{sec:recoil}.

The sensitivity of the low temperature nuclear orientation (LTNO) method \cite{Stone1986} for this type of measurement has been demonstrated recently in experiments with $^{60}$Co \cite{Wauters2010} and $^{114}$In \cite{Wauters2009}, which constitute the most precise measurements of this type in nuclear $\beta$ decay to date (see also Ref.~\cite{Severijns2005} for earlier work). Here we present the results of an experiment using the same method for the pure Gamow-Teller $\beta^{-}$ decay of $^{67}$Cu. The isotope $^{68}$Cu with a much higher $\beta$-endpoint energy, and therefore a much smaller sensitivity to tensor-type weak currents, was used for normalization purposes.

\section{Experimental method}
\subsection{The isotopes $^{67}$Cu and $^{68}$Cu}

\begin{figure}
\includegraphics[width=0.4\textwidth]{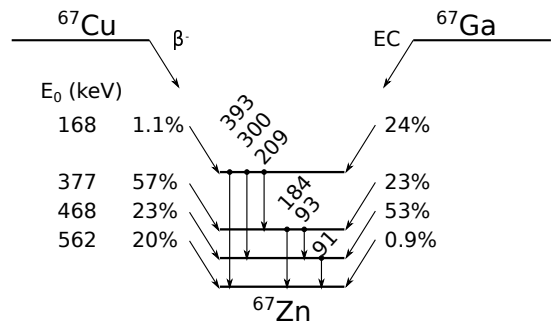}%
\caption{\label{fig:decay67Cu} Simplified decay scheme of $^{67}$Cu and $^{67}$Ga. For $^{67}$Cu the $\beta$ intensities and endpoint energies ($E_0$) are shown, while for $^{67}$Ga the EC transition intensities are indicated. The accompanying $\gamma$ rays in $^{67}$Zn are also shown. }
\end{figure}

\begin{figure}
\includegraphics[width=0.4\textwidth]{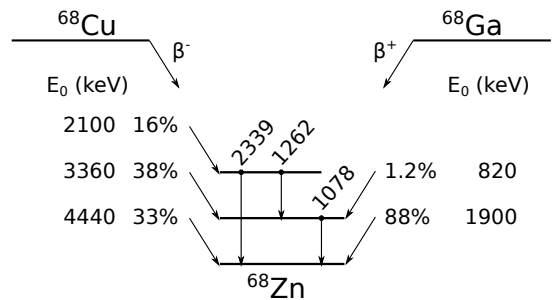}%
\caption{\label{fig:decay68Cu} Simplified decay scheme of $^{68}$Cu and $^{68}$Ga with $\beta$ branch intensities, $\beta$-endpoint energies ($E_0$) and accompanying $\gamma$ rays being shown. For clarity, two weak branches in the decay of $^{68}$Cu (with total intensity of 5\%) decaying to levels in $^{68}$Zn between the two higher lying ones, as well as other, low-endpoint branches are not shown.}
\end{figure}

The isotope $^{67}$Cu has a spin-parity $I^\pi = 3/2^-$ and a magnetic moment
$\mu = +2.5142(6) \mu_{\text{N}}$ \cite{Vingerhoets2010}, with a half-life of 61.83(12)\,h
and a $\beta$-endpoint energy of 561.7\,keV ($\log{ft}$~=~6.3) \cite{Junde2005}. It decays via a pure Gamow-Teller transition to the ground state of  $^{67}$Zn. A simplified
decay scheme is shown in Figure~\ref{fig:decay67Cu}.
The isotope $^{68}$Cu has a spin-parity $I^\pi=1^+$ and a magnetic moment
$\mu = 2.3933(6)\mu_{\text{N}}$ \cite{Vingerhoets2010}, with a half-life of
31.1(15)\,s and a $\beta$-endpoint energy of 4.440\,MeV \cite{McCutchan2012}.  It decays via a pure Gamow-Teller transition to the ground state of  $^{68}$Zn.
A simplified decay scheme is shown in Figure~\ref{fig:decay68Cu}.

\subsection{Sample preparation}

Both isotopes were produced at ISOLDE (CERN)
with a 1.4\,GeV proton beam from the Proton Synchrotron Booster 
bombarding a ZrO$_{2}$ felt target (6.3 g Zr/cm$^{2}$) \cite{Koster2003}
connected to RILIS \cite{Koster2000,Weissman2002a}, which provided
the required element selectivity.
After ionization and acceleration to 60 keV, the ion beams
were mass-separated by the General Purpose Separator, transported
through the beam distribution system, and implanted into a
polished and annealed 99.99\% pure Fe foil (size 13.1\,mm~$\times$~15.5\,mm,
original thickness 250\,$\mu$m) soldered onto the cold finger of the NICOLE
$^{3}$He-$^{4}$He dilution refrigerator \cite{Schloesser1988, Wouters1990}.
The foil had been polished with diamond-base paste with grain sizes of
3\,$\mu$m and 1\,$\mu$m. It is estimated that this procedure
reduced the thickness of the foil to 90~$\pm$~20~$\mu$m.
The average implantation depth of $^{67,68}$Cu ions with an energy
of 60\,keV in this foil was calculated to be approximately 20\,nm.
The corresponding energy loss for $\beta$ particles leaving the foil is
on the order of 100\,eV, which is negligible in comparison to the
$\beta$-endpoint energies of the isotopes studied here. Prior to the experiment
the foil was magnetized to saturation in a 0.5\,T horizontal external
magnetic field generated by a superconducting split-coil magnet.
During the experiment this field was reduced to
$B_\text{app} = 0.100(2)$\,T so as to minimize its
influence on the trajectories of the $\beta$ particles while still 
maintaining the magnetization of the foil.

\subsection{Detection setup}

\begin{figure}
\includegraphics[width=0.4\textwidth]{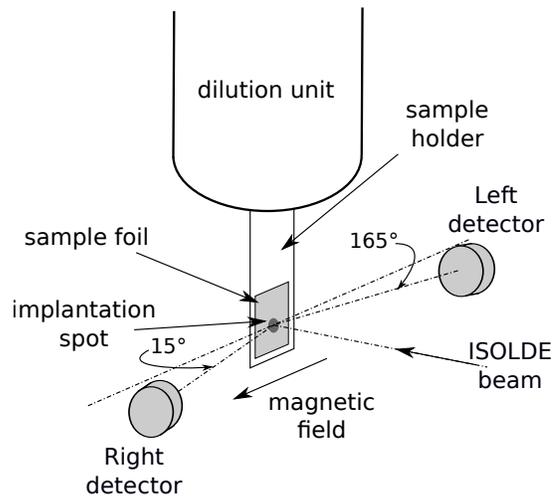}%
\caption{\label{fig:nicole_scheme} Schematic view of the NICOLE low temperature nuclear orientation setup, located at ISOLDE, CERN. The sample foil is soldered to the sample holder, which is in direct thermal contact with the dilution unit. The incoming radioactive beam from ISOLDE is perpendicular to the sample foil. The Left (Right) particle detector is directly facing the sample foil containing the implanted nuclei at an angle of $15^{\circ}$ ($165^{\circ}$).}
\end{figure}

The angular distribution of the electrons emitted during the
$\beta^-$ decay of $^{67}$Cu and $^{68}$Cu was observed with two planar
high purity Ge (HPGe) particle detectors with a sensitive diameter of
16 mm and a thickness of 4 mm produced in the Nuclear Physics
Institute in \v{R}e\v{z}, Prague \cite{Venos2000, Zakoucky2004, Soti2013}.
Their thickness was selected to fully stop the high endpoint energy
$\beta$ electrons from $^{68}$Cu. They were installed at a distance of 37.7~mm from
the sample, inside the 4 K radiation shield of the refrigerator (see Fig.~\ref{fig:nicole_scheme} for details).
The actual operating temperature of the detectors was approximately 10\,K. The fact
that they were looking directly at the radioactive source assured good
counting rates and at the same time avoided effects of scattering or
absorption (or energy loss) of the $\beta$ particles in the material between the detector and the radioactive source.
In order to also minimize effects of scattering in the Fe foil itself the detectors were
mounted to view the foil surface (which was parallel to the magnetic field and perpendicular to the incoming beam direction)
under an angle of 15$^{\circ}$, corresponding to detection angles of
$15^{\circ}$ (Right detector) and $165^{\circ}$ (Left detector)
with respect to the orienting magnetic field.
Thin isolated copper wires ($\sim$13\,cm long)
connected the detectors to the preamplifiers outside the refrigerator,
resulting in an energy resolution of $\sim$3 keV.
Thin wires were used in order to minimize the heat load from room
temperature components to the low temperature parts.

Apart from these particle detectors, large-volume HPGe detectors
for detection of $\gamma$ rays were installed outside the
refrigerator at $0^{\circ}$ and $180^{\circ}$ with
respect to the orientation axis (magnetic field axis). The
energy resolution of these was $\sim$3\,keV for the 1332\,keV
$\gamma$ line of $^{60}$Co.

All data were corrected for the dead
time of the data acquisition system using a precision pulse
generator. The energy calibration of all detectors was performed with the 136.5 and
692.4\,keV $\gamma$ lines of $^{57}$Co and the 184.6\,keV $\gamma$
line of $^{67}$Cu. During the on-line measurement of $^{68}$Cu additional
calibration was performed with the 1077.7\,keV $\gamma$ line of
$^{68}$Cu and the 511\,keV positron annihilation line from the
$\beta^+$ decay of $^{68}$Ga.

\subsection{Angular distributions}

The angular distribution of radiation emitted by oriented nuclei
can be described by the function \cite{Krane1986}
\begin{equation}
    W\left( \theta  \right) = 1 + f
    \sum\limits_\lambda  {B_\lambda
    U_\lambda  A_\lambda  Q_\lambda  P_\lambda  \left( {\cos \theta }
    \right)} ~ .
    \label{eq:GeneralAngularDistribution}
\end{equation}
Here $f$ represents the fraction of nuclei that
experience the full orienting hyperfine interaction $\mu B_{\text{tot}}$ (with $\mu$
the nuclear magnetic moment and $B_{\text{tot}}$ the total magnetic field
the nuclei experience; see Sec.~\ref{sec:magfield}), while
it is supposed that the rest, $(1-f)$, experiences no interaction at all. $B_\lambda$
are the nuclear orientation parameters which depend on $\mu$ and $B_{\text{tot}}$,
but also on the temperature of the sample $T$, the initial spin $I$, the half-life,
and the relaxation constant $C_K$ of the oriented state. The
$U_\lambda$ are deorientation coefficients which account for
the effect of unobserved intermediate radiations, while
$A_\lambda$ are the directional distribution coefficients which
depend on the properties of the observed radiation itself.
Finally, $Q_\lambda$ are solid angle correction factors and
$P_\lambda( \cos \theta )$ are the Legendre polynomials. The angle
$\theta$ is measured with respect to the orientation axis.

For $\gamma$ rays only $\lambda$ even terms occur. For allowed
$\beta$ decays only the $\lambda=1$ term is present
and Eq.~(\ref{eq:GeneralAngularDistribution}) transforms to
\begin{equation}\label{eq:AngularDistributionElectrons}
    W \left( \theta  \right) = 1 + f \frac{v}{c} \tilde{A} P Q_1 \cos
    \theta ~ ,
\end{equation}
where ${v / c}$ is the $\beta$ particle velocity relative to the
speed of light, $\tilde{A}$ is as defined in Eq.~(\ref{eq:asym}) and $P$ is the degree of nuclear polarization. Note that the
product $\frac{v}{c} \tilde{A} P$ is equal to $B_1 A_1$ in Eq.~(\ref{eq:GeneralAngularDistribution}).

Experimentally the angular distribution is obtained as
\begin{equation}\label{eq:ExpCountRatio}
W(\theta) = \frac{N_{\text{cold}}(\theta)}{N_{\text{warm}}(\theta)} ,
\end{equation}
with $N_{\text{cold,warm}}(\theta)$ the count rates when the
sample is ``cold'' (about 10 mK; oriented nuclei) or ``warm'' (about 1K;
unoriented nuclei). Such a ratio is then constructed for each detector.

In on-line experiments, where the count rates vary with beam intensity,
it is customary to construct a double ratio, combining count rates in
two different detectors in order to eliminate effects of beam intensity
fluctuations and avoid the need to correct for the decay of the isotope. 
In the present work the double ratio
\begin{equation}\label{eq:R_doubleratio}
    R =
        \frac {W(15^{\circ})} {W(165^{\circ})}
      = \frac {
          \left[
           \frac { N(15^{\circ})} {N(165^{\circ})}
          \right]_{\text{cold}}}
        { \left[
           \frac { N(15^{\circ})} { N(165^{\circ})}
          \right]_{\text{warm}} }
\end{equation}
was used for the $\beta$ particles of $^{68}$Cu. 

For the $\beta$-asymmetry measurement of $^{67}$Cu, and for the temperature determination with $^{57}$Co, no
double ratios were necessary as these isotopes were not produced on-line and
their half-lives are known with sufficient precision to be
taken into account in the data analysis.

\subsection{\label{sec:magfield}Total magnetic field}

In LTNO experiments the total magnetic field the nuclei feel when
implanted into a ferromagnetic host foil has three components:
\begin{equation}\label{eq:Btot}
B_{\text{tot}}=B_{\text{hf}}+B_{\text{app}}(1+K)-B_{\text{dem}}
\end{equation}
\noindent with $B_{\text{hf}}$ the hyperfine magnetic field,
$B_{\text{app}}$ the externally applied magnetic field,
$B_{\text{dem}}$ the demagnetization field, and $K$ the Knight shift.
In all measurements the external field was set to 0.100(2)\,T. The hyperfine field
of dilute Cu impurities in Fe was recently determined to be
-21.794(10)\,T \cite{Golovko2011}. The demagnetization field for
our foil was calculated to be 0.018(4)\,T \cite{Chikazumi1964}, with a $20\%$ error to
account for the approximations made when deriving the analytical
formulas of the
demagnetization field. The Knight shift for copper in iron has never
been determined at low temperatures, but a conservative upper limit of $5\%$ corresponds to
a 0.005\,T systematic uncertainty on the total magnetic field
\cite{Golovko2011}. The total field the nuclei experience then amounts to
$B_{\text{tot}}$~=~-21.712(12)\,T.

\subsection{\label{sec:thermometry}Thermometry}
The temperature of the oriented sample was maintained in the region between
8\,mK and about 60\,mK and measured by monitoring the intensity of the 136\,keV
$\gamma$ line of a $^{57}$Co\underline{Fe} nuclear orientation thermometer
\cite{Marshak1986} soldered onto the back side of the sample holder,
with two large-volume $\gamma$ HPGe detectors installed outside of the refrigerator.
The $^{57}$Co activity had been diffused into an iron foil similar
to the sample foil and prepared in the same way. Calibration of this
$^{57}$Co\underline{Fe} thermometer against a $^{60}$Co\underline{Co} single crystal thermometer
resulted in a fraction of $94.3(4)$\% of the $^{57}$Co nuclei feeling the full
orienting hyperfine field in the foil.

\subsection{\label{sec:relaxation}Nuclear spin-lattice relaxation}

Due to its short half-life of 31\,s the $^{68}$Cu had to be implanted continuously
and the anisotropy observed in on-line conditions. As a consequence the value for the nuclear polarization
$P = -\sqrt{(I+1)/(3I)} B_1$ results from an equilibrium between
implantation of unoriented (warm) nuclei and the decay of (partially)
relaxed nuclei that
may or may not yet have reached thermal equilibrium (i.e.\ the full
orientation corresponding to the sample temperature). This nuclear spin-lattice relaxation effect
needs to be taken into account when the half-life and the relaxation time are of the same order
of magnitude. For the case of a dominant magnetic-dipole relaxation mechanism, as applies to 3d element impurities
in an Fe host \cite{Funk1999}, the so-called Korringa constant, $C_K$, describes the relaxation process.
Further, if $C_K$ is known for one isotope of a given element in a specific host material
it can be calculated for the other isotopes using the relation \cite{Shaw1989}:
\begin{equation}\label{eq:ck}
\frac{\mu^2 C_K}{I^2} = \text{const}~.
\end{equation}
$C_K$ has been measured in the past for two other Cu isotopes, i.e.\ $^{62}$Cu \cite{Golovko2006} and $^{63}$Cu \cite{Kontani1972}, in iron and at $B_{ext}=0.1$\,T.
The former measurement used the technique of LTNO at milliKelvin temperatures while the latter was a spin-echo experiment at 4.2\,K.
To estimate the $C_K$ of the isotopes of interest in this work we use the result of the $^{62}$Cu measurement,
since the experimental conditions were very similar to ours. The spins, magnetic moments and measured or
calculated $C_K$ values for the different Cu isotopes studied here are listed in Table~\ref{tab:estimated_ck}.
\begin{table}
\caption{\label{tab:estimated_ck} Calculated values of $C_K$ for all Cu isotopes studied here, using Eq.~(\ref{eq:ck}) and the $C_K$ value measured previously for $^{62}$Cu. The uncertainties are dominated by the uncertainty of the measured $C_K$ value for \textsuperscript{62}Cu. All magnetic moment values are from Ref.~\cite{Vingerhoets2010}.}
\begin{ruledtabular}
\begin{tabular}{c c c c c}
~ & {$^{62}$Cu} & {$^{67}$Cu} & {$^{68}$Cu} & {$^{68\text{m}}$Cu} \\
\hline
$\mu$ ($\mu_\textrm{N}$) & 0.3809(12) & 2.5142(6) & 2.3933(6) & 1.1548(6) \\
$I$ ($\hbar$) & 1 & 3/2 & 1 & 6 \\
$T_{\text{int}}$ (mK)\footnote{using the hyperfine field for Cu in an Fe host from \cite{Golovko2011} and Eq.~(\ref{eq:t_int})} & 3.0 & 13.3 & 19.0 & 1.5 \\
$C_K$ (s K) & 4.34(25)\footnote{from Ref.~\cite{Golovko2006}} & 0.225(13) & 0.110(6) & 17.0(10)  \\
\end{tabular}
\end{ruledtabular}
\end{table}
In conventional nuclear magnetic resonance (NMR) at lattice temperatures $T \geq 1$\,K,
spin-lattice relaxation leads to an exponential time dependence of
the signal and a relaxation time $T_1$ can be defined unambiguously. Such experiments are always performed in the high-temperature limit,
i.e.\ $T \gg I T_{\text{int}}$ with the interaction temperature $T_{\text{int}}$ given by the nuclear level splitting:
\begin{equation}\label{eq:t_int}
T_{\text{int}} = |\mu B_{\text{tot}}/k_{\text{B}} I|
\end{equation}
(with $k_{\text{B}}$ the Boltzmann constant), so that for metallic
samples the Korringa law, i.e.\ $C_K \, = \, T_1 ~ T$, is then valid.

The relaxation behavior probed via the $B_1$ orientation
parameter in the $\beta$ decay of $^{67,68}$Cu is the same as in conventional NMR. However, since
$T_{\text{int}}$($^{68}$Cu\underline{Fe}) = 19\,mK and
our data were taken at temperatures between about 18\,mK and
8\,mK, the low temperature limit, $T \leq I T_{\text{int}}$, now applies.
Experimentally it was observed \cite{Klein1986} that the
angular distribution in this case, to first order, still has a single
exponential behavior that can be characterized by a constant ``relaxation''
time, $T_{\mu}$. This $T_{\mu}$ provides a good estimate of the time
required to reach thermal equilibrium and is given by
$T_{\mu} = C_K / I T_{\text{int}}$. Under these conditions the parameter $C_K$ is called
the ``relaxation constant'' to avoid the impression that
the Korringa law has been assumed in the data analysis.

In the low temperature limit one calculates $T_{\mu} \simeq 6$\,s for
$^{68}$Cu\underline{Fe} which is of the same order of magnitude as
the half-life of 31\,s, so that relaxation effects (modifying the $B_1$ parameter) have to be
taken into account in fitting the $\beta$ anisotropy for $^{68}$Cu (see Sec.~\ref{sec:fract68cu}).

For $^{67}$Cu\underline{Fe}, the factor $I T_{\text{int}} =20$\,mK corresponds to
$T_{\mu}\simeq 11$\,s. However, as the $^{67}$Cu measurement was performed off-line,
relaxation phenomena were only an issue when 
the sample temperature was changed during the measurement. As data
were collected throughout the experiment in 300\,s bins,
neglecting the first bin after a change in
sample temperature assured that relaxation effects did not play a role
in the analysis of the data for $^{67}$Cu.

\section{Data taking}
\subsection{$\beta$ spectrum of $^{67}$Cu and beam purity }

\begin{figure}
\includegraphics[width=0.5\textwidth]{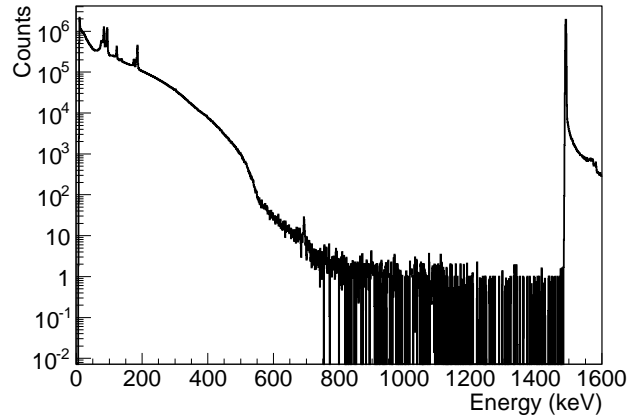}%
\caption{\label{fig:67Cu_beta}The $\beta$ spectrum of $^{67}$Cu as observed by the Left
particle detector. The very weak $\gamma$ line at 692.4\,keV is from the $^{57}$Co thermometer, while the large peak at the end of the spectum is the pulser peak.}
\end{figure}

The observed $\beta$ spectrum of $^{67}$Cu is shown in Fig.~\ref{fig:67Cu_beta}.
The peak at the end of the spectrum is from the pulser that was used to
monitor the dead time of the data acquisition system. Its high-energy tail is the result of detector event pile-up. Because of the rather long
half-life of $^{67}$Cu the measurements with this isotope were performed in
semi-online mode, meaning that a $^{67}$Cu sample was collected
by implanting the ISOLDE beam in the iron sample foil for several hours,
after which the beam line was closed and measurements were started.
To extract the asymmetry parameter for the $3/2^-\rightarrow 5/2^-$ pure 
Gamow-Teller $\beta$ transition of $^{67}$Cu only the upper part
of the spectrum, i.e.\ between 469\,keV and 562\,keV, was considered.
This region does not contain events from the other $\beta$ branches with lower endpoint energy (see Fig.~\ref{fig:decay67Cu}).
The region between 410 and 469\,keV, which also contains counts from the second most energetic $\beta$ branch, was analyzed as well (see Sec.~\ref{sec:twobranches}).

The very weak $\gamma$ line at 692.4\,keV (Fig.~\ref{fig:67Cu_beta}) is from the decay of the $^{57}$Co
nuclear thermometer and does not disturb the $\beta$-asymmetry measurement.
Possible contamination from $^{67}$Ga ($t_{\text{1/2}} = 78.3$\,h) in the beam was not important
since $^{67}$Ga decays purely by electron-capture and has no intense
$\gamma$ rays in its decay that could contaminate the $\beta$ spectrum
in the region of interest (see Fig.~\ref{fig:decay67Cu}).

\subsection{$\beta$ spectrum of $^{68}$Cu and beam purity }

\begin{figure}
\includegraphics[width=0.5\textwidth]{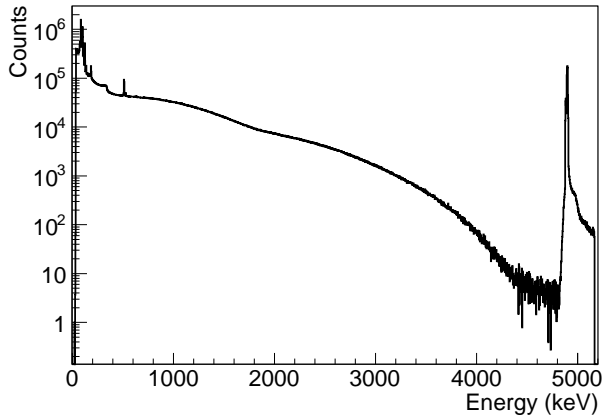}%
\caption{\label{fig:68Cu_beta}The $\beta$ spectrum of $^{68}$Cu as observed by the Right particle detector. The 511\,keV annihilation peak is from the $\beta^+$ decay of $^{68}$Ga. Above the high-energy end of the spectrum the pulser peak is visible. }
\end{figure}

The $\beta$ spectrum of $^{68}$Cu is shown in Figure~\ref{fig:68Cu_beta}. Due to the
short half-life a full on-line measurement was necessary for this
isotope. In this case ions were continuously implanted into the iron foil while
performing the measurement.
The anisotropy of the $\beta$ branch with the largest endpoint energy was used to obtain
the fraction $f$ of Cu nuclei experiencing the full orienting magnetic field via
Eq.~(\ref{eq:AngularDistributionElectrons}). As both $^{67}$Cu and $^{68}$Cu were implanted at low dose
(i.e.\ well below 5~$\times$~10$^{13}$\,atoms/cm$^2$) and at low temperature, a large fraction of atoms are expected to occupy
good lattice sites with no damage to the iron foil (i.e.\ change of the fraction) during
the implantations \cite{Walle1986, Herzog1985, Herzog1987, Dammrich1988}. As both isotopes were, moreover, implanted into the same
iron foil, the fraction obtained for $^{68}$Cu is valid for $^{67}$Cu as well.
As the second most energetic $\beta$ branch of $^{68}$Cu has an endpoint energy
1078\,keV lower than the most energetic one, the region of interest for analysis could be restricted to the upper $\beta$ branch
(i.e.\ from 3.39 to 4.44\,MeV) without a significant loss of statistics.
Further, due to the high endpoint energy of 4.44\,MeV the region of interest is completely free of events coming from other isotopes.

The amount of $^{68}$Ga contamination ($t_{\text{1/2}}= 67.6$\,min) in the beam was estimated from observed $\beta$ spectra 
when no $^{68}$Cu was present in the system, i.e.\ after the beam was stopped, and
was found to be approximately the same intensity as $^{68}$Cu. However, due to its low
$\beta$-endpoint energy of 1.9\,MeV it did not disturb the measurement on the highest-energetic $\beta$ branch of $^{68}$Cu.

The amount of the metastable $^{68\text{m}}$Cu \cite{Koster2000a} in the beam was estimated from the intensity of the 526\,keV
gamma line and found to be around 10\%. With a half-life of 3.75\,min
and a magnetic moment of 1.1548(6)\,$\mu_\text{N}$ \cite{Vingerhoets2010} this isotope
also becomes oriented. It further decays mainly (branching ratio of $86\%$) via
an internal transition to the $^{68}$Cu ground state. The effect of this is discussed in
Sec.~\ref{sec:68cu_errors}. The remaining $16\%$ decay via $\beta^{-}$ decays
that feed the higher-lying excited states in $^{68}$Zn and have $\beta$-endpoint energies below
2.8\,MeV. This is well below the region of interest (i.e.\ from 3.39 to 4.44\,MeV)
for the determination of the fraction $f$ of $^{68}$Cu at good lattice sites.

\subsection{Data taking sequence}
\begin{figure}
\includegraphics[width=0.5\textwidth]{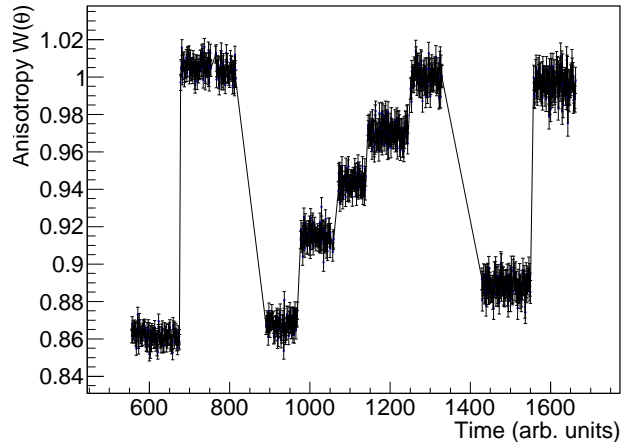}
\caption{\label{fig:67Cu_measurement_2007} Anisotropy of the 136\,keV $\gamma$ line
of the $^{57}$Co thermometer (count rate normalized to isotropic ``warm'' data, Eq.~(\ref{eq:ExpCountRatio})) during the $^{67}$Cu measurement, showing the six ``cold'' and three ``warm'' data sets. Every data point represents a measurement time of 300\,s.}
\end{figure}

The experimental campaign started with a temperature test followed by the online measurement of
$^{68}$Cu, which consisted of one cooling cycle with four different temperature points. After that the $^{67}$Cu
sample was collected. The off-line measurements with this isotope included two cooling cycles with data being collected at six temperatures where the nuclei were oriented
and three measurement periods with isotropic emission (unoriented nuclei). The temperature for each 300\,s measurement was
determined from the anisotropy of the 136\,keV $\gamma$ ray of the $^{57}$Co nuclear thermometer. Figure~\ref{fig:67Cu_measurement_2007}
shows the anisotropy of this $\gamma$ line during the measurements with $^{67}$Cu.

\subsection{Further steps}
After identifying time periods with constant
temperature the experimental spectra recorded at each temperature were summed.
The regions of interest for analysis were divided into several energy bins to check for possible energy-dependent effects in the analysis.
The integrals of the summed experimental spectra in these different energy bins
(providing $N_{\text{cold}}$ for different temperatures, and $N_{\text{warm}}$) were corrected for dead time, pile-up, and total measurement time,
as well as for the decay of $^{67}$Cu during the measurement. The dead time was deduced from the intensity of
the pulser peak and its tail, while the pile-up magnitude (see Sec.~\ref{sec:err_pileup}) was obtained from the pulser tail to peak ratio.

\section{\label{sec:simulations}Simulations}

In the analysis extensive use was made of the Geant4-based Monte Carlo code \cite{Agostinelli2003} that was developed specifically
for this type of experiment \cite{Wauters2009b,Soti2013} and which was modified for the geometry of the
NICOLE low temperature nuclear orientation setup. A detailed description of the setup, as well
as of the magnetic field generated by the split-coil superconducting magnet around the sample,
was implemented in the code. The field map was provided by the magnet manufacturer, Oxford
Instruments\textregistered.

As in previous work \cite{Wauters2009,Wauters2010} the main role of the simulation code was to account for
the effects of electron (back)scattering in the sample foil, on the detectors and the environment,
as well as to deal with the influence of the external magnetic field on the electron trajectories.
It thus provides a value for $\tilde{Q}=\frac{v}{c}Q_1\cos{\theta}$ (see Eq.~\ref{eq:AngularDistributionElectrons}) for each temperature point.

Geant4 has built-in modules to handle the decay of radioactive isotopes which include the full decay scheme.
However, our simulation code had to be used outside this framework, therefore only the three most intense $\beta$ branches together with their accompanying
$\gamma$ rays (see Tab.~\ref{tab:sim_branches}) were included in the simulations. The other $\beta$ branches did not affect the respective
regions of interest and were therefore omitted. Emission asymmetries of  $\gamma$ rays were not included either
as these also do not affect the region of interest in the $\beta$ spectra. The least energetic of the three $\beta$ branches considered for $^{67}$Cu has an
unknown $\tilde{A}$ coefficient (the Fermi/Gamow-Teller mixing ratio is not known) and was therefore simulated
isotropically.
\newcolumntype{y}{D{.}{.}{1.3}}
\begin{table}[]
\caption{\label{tab:sim_branches}The simulated $\beta$ branches for $^{67}$Cu \cite{Junde2005} and $^{68}$Cu \cite{McCutchan2012}.
Endpoint energies ($E_0$), relative intensities (Int.) and the $\tilde{A}$ parameters are shown. The uncertainties on the $\beta$ branch intensities of $^{67}$Cu are based on the corresponding $\gamma$ intensities.}
\begin{ruledtabular}
\begin{tabular}{c c y c c c y}
\multicolumn{3}{c}{$^{68}$Cu} & & \multicolumn{3}{c}{$^{67}$Cu} \\
\cline{1-3} \cline{5-7}
& & & & & & \\[-3mm]

 $E_0$ (MeV) & Int.\ ($\%$) & \multicolumn{1}{c}{$\tilde{A}$} & & $E_0$ (keV) & Int.\ ($\%$)& \multicolumn{1}{c}{$\tilde{A}$} \\
\hline
 4.44 & 33(4) & -1.0 & & 561.7 & 20.0(3) & 0.6 \\
 3.36 & 38(4) & 0.5 & & 468.4 & 21.8(3) & -1.0 \\
 2.10 & 16.0(12) & 0.5 & & 377.1 & 57.1(3) & 0.0\footnote{unknown F/GT mixing ratio, no anisotropy simulated} \\
\end{tabular}
\end{ruledtabular}
\end{table}

\begin{figure}
\includegraphics[width=0.5\textwidth]{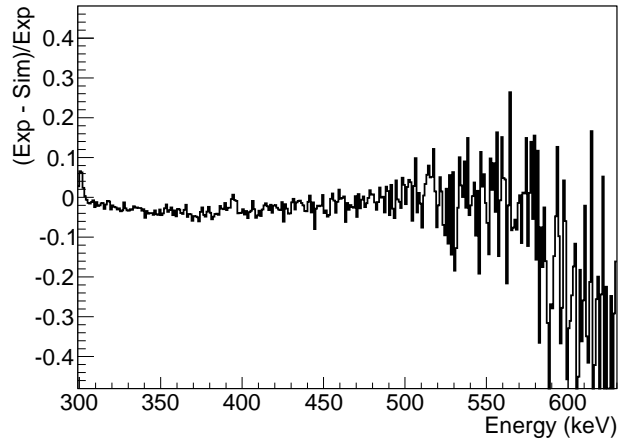}
\caption{\label{fig:67Cu_warm_diff} Difference between the experimental and simulated isotropic spectrum of $^{67}$Cu (Left detector), normalized to the experimental spectrum. The region of interest
is between 470 and 560\,keV, where $\chi^2_{red}=1.24$ for 91 degrees of freedom. }
\end{figure}

Detailed models of the HPGe particle detectors were implemented and verified by test measurements \cite{Wauters2009thesis,Soti2013}.
The reliability of the simulations was further verified by comparing experimental and simulated isotropic data for both
$^{67}$Cu and $^{68}$Cu. For the former, we find agreement between experiment and simulations to
better than 3-4\% over the energy region between 300\,keV and the endpoint at 562\,keV
(Fig.~\ref{fig:67Cu_warm_diff}), and a $\chi^2_{red}=1.24$ for 91 degrees of freedom for the region
of interest, i.e.\ between 470 and 560\,keV.
For $^{68}$Cu the agreement between experiment and simulation is somewhat worse but still better
than about 5\% everywhere from 2.00\,MeV up to about 0.20\,MeV from the endpoint where pile-up events start to dominate.

Simulations for the anisotropic (``cold'') data were performed for each of the six temperature sets, with the
corresponding degree of nuclear polarization and the Standard Model value for the $\beta$-asymmetry
parameter as input values. The value for the fraction of nuclei at good lattice sites was also incorporated
at this stage. Spectra for the nuclear thermometer $^{57}$Co were simulated as well and added to the histograms of $^{67}$Cu.
For $^{68}$Cu the nuclear spin-lattice relaxation was also taken into account (see Sec.~\ref{sec:fract68cu}).
Histograms were then constructed from the simulated data, with the experimentally
determined dead time and detector pulse pile-up taken into account

\section{$^{68}$C\lowercase{u} analysis}

\subsection{\label{sec:fract68cu}Fraction determination}

\begin{figure}
\includegraphics[width=0.5\textwidth]{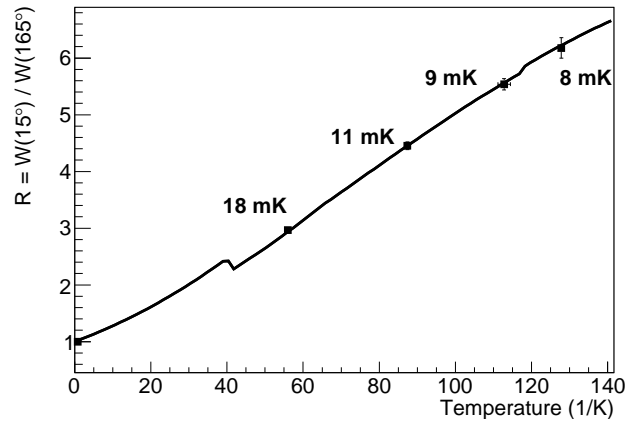}
\caption{\label{fig:68Cu_asymm} Fit of the fraction $f$ to the double ratio $R$ (Eq.~\ref{eq:R_doubleratio}) determined at different temperatures, for the energy range of 3600-3700\,keV. The fit resulted in $f$~=~1.002(4) with $\chi^2_{red}$~=~1.1 for 4 degrees of freedom. The fit function is not continuous because of the different values of $\tilde{Q}$ at different temperature points.}
\end{figure}

Analysis of the $\beta$ spectra for $^{68}$Cu revealed a significant number of events close to and beyond the $\beta$ endpoint as a result of detector event pile-up due to the high count rate (several kHz) during the measurement. This significantly reduced the signal-to-noise ratio near the $\beta$ endpoint, leading us to limit the region of interest (ROI) for analysis to 3400-4000\,keV. For this energy region comparison of experimental and simulated isotropic (``warm'') spectra revealed a $\chi^2_{\text{red}} = 0.97$ for 251 degrees of freedom. This ROI was then divided into six energy bins, each 100\,keV wide, and experimental double ratios $R$ (see Eq.~\ref{eq:R_doubleratio}) were determined for each energy bin and for each temperature point:
\begin{equation}\label{eq:Rdoubleratio}
 R = \frac{W(15^{\circ})}{W(165^{\circ})} = \frac{1+f\tilde{A}B_1(T,C_K)\tilde{Q}(15^{\circ})}{1+f\tilde{A}B_1(T,C_K)\tilde{Q}(165^{\circ})} \text{,}
\end{equation}
(see Fig.~\ref{fig:68Cu_asymm}).

It has been demonstrated with $^{62}$Cu \cite{Golovko2006} that a simultaneous fit of the
fraction $f$ and the relaxation constant $C_K$ (see Sec.~\ref{sec:relaxation}) to the anisotropy curve is possible.
Being independent of temperature, the parameter $f$ determines only the size of
the anisotropy effect $R$, while $B_1$ is temperature dependent and therefore also determines the shape
of the anisotropy $R$ versus temperature. The calculation of $B_1$ takes into account
the fact that upon implantation the nuclei are not oriented. In the time following the 
implantation they relax to thermal equilibrium with
the cold lattice, with the size of $B_1$ being determined by a
competition between nuclear decay and spin-lattice relaxation. In order to take into account the effect of the spin-lattice
relaxation (i.e.\ $C_K$) in Eq.~(\ref{eq:Rdoubleratio})
the orientation parameter $B_1$ was expressed as $B_1(\text{sec}) = \rho_1
B_1(\text{th})$, with $\rho_1$ the ratio of the observed orientation parameter  for
the nuclear ensemble when in secular equilibrium (i.e.\ still relaxing)
and the thermal equilibrium orientation parameter $B_1(\text{th})$. The $\rho_1$
attenuation coefficients were determined according to the
procedure outlined in Ref.~\cite{Venos2003} (see also
Ref.~\cite{Shaw1989}) and taking into account the observed
temperature for each individual data point.

In fitting the experimental data we used $\mu(^{68}$Cu) = 2.3933(6)\,$\mu_{\text{N}}$ (Tab.~\ref{tab:estimated_ck}) and $B_{\text{tot}}=-21.712(12)$\,T (see Sec.~\ref{sec:magfield}). The $\tilde{Q}$ factor was obtained for each bin and for each temperature point using Geant4 simulations, its value ranging from 0.86 to 0.95. The Standard Model value of the $\beta$-asymmetry parameter of $^{68}$Cu, based on the spin sequence only, is $A_{SM}=-1$. However, higher-order corrections (see Sec.~\ref{sec:recoil} for details) modify this value to $-0.9900(12)$, determined at 3700\,keV. The energy dependence of these corrections (a 0.15\% change over the entire energy range of the ROI) was incorporated in the fit procedure.

Two-parameter fits for $f$ and $C_K$ to the double ratio $R$ were performed for all 6 energy bins in the region between 3400 and 4000\,keV. The weighted average of the results obtained for the Korringa relaxation constant yielded $C_K=0.114(29)$\,sK. As the two fit parameters are correlated, their respective uncertainties are significantly larger than expected from the available statistics. Since the fitted value of $C_K$ is in perfect agreement with the value of 0.110(6)\,sK calculated for $^{68}$Cu on the basis of the value previously measured for $^{62}$Cu in an Fe host (see Tab.~\ref{tab:estimated_ck}), $C_K$ was subsequently fixed to 0.110(6)\,sK and only the fraction $f$ fitted (Fig.~\ref{fig:68Cu_asymm}).

The sensitivity of the $\beta$-asymmetry parameter $\tilde{A}$ to possible tensor currents in the decay of $^{68}$Cu is relatively small due to the high energy region used for analysis, see Eq.~(\ref{eq:asym_firstorder}). Nevertheless, this sensitivity can be taken into account by using the notation $\tilde{A}=A_{SM}T_{68}$; see Eqs. (\ref{eq:asym}) and (\ref{eq:asym_firstorder}) for the expression for $T_{68}$. Inserting this expression for $\tilde{A}$ into Eq.~(\ref{eq:Rdoubleratio}) it becomes clear that a fit of the fraction $f$ essentially determines $\tilde{f} = fT_{68}$. This does not pose a problem, as the factor $T_{68}$ can be propagated to the final result obtained with $^{67}$Cu; see Sec.~\ref{sec:67cu_ratio}.

\begin{figure}
\includegraphics[width=0.5\textwidth]{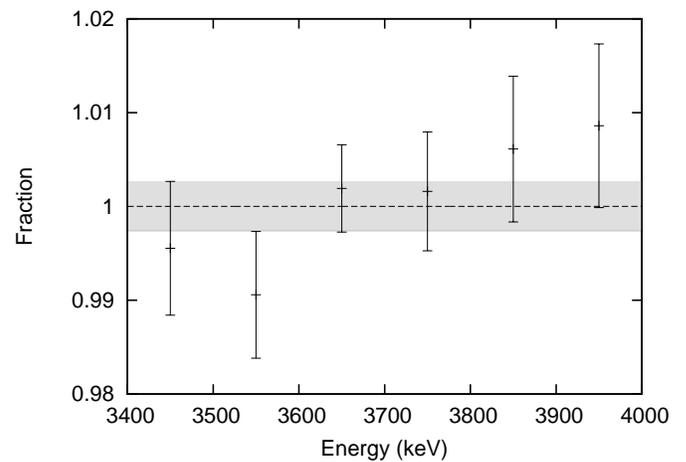}
\caption{\label{fig:68Cu_fract} Results of the fit of the fraction $\tilde{f}$ to the experimental double ratio $R$ in the different energy regions. The weighted average for the total region from 3400 to 4000\,keV is 1.0003(26), which is indicated by the dashed line and the gray band.}
\end{figure}

The results of the fit for the fraction $\tilde{f}$ for each of the 100\,keV wide energy bins are shown in Fig.~\ref{fig:68Cu_fract}, with their statistical error bars increased by a factor of $\sqrt{\chi^2_{red}}$ if $\chi^2_{red} > 1$. The final value for the fraction was determined as the weighted average in the energy region from 3400 to 4000\,keV, yielding $\tilde{f}$~=~1.0003(26), where the error is purely statistical. The observed energy dependence mainly stems from the $\tilde{Q}$ factors obtained by Geant4 simulations, and is discussed in detail in the following section together with the other systematic errors.

\subsection{\label{sec:68cu_errors}Error analysis for the fraction}

The systematic errors related to the determination of the fraction $f$ of nuclei at good lattice sites (i.e.\ feeling the full orienting interaction) are summarized in Tab.~\ref{tab:68cu_fraction_errors} and discussed in detail in the following paragraphs.
\begin{table}[]
\caption{\label{tab:68cu_fraction_errors} Contributions to the systematic uncertainty of the fraction determination from the decay of $^{68}$Cu. }
\begin{ruledtabular}
\begin{tabular}{l  l}
Effect & Value \\
\hline
relaxation constant $C_K$& 0.0072 \\
effective detector thickness & 0.0026 \\
 $^{68\text{m}}$Cu isomeric state & \textendash \\
thermometry & 0.0028 \\
$\mu B$ of $^{68}$Cu & 0.0003 \\
Geant4 & 0.0100 \\
recoil corrections (Sec.~\ref{sec:recoil}) & 0.0012 \\
\hline
\multicolumn{1}{r}{Total:} & 0.0130 \\
\end{tabular}
\end{ruledtabular}
\end{table}

\subsubsection{Relaxation}
The approximately 5\% relative uncertainty related to the calculated value $C_K$~=~0.110(6) used in fitting the $^{68}$Cu data induces a 0.0072 systematic error on the fraction.

\subsubsection{Effective detector thickness}
The thickness of the sensitive area of the planar HPGe $\beta$ detectors is determined as the total physical thickness of the detector (well known) minus the thickness of the rear dead layer which is estimated to be between 0.7 and 0.9\,mm \cite{Venos2000}.
Simulations were performed to check whether this influences the value of the fraction. No effect outside the statistical error bar (0.0026) of these simulations was found. Note that this effect is only relevant when dealing with high-energy electrons which require the full thickness of the detector in order to be stopped. Consequently the results for $^{67}$Cu, with less-energetic $\beta$ particles than $^{68}$Cu, are not affected.

\subsubsection{Orientation of the $^{68\text{m}}$Cu isomeric state}
As mentioned before, the 721\,keV isomeric state of $^{68}$Cu was also present in the beam, with an estimated relative intensity of 10\%. Its relatively long half-life of 3.75\,min and magnetic moment of 1.1548(6)\,$\mu_\text{N}$ \cite{Vingerhoets2010} cause these nuclei to become oriented as well. Since $^{68\text{m}}$Cu decays with a 86\% branching ratio via a $\gamma$ cascade to the ground state, this induces a slight orientation for the $^{68}$Cu nuclei in the sample that were produced by the decay of $^{68\text{m}}$Cu (note that the directly implanted $^{68}$Cu nuclei are initially not oriented). However, due to the interplay of the small interaction temperature ($T_{\text{int}}=1.5$\,mK) and large relaxation constant ($C_K=17.0(10)$\,sK, see Tab.~\ref{tab:estimated_ck}) of $^{68\text{m}}$Cu, the value of the $B_1$ parameter remains below 0.1 even for the lowest temperature of 8\,mK (i.e.\ for the largest degree of orientation). This small induced orientation in the ground state of $^{68}$Cu is further reduced by the de-orientation coefficient $U_1$ (Eq.~(\ref{eq:GeneralAngularDistribution})), which takes into account the $\gamma$ cascade connecting the ground states of $^{68\text{m}}$Cu and $^{68}$Cu and which has to be inserted into Eq.~(\ref{eq:AngularDistributionElectrons}) in this case. Since the M1/E2 mixing ratios of these $\gamma$ transitions are not known we use a worst case scenario, i.e.\ assuming mixing ratios such as to produce the largest possible values. In that case the deorientation coefficient is calculated to be $U_1 = 0.0053$ which, in combination with the calculated $B_1$ value yields induced $\beta$ anisotropies for $^{68}$Cu below $0.04\%$, which is negligible at the present level of precision.

\subsubsection{Thermometry}
Statistical errors related to the temperature determination were included in the fit (horizontal error bars in Fig.~\ref{fig:68Cu_asymm}). Systematic effects are discussed in Sec.~\ref{sec:err_thermo}, and induce a 0.0028 systematic error on the fraction.

\subsubsection{Hyperfine interaction $\mu B$}
The error on the hyperfine interaction experienced by the $^{68}$Cu nuclei has two components. One is from the nuclear magnetic moment $\mu$, the other from the hyperfine field of Cu impurities in the iron host. The relative error on the latter is larger and therefore dominates the total systematic error of 0.0003 from $\mu B$ on the fraction.

\subsubsection{\label{sec:68cu_err_geant4}Geant4}
It was already observed in previous measurements that the differences between simulated and experimental spectra of relatively high-energy electrons (e.g.\ the $\beta$ spectrum of $^{90}$Y) are slightly larger than for low energy electrons (e.g.\ the $\beta$ spectrum of $^{85}$Kr) \cite{Soti2013}. This leads us to associate the apparent energy dependence of the fraction determination (see Fig.~\ref{fig:68Cu_fract}) with remaining deficiencies of the Geant4 simulations for $\beta$ particles with energies in the MeV region. As the two extreme values in Fig.~\ref{fig:68Cu_fract} both differ about 0.01 from the average value, an additional 0.01 systematic error was assigned to this effect.

\subsubsection{Other effects and result for the fraction}
The fact that $R$ is a double ratio assures that the majority of all temperature independent systematic effects are greatly reduced, or even cancel. Among these are geometrical uncertainties, such as the detector positions, together with the uncertainty related to the position of the implanted $^{68}$Cu nuclei in the Fe foil (see Sec.~\ref{sec:err_beamspot}). Other remaining systematic effects were accounted for by increasing the error bars on the fit results for the fraction by the factor $\sqrt{\chi^2_{red}}$ (its value ranging from 0.7 to 3.9) for the measurements performed at different temperatures.

Combining then the systematic and statistical uncertainties one arrives at $\tilde{f}$~=~1.0003(26)$_{stat}$(130)$_{syst}$.

\section{$^{67}$C\lowercase{u} analysis}

In order to be able to check for possible energy-dependent systematic effects in the analysis of the data taken with $^{67}$Cu, the region of interest (ROI) was also divided into several energy bins. Analysis showed that, similar to the case of $^{68}$Cu, the ROI had to be reduced to 470-510\,keV because of a significant number of detector pile-up events near the $\beta$ endpoint. The lower edge of the ROI is limited by the presence of another $\beta$ branch with endpoint energy $E_0=469$\,keV (see Fig.~\ref{fig:decay67Cu} and Tab.~\ref{tab:sim_branches}). The energy region between 410 and 469\,keV has been analyzed as well, but the precision obtained in this case is worse due to the presence of counts from the other $\beta$ branch. This will be discussed in detail in Sec.~\ref{sec:twobranches}.

\subsection{\label{sec:67cu_ratio}Extraction of the beta-asymmetry parameter}

Having obtained both the experimental and simulated integrals ($N_{\text{cold,warm}}$) for all energy bins, the ratio
\begin{equation}\label{eq:W-ratio}
 \frac{W^{exp}-1}{W^{sim}-1} = \frac{f B_1 \tilde{A} \tilde{Q}}{f B_1 A_{SM} \tilde{Q}}
\end{equation}
can be calculated for every bin and for each sample temperature. Assuming that the Geant4 simulations duly account for the ``geometrical'' factors $\tilde{Q}$ they cancel in this ratio. Note that the simulations used the values of the orientation coefficients $B_1$ corresponding to the sample temperatures determined with the $^{57}$Co nuclear thermometer (see Sec.~\ref{sec:thermometry}). Further, the fraction that was obtained from the asymmetry measurement with $^{68}$Cu was in fact $\tilde{f} \equiv f T_{68}$ (see Sec.~\ref{sec:fract68cu}). Writing $\tilde{A}_{67} = A_{SM,67} T_{67}$, similar to the case of $^{68}$Cu, Eq.~(\ref{eq:W-ratio}) for $^{67}$Cu determines in fact the ratio of the tensor current contributions of the two isotopes:
\begin{equation}\label{eq:67cu_ratio}
 \frac{W^{exp}-1}{W^{sim}-1} = \frac{f B_1 \tilde{A}_{67} \tilde{Q}}{f T_{68} B_1 A_{SM,67} \tilde{Q}} = \frac{T_{67}}{T_{68}} ~ \rm{.}
\end{equation}
\noindent For ease of notation we will further denote this ratio as $\tilde{A}/A_{SM}$. This ratio is constructed for the different energy bins considered in the experiment with $^{67}$Cu. When changing over from measuring $^{68}$Cu to $^{67}$Cu we unfortunately lost electrical contact to the Right $\beta$ detector. The $^{67}$Cu analysis therefore includes only data from the Left $\beta$ detector.

\begin{figure*}[htb!]
\includegraphics[width=1.0\textwidth]{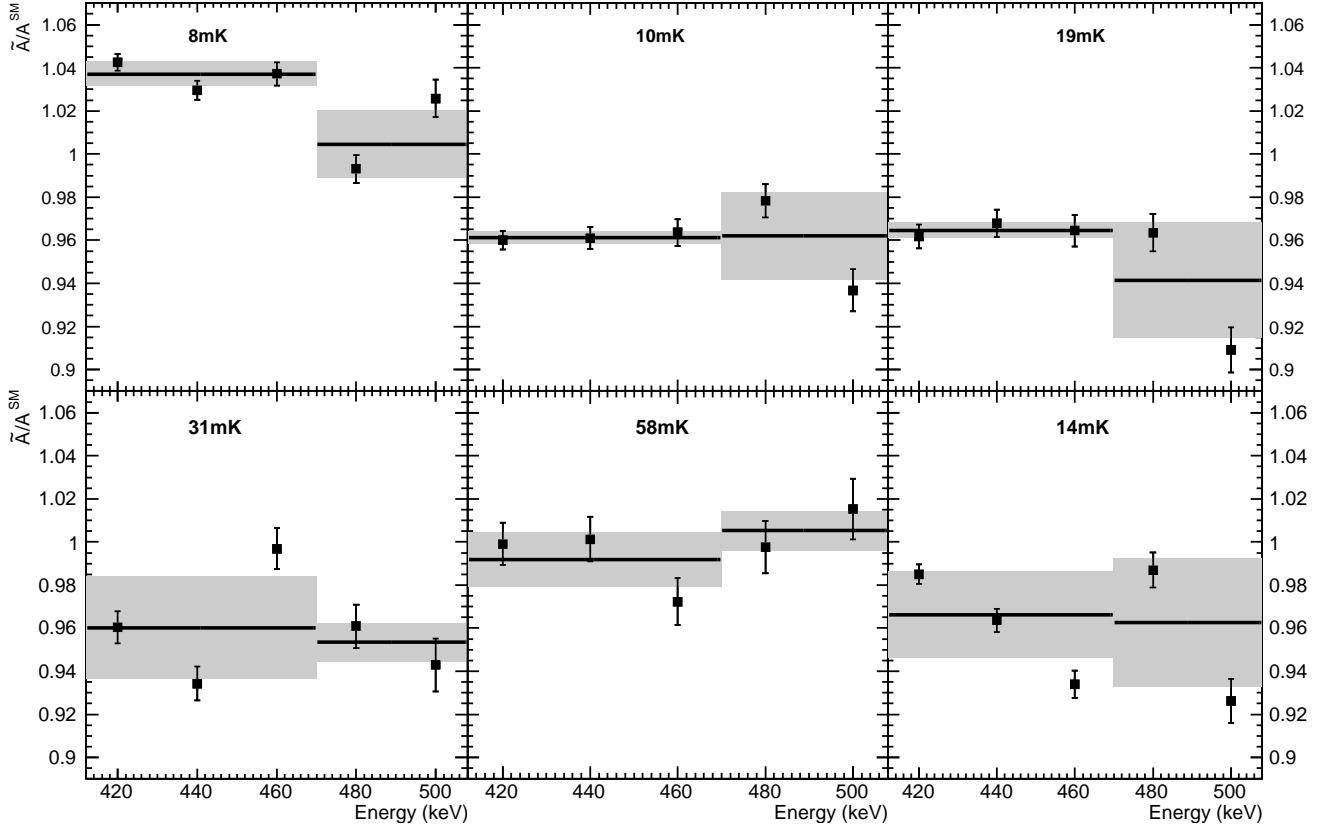}
\caption{\label{fig:67Cu_ratio_ene_dep} Ratio $\tilde{A}/A_{SM}$ as a function of energy in 20\,keV wide energy bins for the different sample temperatures. Weighted averages are shown as horizontal lines (with the gray bands indicating the $1\sigma$ error bar) in the regions from 410 to 470\,keV and from 470 to 510\,keV. The energy region between 410 and 470\,keV also contains events from the second most energetic $\beta$ branch in the decay of $^{67}$Cu (see Fig.~\ref{fig:decay67Cu} and Tab.~\ref{tab:sim_branches}). No temperature or time dependent effects are observed (the temperature points are ordered chronologically).}
\end{figure*}

An overview of the values obtained for $\tilde{A}/A_{SM}$ in 20\,keV wide energy bins in the region from 410 to 510\,keV, and for the different temperatures (viz. degrees of nuclear orientation), is given in Fig.~\ref{fig:67Cu_ratio_ene_dep}. The weighted averages for the energy regions from 470 to 510\,keV (including only counts from the highest energetic branch) and from 410 to 470\,keV (with counts from the two highest $Q_\beta$ branches) are also shown. As can be seen, the weighted averages for both energy regions are always in good agreement, indicating the good quality of the Geant4 simulations of the $\beta$ spectra over the entire energy region considered here. Indeed, many of the effects Geant4 has to take into account are energy dependent so that their improper treatment in the simulations would cause an energy dependence of $\tilde{A}/A_{SM}$. Also, no temperature or time dependence of $\tilde{A}/A_{SM}$ was observed.

In order to obtain a final value of $\tilde{A}/A_{SM}$ in the energy region with only one $\beta$ branch (i.e. from 470 to 510\,keV), only a single 40\,keV wide energy bin was constructed. The ratio $\tilde{A}/A_{SM}$ was then determined as the weighted average of the values obtained at the different temperatures, yielding 0.9709(27), where the error is purely statistical. This still needs to be increased by a factor of $\sqrt{\chi^2_{\text{red}}}$~=~3.6 (see Sec.~\ref{sec:err_geant4} for details), finally arriving at a value of 0.971(10)$_{\text{stat}}$.

\subsection{\label{sec:67cu_errors}Error analysis}
The systematic errors associated with the ratio $\tilde{A}/A_{SM}$ for $^{67}$Cu are summarized in Tab.~\ref{tab:67cu_syst_errors} and discussed in the rest of this section.
\begin{table}[htb]
\caption{\label{tab:67cu_syst_errors} Systematic error budget of the ratio $\tilde{A}/A_{SM}$ measured in the decay of $^{67}$Cu.}
\begin{ruledtabular}
\begin{tabular}{l  l}
Effect & Value \\
\hline
beam spot position & 0.0027 \\
thermometry\footnote{linear sum of the thermometry related uncertainties of both isotopes } &  0.0087 \\
pile-up & 0.0028 \\
half-life of $^{67}$Cu & 0.0068 \\
geometry & 0.0032 \\
hyperfine interaction $\mu B$\footnote{linear sum of the uncertainties related to the hyperfine interaction of both isotopes} & 0.0006 \\
recoil corrections (Sec.~\ref{sec:recoil})& 0.0002 \\
fraction from $^{68}$Cu\footnote{this contains the statistical error, but not the error related to thermometry} & 0.0130 \\
\hline
\multicolumn{1}{r}{Total:} & 0.0178 \\
\end{tabular}
\end{ruledtabular}
\end{table}

\subsubsection{\label{sec:err_beamspot}Beam spot position}
In this experiment the radioactive ions from ISOLDE were implanted into a cold sample foil which was soldered onto the copper sample holder. However, it was noticed in previous experiments that the resulting radioactive beam spot is not perfectly centered on the surface of the sample holder. Consequently the Right and Left detectors have different count rates even when the nuclei are not oriented. To take this into account in the simulations the position of the implantation spot was verified in an additional auxiliary measurement with an adhesive tape mounted on the sample holder. The NICOLE setup was cooled to 77\,K to achieve experimental conditions and a stable beam of $^{85}$Rb was then implanted for 5 hours with a beam current of approximately 8\,nA. After the implantation a discolored spot was visible on the tape clearly indicating the position of the beam spot with respect to the sample holder. Two different methods to measure the position of this spot yielded values of 0.85\,mm and 1.1\,mm for the offset toward the Right detector. To assess the effect of the beam spot position on the final result, all simulations were performed for both offset values and their average was then used. As all results agreed with each other within the statistical error of 0.0027 of the simulations, the latter value was used for the systematic uncertainty related to the beam spot position.

For the fraction $f$ obtained from the $^{68}$Cu data, the exact beam spot position does not cause a significant change. This is expected, since the usage of the double ratios $R$ (Eq.~\ref{eq:R_doubleratio}) reduces any effect of the beam spot not being centered to a second order effect, which is in this case negligible. Nevertheless, the analysis was still performed with simulated values of the $\tilde{Q}$ factors for both beam spot offset values, with the average then being used as a final result.

\subsubsection{\label{sec:err_thermo}Thermometry}
The distance between the $^{57}$Co\underline{Fe} nuclear thermometer and the HPGe $\gamma$ detectors to measure the temperature was determined with a 1\,mm precision, which also includes the uncertainty of the position of the nuclear thermometer within the setup. This induces a 0.04\,mK systematic error on the temperature determination. Further, the fraction of the $^{57}$Co nuclei in the nuclear orientation thermometer that feel the full orienting interaction, which was determined to be $0.943(4)$ (see Sec.~\ref{sec:thermometry}), induces a 0.07\,mK error in the temperature determination. These errors are identical in the analysis of both $^{68}$Cu and $^{67}$Cu, and the final uncertainties were propagated accordingly.

In the analysis of the $^{67}$Cu data the statistical error on the temperature was added quadratically to the statistical error of the simulated asymmetry and was thus taken into account in the fit via the y-axis error bar.

\subsubsection{\label{sec:err_pileup}Pile-up and dead time}
The relatively high count rate of several kHz during the experiment caused non-negligible pile-up and dead time in the electronics chains. The dead time of the system was monitored with a precision pulser signal fed to the preamplifiers. The variation in the pile-up magnitude was determined from the ratio of the number of counts in the pulser peak to the number of counts in the tail at its right hand side, which contains the events that piled-up with the pulser peak (Fig.~\ref{fig:67Cu_beta}). The value of this tail to peak ratio at the beginning of the measurement was around 6\%, and it decreased down to about 3\% together with the decrease of the $^{67}$Cu intensity. 

During the generation of the simulated spectra, the pile-up probability for two events in the beta spectrum is taken to be proportional to the experimental pulser tail to peak ratio. This proportionality factor can not be calculated based on first principles, since \begin{inparaenum}[(i)]
\item the tail of the pulser peak was not fully contained within the recorded spectra, 
\item the beta-particle count rate and the pulser rate were not the same, etc. 
\end{inparaenum}
Therefore, this factor was determined by searching for the best match between simulated and experimental warm data, yielding the values of 4.98, 4.99, and 4.98 for the three warm spectra. Their average value of 4.98 was then applied to all warm and cold simulated data.

To estimate the systematic error related to the pile-up correction the proportionality factor was varied by 10\%, in both directions, when simulating isotropic (warm) and anisotropic (cold) spectra. Such a large variation in the proportionality factor leads to a disagreement between experiment and simulation which is clearly distinguishable by eye, and yields a generous 0.0028 systematic uncertainty on the $\tilde{A}/A_{SM}$ ratio.

For the case of $^{68}$Cu the amount of pile-up (again extracted from the pulser tail to peak ratio) was highest in the warm data, indicating that this effect reduces the observed anisotropy. As the region of interest was reduced to the energy range between 3400 and 4000 keV the uncertainties related to pile-up were absorbed in the apparent energy dependence of the fraction (see Sec.~\ref{sec:68cu_err_geant4}).

\subsubsection{Half-life of $^{67}$Cu}
The experimental spectra of $^{67}$Cu had to be corrected for the half-life of this isotope. The literature value is $T_{1/2}=61.83(12)$\,h \cite{Junde2005}, and this uncertainty induces a 0.0011 systematic error on the ratio $\tilde{A}/A_{SM}$. 

The half-life can also be obtained from an exponential fit to the three groups of isotropic (warm) data taken throughout our measuring campaign (see Fig.~\ref{fig:67Cu_measurement_2007}), resulting in $T_{1/2}=61.07(12)$\,h. Geant4 simulations indicate that this difference is dominated by pile-up. Indeed, with the source intensity decreasing the amount of pile-up (being proportional to the square of the number of counts) decreases at a faster rate, leading to a seemingly lower value for the half-life. The difference between the two values induces a systematic shift of 0.0067 in the ratio $\tilde{A}/A_{SM}$, which is used as a systematic error related to the half-life. 

The final value for the uncertainty related to the half-life of $^{67}$Cu is then obtained as the square sum of this uncertainty and the uncertainty on the half-life value itself, leading to 0.0068.

\subsubsection{Geometry}
The geometrical uncertainties related to the experimental setup (e.g.\ detector position) also induce systematic errors on the simulated spectra of $^{67}$Cu. Varying the position of the detectors within their uncertainties a systematic error of 0.0032 can be assigned to the ratio $\tilde{A}/A_{SM}$.

\subsubsection{Hyperfine interaction $\mu B$}
The uncertainties related to the magnetic moment $\mu$ and the total magnetic field $B$ yield a total systematic error of 0.0003 on the ratio $\tilde{A}/A_{SM}$.
Further, similar to the case of $^{68}$Cu, the error on the hyperfine interaction experienced by the $^{67}$Cu nuclei is dominated by the uncertainty related to the total magnetic field (see Sec.~\ref{sec:magfield}). The corresponding errors were therefore propagated by considering them to be fully correlated, yielding a total systematic error of 0.0006.

\subsubsection{\label{sec:err_geant4}Quality of Geant4 simulations}
Previous studies \cite{Soti2013} concerning the HPGe $\beta$ detectors used in this experiment indicate that the difference between simulated and experimental spectra should not exceed the level of a few percent. The relatively low endpoint energy of 561.7\,keV is comparable to the case of $^{85}$Kr from Ref.~\cite{Soti2013}. The findings of Ref.~\cite{Soti2013} indicate that at this level of precision the effect of the different Geant4 models and simulation parameter values will be negligible. The simulated spectra of warm (isotropic) data measured at 1.2\,K agree with the experimental spectra at the level of 2-3\% (see Figure~\ref{fig:67Cu_warm_diff}), with an absence of any systematic trends. Furthermore, results obtained in the lower energy region (i.e.\ 410-470\,keV) indicate the absence of an energy dependence of the ratio $\tilde{A}/A_{SM}$ (see Fig.~\ref{fig:67Cu_ratio_ene_dep}). Finally, during the analysis of this experiment, ratios of count rates obtained for identical geometries (i.e.\ $N_{\text{cold}}/N_{\text{warm}}$) were constructed, so that many temperature independent systematic effects cancel, while others are reduced in size.

However, the observed scatter of the ratio $\tilde{A}/A_{SM}$ within the different temperature groups (Fig.~\ref{fig:67Cu_ratio_ene_dep}) might indicate the presence of a small difference between simulated and experimental spectra. The statistical uncertainty of the weighted average for each temperature point was therefore increased by a factor of $\sqrt{\chi^2_{red}}$ if $\chi^2_{red} > 1$, which we consider as a systematic error related to the Geant4 simulations. As a result, full separation of ``statistical'' and ``systematic'' uncertainties becomes impossible at this stage.

\subsection{\label{sec:twobranches}The $\beta$ asymmetry parameter in an energy region with two $\beta$ branches}
During the analysis of the data recorded with $^{67}$Cu we initially restricted ourselves to the relatively narrow energy region where only the highest endpoint-energy $\beta$-branch is present. However, with some changes, this analysis can be extended to lower energy regions where events from the two most-energetic $\beta$-branches are present. In that case, the ratio of cold and warm experimental count rates can be written as
\begin{equation}
W = \frac{N_{\text{cold}}^1 + N_{\text{cold}}^2}{N_{\text{warm}}^1 + N_{\text{warm}}^2} ,
\end{equation}
where $N^{1,2}$ represent the count rates of the two branches. After some rearranging and by the use of Eqs.~(\ref{eq:AngularDistributionElectrons}) and (\ref{eq:ExpCountRatio}), the expression for $W$ becomes
\begin{equation}
W = \frac{1+\eta + fP\tilde{Q}(\tilde{A}^1+\eta\tilde{A}^2)}{1+\eta} ,
\end{equation}
where $\eta = N_{\text{warm}}^2/N_{\text{warm}}^1$ and $A^{1,2}$ are the asymmetry parameters of the two branches. Constructing then the ratio of experiment and simulation, while keeping in mind that the fraction used in the simulations is in fact $f T_{68}$ (see Sec.~\ref{sec:fract68cu}), one arrives to
\begin{equation}
\frac{W_{\text{exp}}-1}{W_{\text{sim}}-1} = \frac{(A^1_{\text{SM}}+\eta A^2_{\text{SM}})/(1+\eta)}{(A^1_{\text{SM}}+\eta A^2_{\text{SM}})/(1+\eta)} \ \frac{T_{67}}{T_{68}}.
\end{equation}
Note that the value of the ratio multiplying $T_{67}/T_{68}$ is equal to unity. However, the ratio of the warm count rates, $\eta$, includes an uncertainty related to the branching ratios (see Tab.~\ref{tab:sim_branches}) and is energy dependent via the different count rates for the two branches in a given energy bin. These uncertainties are given in Tab.~\ref{tab:67Cu_twobrancherr}.
\begin{table}[t]
\caption{\label{tab:67Cu_twobrancherr} Value of the systematic uncertainty of $\tilde{A}/A_{SM}$ related to the branching ratio (B.R. error) in different energy regions.}
\begin{ruledtabular}
\begin{tabular}{c c }
Energy region (keV) & B.R. error \\
\hline
410 - 430 & 0.0129 \\
430 - 450 & 0.0049 \\
450 - 470 & 0.0009 \\
\end{tabular}
\end{ruledtabular}
\end{table}
Below 430\,keV the uncertainty related to the branching ratio is already above the 1\% level, therefore we restrict ourselves to the energy range from 430 to 470 keV. For a final result again 40\,keV wide bins were constructed and the value of $\tilde{A}/A_{SM}$ determined for each temperature point. The weighted average of the values obtained for $\tilde{A}/A_{SM}$ is $0.982(16)_{\text{stat}}$. The systematic error related to the branching ratios can be obtained as a linear sum of the relevant values given in Tab.~\ref{tab:67Cu_twobrancherr} as they are considered to be fully correlated, while the remaining systematic errors are identical to the ones discussed in Sec.~\ref{sec:67cu_errors} as they are not energy dependent. Thus the final value for this energy region is $\tilde{A}/A_{SM} = 0.982(16)_{\text{stat}}(19)_{\text{syst}}$.

\subsection{Total systematic error and final results}
For the highest-energetic $\beta$ branch in the decay of $^{67}$Cu (from 470 to 510\,keV) the final result is $\tilde{A}/A_{SM}$~=~0.971$\pm$0.010$_{\text{stat}}\pm$0.018$_{\text{syst}}$ = 0.971(20).

Considering the lower energy range (from 430 to 470\,keV) containing events from two $\beta$ branches, the value of the ratio $\tilde{A}/A_{SM}$ is found to be 0.982$\pm$0.016$_{\text{stat}}\pm$0.019$_{\text{syst}}$~=~0.982(25). 

Since the values obtained in the two energy regions are in agreement and systematic errors are identical, except for the small contribution from the branching ratio error (Tab.~\ref{tab:67Cu_twobrancherr}), a fit to the full energy region ranging from 430\,keV to 510\,keV was performed, leading to the final result $\tilde{A}/A_{SM}$~=~0.980$\pm$0.014$_{\text{stat}}\pm$19$_{\text{syst}}$ = 0.980(23).

\section{\label{sec:recoil}Standard Model value for $A_{SM}$}

The Standard Model value of the correlation coefficients is modified up to the percent level by higher order corrections. Considering the recoil corrections in the formalism developed by Holstein \cite{Holstein1974,*Holstein1976}, the correlation coefficients in nuclear decay are expressed as ratios of spectral functions $F$. Holstein also provided expressions for the Coulomb corrections to the spectral functions, which are noted as $\Delta F$ \cite{Holstein1974,*Holstein1976}. Furthermore, including the radiative corrections ($\Delta R_2$ and $\Delta R_3$) as defined by Yokoo and Morita \cite{Yokoo1973}, the $\beta$-asymmetry parameter can be written as \cite{DeLeebeeck2014}:
\begin{equation}\label{eq:Asm_recoil}
 A_{SM} = \frac{F_4 + F_7/3 + \Delta F_4 + \Delta F_7 /3 + \Delta R_3 }{F_1 + \Delta F_1 + \Delta R_2}~\rm{.}
\end{equation}
By using the impulse approximation all recoil corrections can be expressed in terms of the nuclear matrix elements involved in the decay. For a pure Gamow-Teller decay these are mainly the $M_{GT}$,
 $M_L$, $M_{\sigma L}$ and $M_{1y}$ matrix elements \cite{Calaprice1976}. The Gamow-Teller matrix element $M_{GT}$ can be obtained from the relation \cite{Calaprice1976}:
\begin{equation}\label{eq:c_logft}
 g^2_A M_{GT}^2 \simeq c^2 \simeq \frac{2 {\cal{F}} t^{0^+ \rightarrow 0^+}}{(1+\delta'_R)ft}
\end{equation}
with ${\cal{F}} t^{0^+ \rightarrow 0^+}$~=~3071.8(8)\,s the corrected $ft$ value of the superallowed $0^+ \rightarrow 0^+$ transitions \cite{Hardy2009, Severijns2011}, $ft$ the $ft$ value of the transition under investigation, and $\delta'_R$ the nucleus-dependent part of the ``outer'' radiative correction which depends only on the electron energy and the atomic number $Z$ of the daughter nucleus. Further, $g_A$ is the axial-vector coupling constant whose value in neutron decay is $g_A=1.27$. Since in finite nuclei the value of $g_A$ is known to be quenched (e.g.\ \cite{Towner1987,Martinez1997,Siiskonen2001}), we adopt here the conventional choice of $g_A=1$. For the case of a retarded Gamow-Teller transition, where $M_{GT}$ is small, this relation is not accurate enough, and a shape correction factor $S(Z,W)$ needs to be included in the statistical rate function $f$. The expression for $f$ then becomes
\begin{equation}\label{eq:statrate_exact}
 f_{\text{exact}} = \int_1^{W_0}pW(W_0-W)^2 F(Z,W) S(Z,W) \text{d}W
\end{equation}
where $W$ ($W_0$) is the total electron (endpoint) energy in units of the electron rest mass $m_ec^2$, $p$ is the electron momentum, $F(Z,W)$ is the Fermi function and $S(Z,W)=F_1(W)/c^2$ where $F_1(W)$ is one of  the spectral functions defined by Holstein \cite{Holstein1974}.

All matrix elements mentioned above can also be computed using shell model calculations from the expression:
\begin{equation}\label{eq:obdme}
 M = \sum_{j_1,j_2} \langle f | \left[ a^\dag_{j_1} a_{j_2} \right] ^K  | i \rangle  \langle j_1 | {\cal{O}}^K | j_2 \rangle
\end{equation}
where the first factor is the one-body density matrix element (OBDME), which is the expectation value of the creation and annihilation operators for $j_1$ and $j_2$ orbitals evaluated in the many-body initial and final states, $|i\rangle$ and $|f\rangle$. The OBDME depends only on the rank $K$ of the operator being evaluated, and the ones of interest here all have $K=1$. The second factor is a single-particle matrix element and depends on the operator but not on the many-body physics included in the initial and final wave functions.

Shell model calculations were performed with the GXPF1A \cite{Honma2005} effective interaction in a truncated model space that allowed at most only one hole in a closed $f_{7/2}$ orbital. Since this is not the full $pf$-shell model space, the single-particle energy part of the effective interaction was readjusted to reproduce the excitation energies of the low-lying levels in $^{57}$Ni in the truncated model space. For $^{67}$Cu the computed $M_{GT}$ value is 0.045, which differs from the experimental value of 0.060 deduced from Eq.~(\ref{eq:c_logft}) with the statistical rate function $f$ computed from Eq.~(\ref{eq:statrate_exact}). Recall, however, that $f_{\text{exact}}$ depends on the shape-correction factor $S(Z,W)$, which in turn depends on the shell-model calculated value of $M_{GT}$. To bring the shell-model and experimental values of $M_{GT}$ in closer agreement we slightly adjust the shell-model calculation by multiplying one of the OBDME in Eq.~(\ref{eq:obdme}) by a scaling factor, $\alpha$. This sets up the following iterative scheme: adjust one OBDME, recompute $M_{GT}$, insert in $S(Z,W)$, recompute $f_{\text{exact}}$, and obtain an experimental $M_{GT}$ from Eq.~(\ref{eq:c_logft}). Adjust $\alpha$ until the shell-model input $M_{GT}$ value agrees with the experimental output value. There are five OBDME with multipolarity $K=1$ that can contribute to $M_{GT}$, thus there are five different ways we can do this adjustment. The other matrix elements, $M_L$, $M_{\sigma L}$ and $M_{1y}$ , that depend on the multipolarity-one OBDME will be altered by this adjustment in the OBDME. The error assigned to these matrix elements in Tab.~\ref{tab:Cu_matrixelements} reflects the range of results obtained in the five different ways of performing the adjustment.

\begin{table}[h]
\caption{\label{tab:Cu_matrixelements} Relevant nuclear matrix elements together with the form factors for the decay of $^{67}$Cu and $^{68}$Cu. The value of $M_{GT}$ was obtained experimentally using Eq.~(\ref{eq:c_logft}). The other matrix elements were calculated within the shell model employing the GXPF1A interaction, with one hole in the $\pi f_{7/2}$ shell. The uncertainty connected to the values of the matrix elements is based on the spread observed during the iterative process (see text). The form factors (with $A$ being the mass of the nucleus) were calculated according to the impulse approximation described in Ref.~\cite{Holstein1974}. Note that for $^{68}$Cu the spin sequence ($1^+ \rightarrow 0^+$) ensures that the $f,g,j_2,j_3$ form factors are equal to zero. The $\beta$-asymmetry parameter $A_{SM}$ is obtained using Eq.~(\ref{eq:Asm_recoil}) for $^{67}$Cu and $^{68}$Cu at 470\,keV and 3700\,keV, respectively.}
\begin{ruledtabular}
\begin{tabular}{c c c}
~ & $^{67}$Cu & $^{68}$Cu\\
\hline
$M_{GT}$ & 0.0600(3) & 0.102(6) \\
$M_L$ & 0.213(37) & 0.349(15) \\
$M_{\sigma L}$ & -0.257(22) & -0.814(8) \\
$M_{1y}$ & 7.91(41) & 1.28(16) \\
\hline
$b/Ac$ & 1.13(61) & 1.26(14)\\
$d/Ac$ & 4.32(37) & 8.06(8)\\
$f/Ac$ & -0.89 & 0\\
$g/Ac$ & 8.5$\times10^4$ & 0\\
$h/Ac$ & 1.39(6)$\times10^5$ & 2.47(16)$\times 10^4$\\
$j_2/Ac$ & -1.5$\times10^5$ & 0\\
$j_3/Ac$ & 5.9$\times10^4$ & 0\\
\hline
$A_{SM}$ & 0.5991(2) & -0.9900(12)\\
\end{tabular}
\end{ruledtabular}
\end{table}

The results for the matrix elements obtained for both $^{67}$Cu and $^{68}$Cu with this iterative procedure are summarised in Table~\ref{tab:Cu_matrixelements}. The form factors occurring in the spectral functions $F_{1,4,7}(W)$ were calculated using the impulse-approximation formulae of Holstein~\cite{Holstein1974}. In the last line the values for the $\beta$-asymmetry parameter $A_{SM}$ computed from Eq.~(\ref{eq:Asm_recoil}) are listed. The errors are based on the uncertainties of the matrix elements listed in Table~\ref{tab:Cu_matrixelements}. The relatively large change of about 1\% for the $A_{SM}$ value of $^{68}$Cu from the value of -1 when recoil corrections are neglected is due to the high energy region that is used for analysis. At lower energies this change becomes significantly smaller (i.e.\ 0.0012 at 500\,keV).

\section{Discussion}

Our experimental result for the $\beta$ asymmetry parameter of $^{67}$Cu relative to $^{68}$Cu, i.e.
\begin{equation}
\frac{\tilde{A_{67}}} {A_{SM,67} T_{68}} \equiv \frac{T_{67}}{T_{68}} = 0.980(23)
\end{equation}
\noindent
constitutes one of the most accurate values for the $\beta$-asymmetry parameter in a nuclear transition to date and is in agreement with the Standard Model value of unity within $1\sigma$. Combining this with the Standard Model prediction of 0.5991(2) yields $\tilde{A}$~=~0.587(14).

\begin{figure}[htb]
\includegraphics[width=0.5\textwidth]{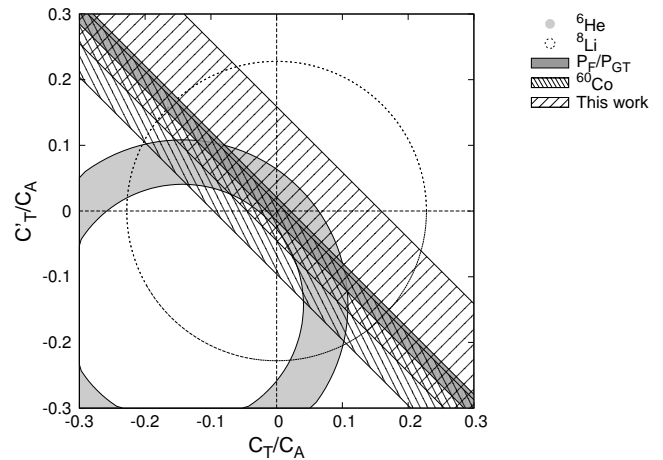}%
\caption{\label{fig:limits_ct_ca} Limits (90\% C.L.) on time reversal invariant tensor type coupling constants $C_T$ and $C'_T$ normalized to the axial-vector coupling constant $C_A$. The result of this work is shown together with results from other experiments in nuclear $\beta$ decay: the $a_{\beta\nu}$ measurement of $^6$He \cite{Johnson1963, Gluck1998}, the $\alpha$-$\beta$-$\nu$ correlation of $^8$Li \cite{Li2013}, the longitudinal positron polarization in the decays of $^{14}$O and $^{10}$C ($P_F/P_{GT}$) \cite{Carnoy1991}, and the $\tilde{A}$ measurement of $^{60}$Co \cite{Wauters2010}. }
\end{figure}

To deduce limits on possible charged current weak tensor couplings the following equation can be used
\begin{equation}
\label{eq:67cu_ratio-2}
\begin{split}
\frac{\tilde{A}} {A_{SM}} &\equiv \frac{\tilde{A_{67}}} {{A_{SM,67} T_{68}}} \\
&\simeq ~ 1 + \left( \frac{\gamma m}{\langle E_{68} \rangle} - \frac{\gamma m}{\langle E_{67} \rangle} \right) \left( \frac{C_T + C_T^\prime}{C_A} \right) \\
&\simeq ~ 1 - 0.395 ~~ \frac{C_T + C_T^\prime}{C_A} ~ \rm{,}
\end{split}
\end{equation}
where the factor $\gamma m/ \langle E \rangle$ in the ROI is equal to 0.514 for $^{67}$Cu (ROI between 430 and 510\,keV), and 0.119 for $^{68}$Cu (ROI between 3.4 and 4.0\,MeV). The limits obtained from this measurement are thus -0.045 $< (C_T+C'_T)/C_A <$ 0.159 (90\% C.L.). A comparison of these limits with limits obtained from other recent precision measurements of correlation coefficients in nuclear $\beta$ decay is shown in Fig.~\ref{fig:limits_ct_ca}. As can be seen all measurements are consistent with the tensor coupling constants $C_T$ and $C'_T$ being equal to zero. Note that when the full second-order approximation for the tensor current expression $T$, given by Eq.~(\ref{eq:asym}), is used instead of Eq.~(\ref{eq:asym_firstorder}) the difference in the obtained limits is smaller than the line thickness used in Fig.~\ref{fig:limits_ct_ca} .

\section{Conclusions}
A measurement of the $\beta$-asymmetry parameter in the decay of $^{67}$Cu was presented. The technique of low temperature nuclear orientation was used in combination with Geant4 simulations to account for systematic effects such as electron scattering. An on-line measurement of the $\beta$-asymmetry parameter for $^{68}$Cu was also performed for normalization purposes.

The experiment yielded a value of $\tilde{A}$~=~0.587(14), one of the most precise determinations of the $\beta$-asymmetry parameter in a nuclear transition to date and in agreement with the Standard Model prediction of $A_{SM}$~=~0.5991(2) which includes recoil, radiative and Coulomb corrections.

The results were interpreted in terms of possible time reversal invariant tensor currents in the weak interaction Hamiltonian. The limits obtained are competitive with limits from other state-of-the-art correlations measurements in nuclear $\beta$ decay.

Comparing the error budgets of this and previous $\beta$ asymmetry parameter measurements that used the low temperature nuclear orientation technique, \cite{Wauters2009, Wauters2010} it is clear that, besides some specific effects which can be reduced under favorable conditions, the main sources of error are the fraction determination, Geant4-related uncertainties and thermometry. As these effects are inherent to the present LTNO technology they will be limiting the accuracy of any future experiment of this type to a level of around 1\%.

Recent reviews \cite{Bhattacharya2012, Naviliat2013a} have shown that low-energy weak interaction experiments in neutron decay and nuclear $\beta$ decay focusing on the Fierz term and aiming for precision at the 10$^{-3}$ level remain competitive with ongoing searches for new bosons at the TeV mass scale at the LHC. This pursuit would require major modifications to the LTNO method as presently used, or application of a different type of method. In this respect we are setting up a new $\beta$ spectrometer based on a combination of energy sensitive detectors and a multi-wire drift chamber for $\beta$ particles as described in Refs.~\cite{Lojek2009, Ban2009a}. In addition, measurements of the asymmetry parameter using laser optical pumping to polarize the nuclei are being prepared as well \cite{Severijns2013}.

\begin{acknowledgments}
This work was supported by the Fund for Scientific Research Flanders (FWO), Projects GOA/2004/03 and GOA/2010/10 of the KU Leuven, the Interuniversity Attraction Poles Programme, Belgian State Belgian Science Policy (BriX network P6/23), and grants LA08015 and LG13031 of the Ministry of Education of the Czech Republic.
\end{acknowledgments}


\begin{thebibliography}{69}%
\makeatletter
\providecommand \@ifxundefined [1]{%
 \@ifx{#1\undefined}
}%
\providecommand \@ifnum [1]{%
 \ifnum #1\expandafter \@firstoftwo
 \else \expandafter \@secondoftwo
 \fi
}%
\providecommand \@ifx [1]{%
 \ifx #1\expandafter \@firstoftwo
 \else \expandafter \@secondoftwo
 \fi
}%
\providecommand \natexlab [1]{#1}%
\providecommand \enquote  [1]{``#1''}%
\providecommand \bibnamefont  [1]{#1}%
\providecommand \bibfnamefont [1]{#1}%
\providecommand \citenamefont [1]{#1}%
\providecommand \href@noop [0]{\@secondoftwo}%
\providecommand \href [0]{\begingroup \@sanitize@url \@href}%
\providecommand \@href[1]{\@@startlink{#1}\@@href}%
\providecommand \@@href[1]{\endgroup#1\@@endlink}%
\providecommand \@sanitize@url [0]{\catcode `\\12\catcode `\$12\catcode
  `\&12\catcode `\#12\catcode `\^12\catcode `\_12\catcode `\%12\relax}%
\providecommand \@@startlink[1]{}%
\providecommand \@@endlink[0]{}%
\providecommand \url  [0]{\begingroup\@sanitize@url \@url }%
\providecommand \@url [1]{\endgroup\@href {#1}{\urlprefix }}%
\providecommand \urlprefix  [0]{URL }%
\providecommand \Eprint [0]{\href }%
\providecommand \doibase [0]{http://dx.doi.org/}%
\providecommand \selectlanguage [0]{\@gobble}%
\providecommand \bibinfo  [0]{\@secondoftwo}%
\providecommand \bibfield  [0]{\@secondoftwo}%
\providecommand \translation [1]{[#1]}%
\providecommand \BibitemOpen [0]{}%
\providecommand \bibitemStop [0]{}%
\providecommand \bibitemNoStop [0]{.\EOS\space}%
\providecommand \EOS [0]{\spacefactor3000\relax}%
\providecommand \BibitemShut  [1]{\csname bibitem#1\endcsname}%
\let\auto@bib@innerbib\@empty
\bibitem [{\citenamefont {Lee}\ and\ \citenamefont {Yang}(1956)}]{Lee1956}%
  \BibitemOpen
  \bibfield  {author} {\bibinfo {author} {\bibfnamefont {T.}~\bibnamefont
  {Lee}}\ and\ \bibinfo {author} {\bibfnamefont {C.}~\bibnamefont {Yang}},\
  }\href {\doibase 10.1103/PhysRev.104.254} {\bibfield  {journal} {\bibinfo
  {journal} {Physical Review}\ }\textbf {\bibinfo {volume} {104}},\ \bibinfo
  {pages} {254} (\bibinfo {year} {1956})}\BibitemShut {NoStop}%
\bibitem [{\citenamefont {Severijns}\ \emph {et~al.}(2006)\citenamefont
  {Severijns}, \citenamefont {Beck},\ and\ \citenamefont
  {Naviliat-Cuncic}}]{Severijns2006}%
  \BibitemOpen
  \bibfield  {author} {\bibinfo {author} {\bibfnamefont {N.}~\bibnamefont
  {Severijns}}, \bibinfo {author} {\bibfnamefont {M.}~\bibnamefont {Beck}}, \
  and\ \bibinfo {author} {\bibfnamefont {O.}~\bibnamefont {Naviliat-Cuncic}},\
  }\href {\doibase 10.1103/RevModPhys.78.991} {\bibfield  {journal} {\bibinfo
  {journal} {Reviews of Modern Physics}\ }\textbf {\bibinfo {volume} {78}},\
  \bibinfo {pages} {991} (\bibinfo {year} {2006})}\BibitemShut {NoStop}%
\bibitem [{\citenamefont {Wauters}\ \emph {et~al.}(2014)\citenamefont
  {Wauters}, \citenamefont {Garc\'ia},\ and\ \citenamefont
  {Hong}}]{Wauters2014}%
  \BibitemOpen
  \bibfield  {author} {\bibinfo {author} {\bibfnamefont {F.}~\bibnamefont
  {Wauters}}, \bibinfo {author} {\bibfnamefont {A.}~\bibnamefont {Garc\'ia}}, \
  and\ \bibinfo {author} {\bibfnamefont {R.}~\bibnamefont {Hong}},\ }\href
  {\doibase 10.1103/PhysRevC.89.025501} {\bibfield  {journal} {\bibinfo
  {journal} {Physical Review C}\ }\textbf {\bibinfo {volume} {89}},\ \bibinfo
  {pages} {025501} (\bibinfo {year} {2014})}\BibitemShut {NoStop}%
\bibitem [{\citenamefont {Abele}(2008)}]{Abele2008}%
  \BibitemOpen
  \bibfield  {author} {\bibinfo {author} {\bibfnamefont {H.}~\bibnamefont
  {Abele}},\ }\href {\doibase 10.1016/j.ppnp.2007.05.002} {\bibfield  {journal}
  {\bibinfo  {journal} {Progress in Particle and Nuclear Physics}\ }\textbf
  {\bibinfo {volume} {60}},\ \bibinfo {pages} {1} (\bibinfo {year}
  {2008})}\BibitemShut {NoStop}%
\bibitem [{\citenamefont {Konrad}\ \emph {et~al.}(2010)\citenamefont {Konrad},
  \citenamefont {Heil}, \citenamefont {Bae{\ss}ler}, \citenamefont {Pocanic},\
  and\ \citenamefont {Gl\"uck}}]{Konrad2010}%
  \BibitemOpen
  \bibfield  {author} {\bibinfo {author} {\bibfnamefont {G.}~\bibnamefont
  {Konrad}}, \bibinfo {author} {\bibfnamefont {W.}~\bibnamefont {Heil}},
  \bibinfo {author} {\bibfnamefont {S.}~\bibnamefont {Bae{\ss}ler}}, \bibinfo
  {author} {\bibfnamefont {D.}~\bibnamefont {Pocanic}}, \ and\ \bibinfo
  {author} {\bibfnamefont {F.}~\bibnamefont {Gl\"uck}},\ }in\ \href {\doibase
  10.1142/9789814340861_0061} {\emph {\bibinfo {booktitle} {Physics Beyond the
  Standard Models of Particles, Cosmology and Astrophysics (Proc. 5th Int.
  Conf., Beyond 2010)}}},\ \bibinfo {editor} {edited by\ \bibinfo {editor}
  {\bibfnamefont {R.~V.}\ \bibnamefont {H.~V. Klapdor-Kleingrothaus},
  \bibfnamefont {I.~V.~Krivosheina}}}\ (\bibinfo {year} {2010})\ \Eprint
  {http://arxiv.org/abs/1007.3027} {arXiv:1007.3027 [nucl-ex]} \BibitemShut
  {NoStop}%
\bibitem [{\citenamefont {Dubbers}\ and\ \citenamefont
  {Schmidt}(2011)}]{Dubbers2011}%
  \BibitemOpen
  \bibfield  {author} {\bibinfo {author} {\bibfnamefont {D.}~\bibnamefont
  {Dubbers}}\ and\ \bibinfo {author} {\bibfnamefont {M.~G.}\ \bibnamefont
  {Schmidt}},\ }\href {\doibase 10.1103/RevModPhys.83.1111} {\bibfield
  {journal} {\bibinfo  {journal} {Reviews of Modern Physics}\ }\textbf
  {\bibinfo {volume} {83}},\ \bibinfo {pages} {1111} (\bibinfo {year}
  {2011})}\BibitemShut {NoStop}%
\bibitem [{\citenamefont {Mumm}\ \emph {et~al.}(2011)\citenamefont {Mumm},
  \citenamefont {Chupp}, \citenamefont {Cooper}, \citenamefont {Coulter},
  \citenamefont {Freedman}, \citenamefont {Fujikawa}, \citenamefont
  {Garc\'{\i}a}, \citenamefont {Jones}, \citenamefont {Nico}, \citenamefont
  {Thompson}, \citenamefont {Trull}, \citenamefont {Wilkerson},\ and\
  \citenamefont {Wietfeldt}}]{Mumm2011}%
  \BibitemOpen
  \bibfield  {author} {\bibinfo {author} {\bibfnamefont {H.~P.}\ \bibnamefont
  {Mumm}}, \bibinfo {author} {\bibfnamefont {T.~E.}\ \bibnamefont {Chupp}},
  \bibinfo {author} {\bibfnamefont {R.~L.}\ \bibnamefont {Cooper}}, \bibinfo
  {author} {\bibfnamefont {K.~P.}\ \bibnamefont {Coulter}}, \bibinfo {author}
  {\bibfnamefont {S.~J.}\ \bibnamefont {Freedman}}, \bibinfo {author}
  {\bibfnamefont {B.~K.}\ \bibnamefont {Fujikawa}}, \bibinfo {author}
  {\bibfnamefont {A.}~\bibnamefont {Garc\'{\i}a}}, \bibinfo {author}
  {\bibfnamefont {G.~L.}\ \bibnamefont {Jones}}, \bibinfo {author}
  {\bibfnamefont {J.~S.}\ \bibnamefont {Nico}}, \bibinfo {author}
  {\bibfnamefont {A.~K.}\ \bibnamefont {Thompson}}, \bibinfo {author}
  {\bibfnamefont {C.~A.}\ \bibnamefont {Trull}}, \bibinfo {author}
  {\bibfnamefont {J.~F.}\ \bibnamefont {Wilkerson}}, \ and\ \bibinfo {author}
  {\bibfnamefont {F.~E.}\ \bibnamefont {Wietfeldt}},\ }\href {\doibase
  10.1103/PhysRevLett.107.102301} {\bibfield  {journal} {\bibinfo  {journal}
  {Physical Review Letters}\ }\textbf {\bibinfo {volume} {107}},\ \bibinfo
  {pages} {102301} (\bibinfo {year} {2011})}\BibitemShut {NoStop}%
\bibitem [{\citenamefont {Kozela}\ \emph {et~al.}(2009)\citenamefont {Kozela},
  \citenamefont {Ban}, \citenamefont {Bia\l{}ek}, \citenamefont {Bodek},
  \citenamefont {Gorel}, \citenamefont {Kirch}, \citenamefont {Kistryn},
  \citenamefont {Ku\'{z}niak}, \citenamefont {Naviliat-Cuncic}, \citenamefont
  {Pulut}, \citenamefont {Severijns}, \citenamefont {Stephan},\ and\
  \citenamefont {Zejma}}]{Kozela2009}%
  \BibitemOpen
  \bibfield  {author} {\bibinfo {author} {\bibfnamefont {A.}~\bibnamefont
  {Kozela}}, \bibinfo {author} {\bibfnamefont {G.}~\bibnamefont {Ban}},
  \bibinfo {author} {\bibfnamefont {A.}~\bibnamefont {Bia\l{}ek}}, \bibinfo
  {author} {\bibfnamefont {K.}~\bibnamefont {Bodek}}, \bibinfo {author}
  {\bibfnamefont {P.}~\bibnamefont {Gorel}}, \bibinfo {author} {\bibfnamefont
  {K.}~\bibnamefont {Kirch}}, \bibinfo {author} {\bibfnamefont
  {S.}~\bibnamefont {Kistryn}}, \bibinfo {author} {\bibfnamefont
  {M.}~\bibnamefont {Ku\'{z}niak}}, \bibinfo {author} {\bibfnamefont
  {O.}~\bibnamefont {Naviliat-Cuncic}}, \bibinfo {author} {\bibfnamefont
  {J.}~\bibnamefont {Pulut}}, \bibinfo {author} {\bibfnamefont
  {N.}~\bibnamefont {Severijns}}, \bibinfo {author} {\bibfnamefont
  {E.}~\bibnamefont {Stephan}}, \ and\ \bibinfo {author} {\bibfnamefont
  {J.}~\bibnamefont {Zejma}},\ }\href {\doibase 10.1103/PhysRevLett.102.172301}
  {\bibfield  {journal} {\bibinfo  {journal} {Physical Review Letters}\
  }\textbf {\bibinfo {volume} {102}},\ \bibinfo {pages} {172301} (\bibinfo
  {year} {2009})}\BibitemShut {NoStop}%
\bibitem [{\citenamefont {Kozela}\ \emph {et~al.}(2012)\citenamefont {Kozela},
  \citenamefont {Ban}, \citenamefont {Bia\l{}ek}, \citenamefont {Bodek},
  \citenamefont {Gorel}, \citenamefont {Kirch}, \citenamefont {Kistryn},
  \citenamefont {Naviliat-Cuncic}, \citenamefont {Severijns}, \citenamefont
  {Stephan},\ and\ \citenamefont {Zejma}}]{Kozela2012}%
  \BibitemOpen
  \bibfield  {author} {\bibinfo {author} {\bibfnamefont {A.}~\bibnamefont
  {Kozela}}, \bibinfo {author} {\bibfnamefont {G.}~\bibnamefont {Ban}},
  \bibinfo {author} {\bibfnamefont {A.}~\bibnamefont {Bia\l{}ek}}, \bibinfo
  {author} {\bibfnamefont {K.}~\bibnamefont {Bodek}}, \bibinfo {author}
  {\bibfnamefont {P.}~\bibnamefont {Gorel}}, \bibinfo {author} {\bibfnamefont
  {K.}~\bibnamefont {Kirch}}, \bibinfo {author} {\bibfnamefont
  {S.}~\bibnamefont {Kistryn}}, \bibinfo {author} {\bibfnamefont
  {O.}~\bibnamefont {Naviliat-Cuncic}}, \bibinfo {author} {\bibfnamefont
  {N.}~\bibnamefont {Severijns}}, \bibinfo {author} {\bibfnamefont
  {E.}~\bibnamefont {Stephan}}, \ and\ \bibinfo {author} {\bibfnamefont
  {J.}~\bibnamefont {Zejma}},\ }\href {\doibase 10.1103/PhysRevC.85.045501}
  {\bibfield  {journal} {\bibinfo  {journal} {Physical Review C}\ }\textbf
  {\bibinfo {volume} {85}},\ \bibinfo {pages} {045501} (\bibinfo {year}
  {2012})}\BibitemShut {NoStop}%
\bibitem [{\citenamefont {Mund}\ \emph {et~al.}(2013)\citenamefont {Mund},
  \citenamefont {M\"{a}rkisch}, \citenamefont {Deissenroth}, \citenamefont
  {Krempel}, \citenamefont {Schumann}, \citenamefont {Abele}, \citenamefont
  {Petoukhov},\ and\ \citenamefont {Soldner}}]{Mund2013}%
  \BibitemOpen
  \bibfield  {author} {\bibinfo {author} {\bibfnamefont {D.}~\bibnamefont
  {Mund}}, \bibinfo {author} {\bibfnamefont {B.}~\bibnamefont {M\"{a}rkisch}},
  \bibinfo {author} {\bibfnamefont {M.}~\bibnamefont {Deissenroth}}, \bibinfo
  {author} {\bibfnamefont {J.}~\bibnamefont {Krempel}}, \bibinfo {author}
  {\bibfnamefont {M.}~\bibnamefont {Schumann}}, \bibinfo {author}
  {\bibfnamefont {H.}~\bibnamefont {Abele}}, \bibinfo {author} {\bibfnamefont
  {A.}~\bibnamefont {Petoukhov}}, \ and\ \bibinfo {author} {\bibfnamefont
  {T.}~\bibnamefont {Soldner}},\ }\href {\doibase
  10.1103/PhysRevLett.110.172502} {\bibfield  {journal} {\bibinfo  {journal}
  {Physical Review Letters}\ }\textbf {\bibinfo {volume} {110}},\ \bibinfo
  {pages} {172502} (\bibinfo {year} {2013})},\ \Eprint
  {http://arxiv.org/abs/1204.0013} {arXiv:1204.0013} \BibitemShut {NoStop}%
\bibitem [{\citenamefont {Severijns}\ and\ \citenamefont
  {Naviliat-Cuncic}(2011)}]{Severijns2011}%
  \BibitemOpen
  \bibfield  {author} {\bibinfo {author} {\bibfnamefont {N.}~\bibnamefont
  {Severijns}}\ and\ \bibinfo {author} {\bibfnamefont {O.}~\bibnamefont
  {Naviliat-Cuncic}},\ }\href {\doibase 10.1146/annurev-nucl-102010-130410}
  {\bibfield  {journal} {\bibinfo  {journal} {Annual Review of Nuclear and
  Particle Science}\ }\textbf {\bibinfo {volume} {61}},\ \bibinfo {pages} {23}
  (\bibinfo {year} {2011})}\BibitemShut {NoStop}%
\bibitem [{\citenamefont {Pitcairn}\ \emph {et~al.}(2009)\citenamefont
  {Pitcairn}, \citenamefont {Roberge}, \citenamefont {Gorelov}, \citenamefont
  {Ashery}, \citenamefont {Aviv}, \citenamefont {Behr}, \citenamefont
  {Bricault}, \citenamefont {Dombsky}, \citenamefont {Holt}, \citenamefont
  {Jackson}, \citenamefont {Lee}, \citenamefont {Pearson}, \citenamefont
  {Gaudin}, \citenamefont {Dej}, \citenamefont {H\"{o}hr}, \citenamefont
  {Gwinner},\ and\ \citenamefont {Melconian}}]{Pitcairn2009}%
  \BibitemOpen
  \bibfield  {author} {\bibinfo {author} {\bibfnamefont {J.~R.~A.}\
  \bibnamefont {Pitcairn}}, \bibinfo {author} {\bibfnamefont {D.}~\bibnamefont
  {Roberge}}, \bibinfo {author} {\bibfnamefont {A.}~\bibnamefont {Gorelov}},
  \bibinfo {author} {\bibfnamefont {D.}~\bibnamefont {Ashery}}, \bibinfo
  {author} {\bibfnamefont {O.}~\bibnamefont {Aviv}}, \bibinfo {author}
  {\bibfnamefont {J.~A.}\ \bibnamefont {Behr}}, \bibinfo {author}
  {\bibfnamefont {P.~G.}\ \bibnamefont {Bricault}}, \bibinfo {author}
  {\bibfnamefont {M.}~\bibnamefont {Dombsky}}, \bibinfo {author} {\bibfnamefont
  {J.~D.}\ \bibnamefont {Holt}}, \bibinfo {author} {\bibfnamefont {K.~P.}\
  \bibnamefont {Jackson}}, \bibinfo {author} {\bibfnamefont {B.}~\bibnamefont
  {Lee}}, \bibinfo {author} {\bibfnamefont {M.~R.}\ \bibnamefont {Pearson}},
  \bibinfo {author} {\bibfnamefont {A.}~\bibnamefont {Gaudin}}, \bibinfo
  {author} {\bibfnamefont {B.}~\bibnamefont {Dej}}, \bibinfo {author}
  {\bibfnamefont {C.}~\bibnamefont {H\"{o}hr}}, \bibinfo {author}
  {\bibfnamefont {G.}~\bibnamefont {Gwinner}}, \ and\ \bibinfo {author}
  {\bibfnamefont {D.}~\bibnamefont {Melconian}},\ }\href {\doibase
  10.1103/PhysRevC.79.015501} {\bibfield  {journal} {\bibinfo  {journal}
  {Physical Review C}\ }\textbf {\bibinfo {volume} {79}},\ \bibinfo {pages}
  {015501} (\bibinfo {year} {2009})}\BibitemShut {NoStop}%
\bibitem [{\citenamefont {Fl\'{e}chard}\ \emph {et~al.}(2011)\citenamefont
  {Fl\'{e}chard}, \citenamefont {Velten}, \citenamefont {Li\'{e}nard},
  \citenamefont {M\'{e}ry}, \citenamefont {Rodr\'{\i}guez}, \citenamefont
  {Ban}, \citenamefont {Durand}, \citenamefont {Mauger}, \citenamefont
  {Naviliat-Cuncic},\ and\ \citenamefont {Thomas}}]{Flechard2011}%
  \BibitemOpen
  \bibfield  {author} {\bibinfo {author} {\bibfnamefont {X.}~\bibnamefont
  {Fl\'{e}chard}}, \bibinfo {author} {\bibfnamefont {P.}~\bibnamefont
  {Velten}}, \bibinfo {author} {\bibfnamefont {E.}~\bibnamefont {Li\'{e}nard}},
  \bibinfo {author} {\bibfnamefont {A.}~\bibnamefont {M\'{e}ry}}, \bibinfo
  {author} {\bibfnamefont {D.}~\bibnamefont {Rodr\'{\i}guez}}, \bibinfo
  {author} {\bibfnamefont {G.}~\bibnamefont {Ban}}, \bibinfo {author}
  {\bibfnamefont {D.}~\bibnamefont {Durand}}, \bibinfo {author} {\bibfnamefont
  {F.}~\bibnamefont {Mauger}}, \bibinfo {author} {\bibfnamefont
  {O.}~\bibnamefont {Naviliat-Cuncic}}, \ and\ \bibinfo {author} {\bibfnamefont
  {J.~C.}\ \bibnamefont {Thomas}},\ }\href {\doibase
  10.1088/0954-3899/38/5/055101} {\bibfield  {journal} {\bibinfo  {journal}
  {Journal of Physics G: Nuclear and Particle Physics}\ }\textbf {\bibinfo
  {volume} {38}},\ \bibinfo {pages} {055101} (\bibinfo {year}
  {2011})}\BibitemShut {NoStop}%
\bibitem [{\citenamefont {Vetter}\ \emph {et~al.}(2008)\citenamefont {Vetter},
  \citenamefont {Abo-Shaeer}, \citenamefont {Freedman},\ and\ \citenamefont
  {Maruyama}}]{Vetter2008}%
  \BibitemOpen
  \bibfield  {author} {\bibinfo {author} {\bibfnamefont {P.~A.}\ \bibnamefont
  {Vetter}}, \bibinfo {author} {\bibfnamefont {J.~R.}\ \bibnamefont
  {Abo-Shaeer}}, \bibinfo {author} {\bibfnamefont {S.~J.}\ \bibnamefont
  {Freedman}}, \ and\ \bibinfo {author} {\bibfnamefont {R.}~\bibnamefont
  {Maruyama}},\ }\href {\doibase 10.1103/PhysRevC.77.035502} {\bibfield
  {journal} {\bibinfo  {journal} {Physical Review C}\ }\textbf {\bibinfo
  {volume} {77}},\ \bibinfo {pages} {035502} (\bibinfo {year}
  {2008})}\BibitemShut {NoStop}%
\bibitem [{\citenamefont {Wauters}\ \emph
  {et~al.}(2009{\natexlab{a}})\citenamefont {Wauters}, \citenamefont
  {De~Leebeeck}, \citenamefont {Kraev}, \citenamefont {Tandecki}, \citenamefont
  {Traykov}, \citenamefont {Gorp}, \citenamefont {Severijns},\ and\
  \citenamefont {Z\'akouck\'y}}]{Wauters2009}%
  \BibitemOpen
  \bibfield  {author} {\bibinfo {author} {\bibfnamefont {F.}~\bibnamefont
  {Wauters}}, \bibinfo {author} {\bibfnamefont {V.}~\bibnamefont
  {De~Leebeeck}}, \bibinfo {author} {\bibfnamefont {I.}~\bibnamefont {Kraev}},
  \bibinfo {author} {\bibfnamefont {M.}~\bibnamefont {Tandecki}}, \bibinfo
  {author} {\bibfnamefont {E.}~\bibnamefont {Traykov}}, \bibinfo {author}
  {\bibfnamefont {S.~V.}\ \bibnamefont {Gorp}}, \bibinfo {author}
  {\bibfnamefont {N.}~\bibnamefont {Severijns}}, \ and\ \bibinfo {author}
  {\bibfnamefont {D.}~\bibnamefont {Z\'akouck\'y}},\ }\href {\doibase
  10.1103/PhysRevC.80.062501} {\bibfield  {journal} {\bibinfo  {journal}
  {Physical Review C}\ }\textbf {\bibinfo {volume} {80}},\ \bibinfo {pages}
  {062501} (\bibinfo {year} {2009}{\natexlab{a}})}\BibitemShut {NoStop}%
\bibitem [{\citenamefont {Wauters}\ \emph {et~al.}(2010)\citenamefont
  {Wauters}, \citenamefont {Kraev}, \citenamefont {Z\'akouck\'y}, \citenamefont
  {Beck}, \citenamefont {Breitenfeldt}, \citenamefont {De~Leebeeck},
  \citenamefont {Golovko}, \citenamefont {Kozlov}, \citenamefont {Phalet},
  \citenamefont {Roccia}, \citenamefont {Soti}, \citenamefont {Tandecki},
  \citenamefont {Towner}, \citenamefont {Traykov}, \citenamefont {VanGorp},\
  and\ \citenamefont {Severijns}}]{Wauters2010}%
  \BibitemOpen
  \bibfield  {author} {\bibinfo {author} {\bibfnamefont {F.}~\bibnamefont
  {Wauters}}, \bibinfo {author} {\bibfnamefont {I.}~\bibnamefont {Kraev}},
  \bibinfo {author} {\bibfnamefont {D.}~\bibnamefont {Z\'akouck\'y}}, \bibinfo
  {author} {\bibfnamefont {M.}~\bibnamefont {Beck}}, \bibinfo {author}
  {\bibfnamefont {M.}~\bibnamefont {Breitenfeldt}}, \bibinfo {author}
  {\bibfnamefont {V.}~\bibnamefont {De~Leebeeck}}, \bibinfo {author}
  {\bibfnamefont {V.~V.}\ \bibnamefont {Golovko}}, \bibinfo {author}
  {\bibfnamefont {V.~Y.}\ \bibnamefont {Kozlov}}, \bibinfo {author}
  {\bibfnamefont {T.}~\bibnamefont {Phalet}}, \bibinfo {author} {\bibfnamefont
  {S.}~\bibnamefont {Roccia}}, \bibinfo {author} {\bibfnamefont
  {G.}~\bibnamefont {Soti}}, \bibinfo {author} {\bibfnamefont {M.}~\bibnamefont
  {Tandecki}}, \bibinfo {author} {\bibfnamefont {I.~S.}\ \bibnamefont
  {Towner}}, \bibinfo {author} {\bibfnamefont {E.}~\bibnamefont {Traykov}},
  \bibinfo {author} {\bibfnamefont {S.}~\bibnamefont {VanGorp}}, \ and\
  \bibinfo {author} {\bibfnamefont {N.}~\bibnamefont {Severijns}},\ }\href
  {\doibase 10.1103/PhysRevC.82.055502} {\bibfield  {journal} {\bibinfo
  {journal} {Physical Review C}\ }\textbf {\bibinfo {volume} {82}},\ \bibinfo
  {pages} {055502} (\bibinfo {year} {2010})}\BibitemShut {NoStop}%
\bibitem [{\citenamefont {Beck}\ \emph {et~al.}(2011)\citenamefont {Beck},
  \citenamefont {Coeck}, \citenamefont {Kozlov}, \citenamefont {Breitenfeldt},
  \citenamefont {Delahaye}, \citenamefont {Friedag}, \citenamefont {Gl\"{u}ck},
  \citenamefont {Herbane}, \citenamefont {Herlert}, \citenamefont {Kraev},
  \citenamefont {Mader}, \citenamefont {Tandecki}, \citenamefont {{Van Gorp}},
  \citenamefont {Wauters}, \citenamefont {Weinheimer}, \citenamefont
  {Wenander},\ and\ \citenamefont {Severijns}}]{Beck2011}%
  \BibitemOpen
  \bibfield  {author} {\bibinfo {author} {\bibfnamefont {M.}~\bibnamefont
  {Beck}}, \bibinfo {author} {\bibfnamefont {S.}~\bibnamefont {Coeck}},
  \bibinfo {author} {\bibfnamefont {V.~Y.}\ \bibnamefont {Kozlov}}, \bibinfo
  {author} {\bibfnamefont {M.}~\bibnamefont {Breitenfeldt}}, \bibinfo {author}
  {\bibfnamefont {P.}~\bibnamefont {Delahaye}}, \bibinfo {author}
  {\bibfnamefont {P.}~\bibnamefont {Friedag}}, \bibinfo {author} {\bibfnamefont
  {F.}~\bibnamefont {Gl\"{u}ck}}, \bibinfo {author} {\bibfnamefont
  {M.}~\bibnamefont {Herbane}}, \bibinfo {author} {\bibfnamefont
  {a.}~\bibnamefont {Herlert}}, \bibinfo {author} {\bibfnamefont {I.~S.}\
  \bibnamefont {Kraev}}, \bibinfo {author} {\bibfnamefont {J.}~\bibnamefont
  {Mader}}, \bibinfo {author} {\bibfnamefont {M.}~\bibnamefont {Tandecki}},
  \bibinfo {author} {\bibfnamefont {S.}~\bibnamefont {{Van Gorp}}}, \bibinfo
  {author} {\bibfnamefont {F.}~\bibnamefont {Wauters}}, \bibinfo {author}
  {\bibfnamefont {C.}~\bibnamefont {Weinheimer}}, \bibinfo {author}
  {\bibfnamefont {F.}~\bibnamefont {Wenander}}, \ and\ \bibinfo {author}
  {\bibfnamefont {N.}~\bibnamefont {Severijns}},\ }\href {\doibase
  10.1140/epja/i2011-11045-0} {\bibfield  {journal} {\bibinfo  {journal} {The
  European Physical Journal A}\ }\textbf {\bibinfo {volume} {47}},\ \bibinfo
  {pages} {45} (\bibinfo {year} {2011})}\BibitemShut {NoStop}%
\bibitem [{\citenamefont {Li}\ \emph {et~al.}(2013)\citenamefont {Li},
  \citenamefont {Segel}, \citenamefont {Scielzo}, \citenamefont {Bertone},
  \citenamefont {Buchinger}, \citenamefont {Caldwell}, \citenamefont
  {Chaudhuri}, \citenamefont {Clark}, \citenamefont {Crawford}, \citenamefont
  {Deibel}, \citenamefont {Fallis}, \citenamefont {Gulick}, \citenamefont
  {Gwinner}, \citenamefont {Lascar}, \citenamefont {Levand}, \citenamefont
  {Pedretti}, \citenamefont {Savard}, \citenamefont {Sharma}, \citenamefont
  {Sternberg}, \citenamefont {Sun}, \citenamefont {{Van Schelt}}, \citenamefont
  {Yee},\ and\ \citenamefont {Zabransky}}]{Li2013}%
  \BibitemOpen
  \bibfield  {author} {\bibinfo {author} {\bibfnamefont {G.}~\bibnamefont
  {Li}}, \bibinfo {author} {\bibfnamefont {R.}~\bibnamefont {Segel}}, \bibinfo
  {author} {\bibfnamefont {N.~D.}\ \bibnamefont {Scielzo}}, \bibinfo {author}
  {\bibfnamefont {P.~F.}\ \bibnamefont {Bertone}}, \bibinfo {author}
  {\bibfnamefont {F.}~\bibnamefont {Buchinger}}, \bibinfo {author}
  {\bibfnamefont {S.}~\bibnamefont {Caldwell}}, \bibinfo {author}
  {\bibfnamefont {A.}~\bibnamefont {Chaudhuri}}, \bibinfo {author}
  {\bibfnamefont {J.~A.}\ \bibnamefont {Clark}}, \bibinfo {author}
  {\bibfnamefont {J.~E.}\ \bibnamefont {Crawford}}, \bibinfo {author}
  {\bibfnamefont {C.~M.}\ \bibnamefont {Deibel}}, \bibinfo {author}
  {\bibfnamefont {J.}~\bibnamefont {Fallis}}, \bibinfo {author} {\bibfnamefont
  {S.}~\bibnamefont {Gulick}}, \bibinfo {author} {\bibfnamefont
  {G.}~\bibnamefont {Gwinner}}, \bibinfo {author} {\bibfnamefont
  {D.}~\bibnamefont {Lascar}}, \bibinfo {author} {\bibfnamefont {A.~F.}\
  \bibnamefont {Levand}}, \bibinfo {author} {\bibfnamefont {M.}~\bibnamefont
  {Pedretti}}, \bibinfo {author} {\bibfnamefont {G.}~\bibnamefont {Savard}},
  \bibinfo {author} {\bibfnamefont {K.~S.}\ \bibnamefont {Sharma}}, \bibinfo
  {author} {\bibfnamefont {M.~G.}\ \bibnamefont {Sternberg}}, \bibinfo {author}
  {\bibfnamefont {T.}~\bibnamefont {Sun}}, \bibinfo {author} {\bibfnamefont
  {J.}~\bibnamefont {{Van Schelt}}}, \bibinfo {author} {\bibfnamefont {R.~M.}\
  \bibnamefont {Yee}}, \ and\ \bibinfo {author} {\bibfnamefont {B.~J.}\
  \bibnamefont {Zabransky}},\ }\href {\doibase 10.1103/PhysRevLett.110.092502}
  {\bibfield  {journal} {\bibinfo  {journal} {Physical Review Letters}\
  }\textbf {\bibinfo {volume} {110}},\ \bibinfo {pages} {092502} (\bibinfo
  {year} {2013})}\BibitemShut {NoStop}%
\bibitem [{\citenamefont {Jackson}\ \emph {et~al.}(1957)\citenamefont
  {Jackson}, \citenamefont {Treiman},\ and\ \citenamefont
  {Wyld}}]{Jackson1957}%
  \BibitemOpen
  \bibfield  {author} {\bibinfo {author} {\bibfnamefont {J.}~\bibnamefont
  {Jackson}}, \bibinfo {author} {\bibfnamefont {S.}~\bibnamefont {Treiman}}, \
  and\ \bibinfo {author} {\bibfnamefont {J.~H.}\ \bibnamefont {Wyld}},\ }\href
  {\doibase 10.1103/PhysRev.106.517} {\bibfield  {journal} {\bibinfo  {journal}
  {Physical Review}\ }\textbf {\bibinfo {volume} {106}},\ \bibinfo {pages} {517
  } (\bibinfo {year} {1957})}\BibitemShut {NoStop}%
\bibitem [{\citenamefont {Huber}\ \emph {et~al.}(2003)\citenamefont {Huber},
  \citenamefont {Lang}, \citenamefont {Navert}, \citenamefont {Stromicki},
  \citenamefont {Bodek}, \citenamefont {Kistryn}, \citenamefont {Zejma},
  \citenamefont {Naviliat-Cuncic}, \citenamefont {Stephan},\ and\ \citenamefont
  {Haeberli}}]{Huber2003}%
  \BibitemOpen
  \bibfield  {author} {\bibinfo {author} {\bibfnamefont {R.}~\bibnamefont
  {Huber}}, \bibinfo {author} {\bibfnamefont {J.}~\bibnamefont {Lang}},
  \bibinfo {author} {\bibfnamefont {S.}~\bibnamefont {Navert}}, \bibinfo
  {author} {\bibfnamefont {J.}~\bibnamefont {Stromicki}}, \bibinfo {author}
  {\bibfnamefont {K.}~\bibnamefont {Bodek}}, \bibinfo {author} {\bibfnamefont
  {S.}~\bibnamefont {Kistryn}}, \bibinfo {author} {\bibfnamefont
  {J.}~\bibnamefont {Zejma}}, \bibinfo {author} {\bibfnamefont
  {O.}~\bibnamefont {Naviliat-Cuncic}}, \bibinfo {author} {\bibfnamefont
  {E.}~\bibnamefont {Stephan}}, \ and\ \bibinfo {author} {\bibfnamefont
  {W.}~\bibnamefont {Haeberli}},\ }\href {\doibase
  10.1103/PhysRevLett.90.202301} {\bibfield  {journal} {\bibinfo  {journal}
  {Physical Review Letters}\ }\textbf {\bibinfo {volume} {90}},\ \bibinfo
  {pages} {202301} (\bibinfo {year} {2003})}\BibitemShut {NoStop}%
\bibitem [{\citenamefont {Stone}\ and\ \citenamefont
  {Postma}(1986)}]{Stone1986}%
  \BibitemOpen
  \bibfield  {author} {\bibinfo {author} {\bibfnamefont {N.}~\bibnamefont
  {Stone}}\ and\ \bibinfo {author} {\bibfnamefont {H.}~\bibnamefont {Postma}},\
  }\href@noop {} {\emph {\bibinfo {title} {Low-Temperature Nuclear
  Orientation}}}\ (\bibinfo  {publisher} {North-Holland, Amsterdam},\ \bibinfo
  {year} {1986})\BibitemShut {NoStop}%
\bibitem [{\citenamefont {Severijns}\ \emph {et~al.}(2005)\citenamefont
  {Severijns}, \citenamefont {V\'enos}, \citenamefont {Schuurmans},
  \citenamefont {Phalet}, \citenamefont {Honusek}, \citenamefont {Srnka},
  \citenamefont {Vereecke}, \citenamefont {Versyck}, \citenamefont
  {Z\'akouck\'y}, \citenamefont {K\"oster}, \citenamefont {Beck}, \citenamefont
  {Delaur\'e}, \citenamefont {Golovko},\ and\ \citenamefont
  {Kraev}}]{Severijns2005}%
  \BibitemOpen
  \bibfield  {author} {\bibinfo {author} {\bibfnamefont {N.}~\bibnamefont
  {Severijns}}, \bibinfo {author} {\bibfnamefont {D.}~\bibnamefont {V\'enos}},
  \bibinfo {author} {\bibfnamefont {P.}~\bibnamefont {Schuurmans}}, \bibinfo
  {author} {\bibfnamefont {T.}~\bibnamefont {Phalet}}, \bibinfo {author}
  {\bibfnamefont {M.}~\bibnamefont {Honusek}}, \bibinfo {author} {\bibfnamefont
  {D.}~\bibnamefont {Srnka}}, \bibinfo {author} {\bibfnamefont
  {B.}~\bibnamefont {Vereecke}}, \bibinfo {author} {\bibfnamefont
  {S.}~\bibnamefont {Versyck}}, \bibinfo {author} {\bibfnamefont
  {D.}~\bibnamefont {Z\'akouck\'y}}, \bibinfo {author} {\bibfnamefont
  {U.}~\bibnamefont {K\"oster}}, \bibinfo {author} {\bibfnamefont
  {M.}~\bibnamefont {Beck}}, \bibinfo {author} {\bibfnamefont {B.}~\bibnamefont
  {Delaur\'e}}, \bibinfo {author} {\bibfnamefont {V.}~\bibnamefont {Golovko}},
  \ and\ \bibinfo {author} {\bibfnamefont {I.}~\bibnamefont {Kraev}},\
  }\href@noop {} {\bibfield  {journal} {\bibinfo  {journal} {Physical Review
  C}\ }\textbf {\bibinfo {volume} {71}},\ \bibinfo {pages} {064310} (\bibinfo
  {year} {2005})}\BibitemShut {NoStop}%
\bibitem [{\citenamefont {Vingerhoets}\ \emph {et~al.}(2010)\citenamefont
  {Vingerhoets}, \citenamefont {Flanagan}, \citenamefont {Avgoulea},
  \citenamefont {Billowes}, \citenamefont {Bissell}, \citenamefont {Blaum},
  \citenamefont {Brown}, \citenamefont {Cheal}, \citenamefont {De~Rydt},
  \citenamefont {Forest}, \citenamefont {Geppert}, \citenamefont {Honma},
  \citenamefont {Kowalska}, \citenamefont {Kr\"amer}, \citenamefont {Krieger},
  \citenamefont {Man\'e}, \citenamefont {Neugart}, \citenamefont {Neyens},
  \citenamefont {N\"ortersh\"auser}, \citenamefont {Otsuka}, \citenamefont
  {Schug}, \citenamefont {Stroke}, \citenamefont {Tungate},\ and\ \citenamefont
  {Yordanov}}]{Vingerhoets2010}%
  \BibitemOpen
  \bibfield  {author} {\bibinfo {author} {\bibfnamefont {P.}~\bibnamefont
  {Vingerhoets}}, \bibinfo {author} {\bibfnamefont {K.~T.}\ \bibnamefont
  {Flanagan}}, \bibinfo {author} {\bibfnamefont {M.}~\bibnamefont {Avgoulea}},
  \bibinfo {author} {\bibfnamefont {J.}~\bibnamefont {Billowes}}, \bibinfo
  {author} {\bibfnamefont {M.~L.}\ \bibnamefont {Bissell}}, \bibinfo {author}
  {\bibfnamefont {K.}~\bibnamefont {Blaum}}, \bibinfo {author} {\bibfnamefont
  {B.~A.}\ \bibnamefont {Brown}}, \bibinfo {author} {\bibfnamefont
  {B.}~\bibnamefont {Cheal}}, \bibinfo {author} {\bibfnamefont
  {M.}~\bibnamefont {De~Rydt}}, \bibinfo {author} {\bibfnamefont {D.~H.}\
  \bibnamefont {Forest}}, \bibinfo {author} {\bibfnamefont {C.}~\bibnamefont
  {Geppert}}, \bibinfo {author} {\bibfnamefont {M.}~\bibnamefont {Honma}},
  \bibinfo {author} {\bibfnamefont {M.}~\bibnamefont {Kowalska}}, \bibinfo
  {author} {\bibfnamefont {J.}~\bibnamefont {Kr\"amer}}, \bibinfo {author}
  {\bibfnamefont {A.}~\bibnamefont {Krieger}}, \bibinfo {author} {\bibfnamefont
  {E.}~\bibnamefont {Man\'e}}, \bibinfo {author} {\bibfnamefont
  {R.}~\bibnamefont {Neugart}}, \bibinfo {author} {\bibfnamefont
  {G.}~\bibnamefont {Neyens}}, \bibinfo {author} {\bibfnamefont
  {W.}~\bibnamefont {N\"ortersh\"auser}}, \bibinfo {author} {\bibfnamefont
  {T.}~\bibnamefont {Otsuka}}, \bibinfo {author} {\bibfnamefont
  {M.}~\bibnamefont {Schug}}, \bibinfo {author} {\bibfnamefont {H.~H.}\
  \bibnamefont {Stroke}}, \bibinfo {author} {\bibfnamefont {G.}~\bibnamefont
  {Tungate}}, \ and\ \bibinfo {author} {\bibfnamefont {D.~T.}\ \bibnamefont
  {Yordanov}},\ }\href {\doibase 10.1103/PhysRevC.82.064311} {\bibfield
  {journal} {\bibinfo  {journal} {Physical Review C}\ }\textbf {\bibinfo
  {volume} {82}},\ \bibinfo {pages} {064311} (\bibinfo {year}
  {2010})}\BibitemShut {NoStop}%
\bibitem [{\citenamefont {Junde}\ \emph {et~al.}(2005)\citenamefont {Junde},
  \citenamefont {Xiaolong},\ and\ \citenamefont {Tuli}}]{Junde2005}%
  \BibitemOpen
  \bibfield  {author} {\bibinfo {author} {\bibfnamefont {H.}~\bibnamefont
  {Junde}}, \bibinfo {author} {\bibfnamefont {H.}~\bibnamefont {Xiaolong}}, \
  and\ \bibinfo {author} {\bibfnamefont {J.}~\bibnamefont {Tuli}},\ }\href
  {\doibase 10.1016/j.nds.2005.10.006} {\bibfield  {journal} {\bibinfo
  {journal} {Nuclear Data Sheets}\ }\textbf {\bibinfo {volume} {106}},\
  \bibinfo {pages} {159} (\bibinfo {year} {2005})}\BibitemShut {NoStop}%
\bibitem [{\citenamefont {McCutchan}(2012)}]{McCutchan2012}%
  \BibitemOpen
  \bibfield  {author} {\bibinfo {author} {\bibfnamefont {E.}~\bibnamefont
  {McCutchan}},\ }\href {\doibase 10.1016/j.nds.2012.06.002} {\bibfield
  {journal} {\bibinfo  {journal} {Nuclear Data Sheets}\ }\textbf {\bibinfo
  {volume} {113}},\ \bibinfo {pages} {1735} (\bibinfo {year}
  {2012})}\BibitemShut {NoStop}%
\bibitem [{\citenamefont {K\"{o}ster}\ \emph {et~al.}(2003)\citenamefont
  {K\"{o}ster}, \citenamefont {Bergmann}, \citenamefont {Carminati},
  \citenamefont {Catherall}, \citenamefont {Cederk\"{a}ll}, \citenamefont
  {Correia}, \citenamefont {Crepieux}, \citenamefont {Dietrich}, \citenamefont
  {Elder}, \citenamefont {Fedoseyev}, \citenamefont {Fraile}, \citenamefont
  {Franchoo}, \citenamefont {Fynbo}, \citenamefont {Georg}, \citenamefont
  {Giles}, \citenamefont {Joinet}, \citenamefont {Jonsson}, \citenamefont
  {Kirchner}, \citenamefont {Lau}, \citenamefont {Lettry}, \citenamefont
  {Maier}, \citenamefont {Mishin}, \citenamefont {Oinonen}, \citenamefont
  {Per\"{a}j\"{a}rvi}, \citenamefont {Ravn}, \citenamefont {Rinaldi},
  \citenamefont {Santana-Leitner}, \citenamefont {Wahl},\ and\ \citenamefont
  {Weissman}}]{Koster2003}%
  \BibitemOpen
  \bibfield  {author} {\bibinfo {author} {\bibfnamefont {U.}~\bibnamefont
  {K\"{o}ster}}, \bibinfo {author} {\bibfnamefont {U.}~\bibnamefont
  {Bergmann}}, \bibinfo {author} {\bibfnamefont {D.}~\bibnamefont {Carminati}},
  \bibinfo {author} {\bibfnamefont {R.}~\bibnamefont {Catherall}}, \bibinfo
  {author} {\bibfnamefont {J.}~\bibnamefont {Cederk\"{a}ll}}, \bibinfo {author}
  {\bibfnamefont {J.}~\bibnamefont {Correia}}, \bibinfo {author} {\bibfnamefont
  {B.}~\bibnamefont {Crepieux}}, \bibinfo {author} {\bibfnamefont
  {M.}~\bibnamefont {Dietrich}}, \bibinfo {author} {\bibfnamefont
  {K.}~\bibnamefont {Elder}}, \bibinfo {author} {\bibfnamefont
  {V.}~\bibnamefont {Fedoseyev}}, \bibinfo {author} {\bibfnamefont
  {L.}~\bibnamefont {Fraile}}, \bibinfo {author} {\bibfnamefont
  {S.}~\bibnamefont {Franchoo}}, \bibinfo {author} {\bibfnamefont
  {H.}~\bibnamefont {Fynbo}}, \bibinfo {author} {\bibfnamefont
  {U.}~\bibnamefont {Georg}}, \bibinfo {author} {\bibfnamefont
  {T.}~\bibnamefont {Giles}}, \bibinfo {author} {\bibfnamefont
  {a.}~\bibnamefont {Joinet}}, \bibinfo {author} {\bibfnamefont
  {O.}~\bibnamefont {Jonsson}}, \bibinfo {author} {\bibfnamefont
  {R.}~\bibnamefont {Kirchner}}, \bibinfo {author} {\bibfnamefont
  {C.}~\bibnamefont {Lau}}, \bibinfo {author} {\bibfnamefont {J.}~\bibnamefont
  {Lettry}}, \bibinfo {author} {\bibfnamefont {H.}~\bibnamefont {Maier}},
  \bibinfo {author} {\bibfnamefont {V.}~\bibnamefont {Mishin}}, \bibinfo
  {author} {\bibfnamefont {M.}~\bibnamefont {Oinonen}}, \bibinfo {author}
  {\bibfnamefont {K.}~\bibnamefont {Per\"{a}j\"{a}rvi}}, \bibinfo {author}
  {\bibfnamefont {H.}~\bibnamefont {Ravn}}, \bibinfo {author} {\bibfnamefont
  {T.}~\bibnamefont {Rinaldi}}, \bibinfo {author} {\bibfnamefont
  {M.}~\bibnamefont {Santana-Leitner}}, \bibinfo {author} {\bibfnamefont
  {U.}~\bibnamefont {Wahl}}, \ and\ \bibinfo {author} {\bibfnamefont
  {L.}~\bibnamefont {Weissman}},\ }\href {\doibase
  10.1016/S0168-583X(03)00505-6} {\bibfield  {journal} {\bibinfo  {journal}
  {Nuclear Instruments and Methods in Physics Research Section B: Beam
  Interactions with Materials and Atoms}\ }\textbf {\bibinfo {volume} {204}},\
  \bibinfo {pages} {303} (\bibinfo {year} {2003})}\BibitemShut {NoStop}%
\bibitem [{\citenamefont {K\"{o}ster}\ \emph
  {et~al.}(2000{\natexlab{a}})\citenamefont {K\"{o}ster}, \citenamefont
  {Fedoseyev}, \citenamefont {Mishin}, \citenamefont {Weissman}, \citenamefont
  {Huyse}, \citenamefont {Kruglov}, \citenamefont {Mueller}, \citenamefont
  {{Van Duppen}}, \citenamefont {{Van Roosbroeck}}, \citenamefont {Thirolf},
  \citenamefont {Thomas}, \citenamefont {Weisshaar}, \citenamefont {Schulze},
  \citenamefont {Borcea}, \citenamefont {{La Commara}}, \citenamefont {Schatz},
  \citenamefont {Schmidt}, \citenamefont {R\"{o}ttger}, \citenamefont {Huber},
  \citenamefont {Sebastian}, \citenamefont {Kratz}, \citenamefont {Catherall},
  \citenamefont {Georg}, \citenamefont {Lettry}, \citenamefont {Oinonen},
  \citenamefont {Ravn},\ and\ \citenamefont {Simon}}]{Koster2000}%
  \BibitemOpen
  \bibfield  {author} {\bibinfo {author} {\bibfnamefont {U.}~\bibnamefont
  {K\"{o}ster}}, \bibinfo {author} {\bibfnamefont {V.}~\bibnamefont
  {Fedoseyev}}, \bibinfo {author} {\bibfnamefont {V.}~\bibnamefont {Mishin}},
  \bibinfo {author} {\bibfnamefont {L.}~\bibnamefont {Weissman}}, \bibinfo
  {author} {\bibfnamefont {M.}~\bibnamefont {Huyse}}, \bibinfo {author}
  {\bibfnamefont {K.}~\bibnamefont {Kruglov}}, \bibinfo {author} {\bibfnamefont
  {W.}~\bibnamefont {Mueller}}, \bibinfo {author} {\bibfnamefont
  {P.}~\bibnamefont {{Van Duppen}}}, \bibinfo {author} {\bibfnamefont
  {J.}~\bibnamefont {{Van Roosbroeck}}}, \bibinfo {author} {\bibfnamefont
  {P.}~\bibnamefont {Thirolf}}, \bibinfo {author} {\bibfnamefont
  {H.}~\bibnamefont {Thomas}}, \bibinfo {author} {\bibfnamefont
  {D.}~\bibnamefont {Weisshaar}}, \bibinfo {author} {\bibfnamefont
  {W.}~\bibnamefont {Schulze}}, \bibinfo {author} {\bibfnamefont
  {R.}~\bibnamefont {Borcea}}, \bibinfo {author} {\bibfnamefont
  {M.}~\bibnamefont {{La Commara}}}, \bibinfo {author} {\bibfnamefont
  {H.}~\bibnamefont {Schatz}}, \bibinfo {author} {\bibfnamefont
  {K.}~\bibnamefont {Schmidt}}, \bibinfo {author} {\bibfnamefont
  {S.}~\bibnamefont {R\"{o}ttger}}, \bibinfo {author} {\bibfnamefont
  {G.}~\bibnamefont {Huber}}, \bibinfo {author} {\bibfnamefont
  {V.}~\bibnamefont {Sebastian}}, \bibinfo {author} {\bibfnamefont
  {K.}~\bibnamefont {Kratz}}, \bibinfo {author} {\bibfnamefont
  {R.}~\bibnamefont {Catherall}}, \bibinfo {author} {\bibfnamefont
  {U.}~\bibnamefont {Georg}}, \bibinfo {author} {\bibfnamefont
  {J.}~\bibnamefont {Lettry}}, \bibinfo {author} {\bibfnamefont
  {M.}~\bibnamefont {Oinonen}}, \bibinfo {author} {\bibfnamefont
  {H.}~\bibnamefont {Ravn}}, \ and\ \bibinfo {author} {\bibfnamefont
  {H.}~\bibnamefont {Simon}},\ }\href {\doibase 10.1016/S0168-583X(99)00552-2}
  {\bibfield  {journal} {\bibinfo  {journal} {Nuclear Instruments and Methods
  in Physics Research Section B: Beam Interactions with Materials and Atoms}\
  }\textbf {\bibinfo {volume} {160}},\ \bibinfo {pages} {528} (\bibinfo {year}
  {2000}{\natexlab{a}})}\BibitemShut {NoStop}%
\bibitem [{\citenamefont {Weissman}\ \emph {et~al.}(2002)\citenamefont
  {Weissman}, \citenamefont {Cederkall}, \citenamefont {\"{A}yst\"{o}},
  \citenamefont {Fynbo}, \citenamefont {Fraile}, \citenamefont {Fedosseyev},
  \citenamefont {Franchoo}, \citenamefont {Jokinen}, \citenamefont
  {K\"{o}ster}, \citenamefont {Mart\'{\i}nez-Pinedo}, \citenamefont {Nilsson},
  \citenamefont {Oinonen}, \citenamefont {Per\"{a}j\"{a}rvi}, \citenamefont
  {Seliverstov},\ and\ \citenamefont {{ISOLDECollaboration}}}]{Weissman2002a}%
  \BibitemOpen
  \bibfield  {author} {\bibinfo {author} {\bibfnamefont {L.}~\bibnamefont
  {Weissman}}, \bibinfo {author} {\bibfnamefont {J.}~\bibnamefont {Cederkall}},
  \bibinfo {author} {\bibfnamefont {J.}~\bibnamefont {\"{A}yst\"{o}}}, \bibinfo
  {author} {\bibfnamefont {H.}~\bibnamefont {Fynbo}}, \bibinfo {author}
  {\bibfnamefont {L.}~\bibnamefont {Fraile}}, \bibinfo {author} {\bibfnamefont
  {V.}~\bibnamefont {Fedosseyev}}, \bibinfo {author} {\bibfnamefont
  {S.}~\bibnamefont {Franchoo}}, \bibinfo {author} {\bibfnamefont
  {A.}~\bibnamefont {Jokinen}}, \bibinfo {author} {\bibfnamefont
  {U.}~\bibnamefont {K\"{o}ster}}, \bibinfo {author} {\bibfnamefont
  {G.}~\bibnamefont {Mart\'{\i}nez-Pinedo}}, \bibinfo {author} {\bibfnamefont
  {T.}~\bibnamefont {Nilsson}}, \bibinfo {author} {\bibfnamefont
  {M.}~\bibnamefont {Oinonen}}, \bibinfo {author} {\bibfnamefont
  {K.}~\bibnamefont {Per\"{a}j\"{a}rvi}}, \bibinfo {author} {\bibfnamefont
  {M.~D.}\ \bibnamefont {Seliverstov}}, \ and\ \bibinfo {author} {\bibnamefont
  {{ISOLDECollaboration}}},\ }\href {\doibase 10.1103/PhysRevC.65.044321}
  {\bibfield  {journal} {\bibinfo  {journal} {Physical Review C}\ }\textbf
  {\bibinfo {volume} {65}},\ \bibinfo {pages} {044321} (\bibinfo {year}
  {2002})}\BibitemShut {NoStop}%
\bibitem [{\citenamefont {Schl\"osser}\ \emph {et~al.}(1988)\citenamefont
  {Schl\"osser}, \citenamefont {Berkes}, \citenamefont {Hagn}, \citenamefont
  {Herzog}, \citenamefont {Niinikoski}, \citenamefont {Postma}, \citenamefont
  {Richard-Serre}, \citenamefont {Rikovska}, \citenamefont {Stone},
  \citenamefont {Vanneste},\ and\ \citenamefont {Zech}}]{Schloesser1988}%
  \BibitemOpen
  \bibfield  {author} {\bibinfo {author} {\bibfnamefont {K.}~\bibnamefont
  {Schl\"osser}}, \bibinfo {author} {\bibfnamefont {I.}~\bibnamefont {Berkes}},
  \bibinfo {author} {\bibfnamefont {E.}~\bibnamefont {Hagn}}, \bibinfo {author}
  {\bibfnamefont {P.}~\bibnamefont {Herzog}}, \bibinfo {author} {\bibfnamefont
  {T.}~\bibnamefont {Niinikoski}}, \bibinfo {author} {\bibfnamefont
  {H.}~\bibnamefont {Postma}}, \bibinfo {author} {\bibfnamefont
  {C.}~\bibnamefont {Richard-Serre}}, \bibinfo {author} {\bibfnamefont
  {J.}~\bibnamefont {Rikovska}}, \bibinfo {author} {\bibfnamefont
  {N.}~\bibnamefont {Stone}}, \bibinfo {author} {\bibfnamefont
  {L.}~\bibnamefont {Vanneste}}, \ and\ \bibinfo {author} {\bibfnamefont
  {E.}~\bibnamefont {Zech}},\ }\href {http://dx.doi.org/10.1007/BF02398294}
  {\bibfield  {journal} {\bibinfo  {journal} {Hyperfine Interactions}\ }\textbf
  {\bibinfo {volume} {43}},\ \bibinfo {pages} {139} (\bibinfo {year} {1988})},\
  \bibinfo {note} {10.1007/BF02398294}\BibitemShut {NoStop}%
\bibitem [{\citenamefont {Wouters}\ \emph {et~al.}(1990)\citenamefont
  {Wouters}, \citenamefont {Severijns}, \citenamefont {Vanhaverbeke},
  \citenamefont {Vanderpoorten},\ and\ \citenamefont {Vanneste}}]{Wouters1990}%
  \BibitemOpen
  \bibfield  {author} {\bibinfo {author} {\bibfnamefont {J.}~\bibnamefont
  {Wouters}}, \bibinfo {author} {\bibfnamefont {N.}~\bibnamefont {Severijns}},
  \bibinfo {author} {\bibfnamefont {J.}~\bibnamefont {Vanhaverbeke}}, \bibinfo
  {author} {\bibfnamefont {W.}~\bibnamefont {Vanderpoorten}}, \ and\ \bibinfo
  {author} {\bibfnamefont {L.}~\bibnamefont {Vanneste}},\ }\href
  {http://dx.doi.org/10.1007/BF02401194} {\bibfield  {journal} {\bibinfo
  {journal} {Hyperfine Interactions}\ }\textbf {\bibinfo {volume} {59}},\
  \bibinfo {pages} {59} (\bibinfo {year} {1990})},\ \bibinfo {note}
  {10.1007/BF02401194}\BibitemShut {NoStop}%
\bibitem [{\citenamefont {V\'enos}\ \emph {et~al.}(2000)\citenamefont
  {V\'enos}, \citenamefont {Van~Assche}, \citenamefont {Severijns},
  \citenamefont {Srnka},\ and\ \citenamefont {Z\'akouck\'y}}]{Venos2000}%
  \BibitemOpen
  \bibfield  {author} {\bibinfo {author} {\bibfnamefont {D.}~\bibnamefont
  {V\'enos}}, \bibinfo {author} {\bibfnamefont {G.}~\bibnamefont {Van~Assche}},
  \bibinfo {author} {\bibfnamefont {N.}~\bibnamefont {Severijns}}, \bibinfo
  {author} {\bibfnamefont {D.}~\bibnamefont {Srnka}}, \ and\ \bibinfo {author}
  {\bibfnamefont {D.}~\bibnamefont {Z\'akouck\'y}},\ }\href {\doibase
  10.1016/S0168-9002(00)00494-0} {\bibfield  {journal} {\bibinfo  {journal}
  {Nuclear Instruments and Methods in Physics Research, Section A:
  Accelerators, Spectrometers, Detectors, and Associated Equipment}\ }\textbf
  {\bibinfo {volume} {454}},\ \bibinfo {pages} {403} (\bibinfo {year}
  {2000})}\BibitemShut {NoStop}%
\bibitem [{\citenamefont {Z\'{a}kouck\'{y}}\ \emph {et~al.}(2004)\citenamefont
  {Z\'{a}kouck\'{y}}, \citenamefont {Srnka}, \citenamefont {V\'{e}nos},
  \citenamefont {Golovko}, \citenamefont {Kraev}, \citenamefont {Phalet},
  \citenamefont {Schuurmans}, \citenamefont {Severijns}, \citenamefont
  {Vereecke},\ and\ \citenamefont {Versyck}}]{Zakoucky2004}%
  \BibitemOpen
  \bibfield  {author} {\bibinfo {author} {\bibfnamefont {D.}~\bibnamefont
  {Z\'{a}kouck\'{y}}}, \bibinfo {author} {\bibfnamefont {D.}~\bibnamefont
  {Srnka}}, \bibinfo {author} {\bibfnamefont {D.}~\bibnamefont {V\'{e}nos}},
  \bibinfo {author} {\bibfnamefont {V.}~\bibnamefont {Golovko}}, \bibinfo
  {author} {\bibfnamefont {I.}~\bibnamefont {Kraev}}, \bibinfo {author}
  {\bibfnamefont {T.}~\bibnamefont {Phalet}}, \bibinfo {author} {\bibfnamefont
  {P.}~\bibnamefont {Schuurmans}}, \bibinfo {author} {\bibfnamefont
  {N.}~\bibnamefont {Severijns}}, \bibinfo {author} {\bibfnamefont
  {B.}~\bibnamefont {Vereecke}}, \ and\ \bibinfo {author} {\bibfnamefont
  {S.}~\bibnamefont {Versyck}},\ }\href {\doibase 10.1016/j.nima.2003.11.226}
  {\bibfield  {journal} {\bibinfo  {journal} {Nuclear Instruments and Methods
  in Physics Research, Section A: Accelerators, Spectrometers, Detectors, and
  Associated Equipment}\ }\textbf {\bibinfo {volume} {520}},\ \bibinfo {pages}
  {80 } (\bibinfo {year} {2004})},\ \bibinfo {note} {proceedings of the 10th
  International Workshop on Low Temperature Detectors}\BibitemShut {NoStop}%
\bibitem [{\citenamefont {Soti}\ \emph {et~al.}(2013)\citenamefont {Soti},
  \citenamefont {Wauters}, \citenamefont {Breitenfeldt}, \citenamefont
  {Finlay}, \citenamefont {Kraev}, \citenamefont {Knecht}, \citenamefont
  {Porobi\'{c}}, \citenamefont {Z\'{a}kouck\'{y}},\ and\ \citenamefont
  {Severijns}}]{Soti2013}%
  \BibitemOpen
  \bibfield  {author} {\bibinfo {author} {\bibfnamefont {G.}~\bibnamefont
  {Soti}}, \bibinfo {author} {\bibfnamefont {F.}~\bibnamefont {Wauters}},
  \bibinfo {author} {\bibfnamefont {M.}~\bibnamefont {Breitenfeldt}}, \bibinfo
  {author} {\bibfnamefont {P.}~\bibnamefont {Finlay}}, \bibinfo {author}
  {\bibfnamefont {I.}~\bibnamefont {Kraev}}, \bibinfo {author} {\bibfnamefont
  {A.}~\bibnamefont {Knecht}}, \bibinfo {author} {\bibfnamefont
  {T.}~\bibnamefont {Porobi\'{c}}}, \bibinfo {author} {\bibfnamefont
  {D.}~\bibnamefont {Z\'{a}kouck\'{y}}}, \ and\ \bibinfo {author}
  {\bibfnamefont {N.}~\bibnamefont {Severijns}},\ }\href {\doibase
  10.1016/j.nima.2013.06.047} {\bibfield  {journal} {\bibinfo  {journal}
  {Nuclear Instruments and Methods in Physics Research Section A: Accelerators,
  Spectrometers, Detectors and Associated Equipment}\ }\textbf {\bibinfo
  {volume} {728}},\ \bibinfo {pages} {11} (\bibinfo {year} {2013})}\BibitemShut
  {NoStop}%
\bibitem [{\citenamefont {Krane}(1986)}]{Krane1986}%
  \BibitemOpen
  \bibfield  {author} {\bibinfo {author} {\bibfnamefont {K.}~\bibnamefont
  {Krane}},\ }in\ \href@noop {} {\emph {\bibinfo {booktitle} {Low-Temperature
  Nuclear Orientation}}},\ \bibinfo {editor} {edited by\ \bibinfo {editor}
  {\bibfnamefont {N.}~\bibnamefont {Stone}}\ and\ \bibinfo {editor}
  {\bibfnamefont {H.}~\bibnamefont {Postma}}}\ (\bibinfo  {publisher}
  {North-Holland, Amsterdam},\ \bibinfo {year} {1986})\ Chap.~\bibinfo
  {chapter} {3}, p.~\bibinfo {pages} {31}\BibitemShut {NoStop}%
\bibitem [{\citenamefont {Golovko}\ \emph {et~al.}(2011)\citenamefont
  {Golovko}, \citenamefont {Wauters}, \citenamefont {Cottenier}, \citenamefont
  {Breitenfeldt}, \citenamefont {De~Leebeeck}, \citenamefont {Roccia},
  \citenamefont {Soti}, \citenamefont {Tandecki}, \citenamefont {Traykov},
  \citenamefont {Van~Gorp}, \citenamefont {Z\'akouck\'y},\ and\ \citenamefont
  {Severijns}}]{Golovko2011}%
  \BibitemOpen
  \bibfield  {author} {\bibinfo {author} {\bibfnamefont {V.~V.}\ \bibnamefont
  {Golovko}}, \bibinfo {author} {\bibfnamefont {F.}~\bibnamefont {Wauters}},
  \bibinfo {author} {\bibfnamefont {S.}~\bibnamefont {Cottenier}}, \bibinfo
  {author} {\bibfnamefont {M.}~\bibnamefont {Breitenfeldt}}, \bibinfo {author}
  {\bibfnamefont {V.}~\bibnamefont {De~Leebeeck}}, \bibinfo {author}
  {\bibfnamefont {S.}~\bibnamefont {Roccia}}, \bibinfo {author} {\bibfnamefont
  {G.}~\bibnamefont {Soti}}, \bibinfo {author} {\bibfnamefont {M.}~\bibnamefont
  {Tandecki}}, \bibinfo {author} {\bibfnamefont {E.}~\bibnamefont {Traykov}},
  \bibinfo {author} {\bibfnamefont {S.}~\bibnamefont {Van~Gorp}}, \bibinfo
  {author} {\bibfnamefont {D.}~\bibnamefont {Z\'akouck\'y}}, \ and\ \bibinfo
  {author} {\bibfnamefont {N.}~\bibnamefont {Severijns}},\ }\href {\doibase
  10.1103/PhysRevC.84.014323} {\bibfield  {journal} {\bibinfo  {journal}
  {Physical Review C}\ }\textbf {\bibinfo {volume} {84}},\ \bibinfo {pages}
  {014323} (\bibinfo {year} {2011})}\BibitemShut {NoStop}%
\bibitem [{\citenamefont {Chikazumi}(1964)}]{Chikazumi1964}%
  \BibitemOpen
  \bibfield  {author} {\bibinfo {author} {\bibfnamefont {S.}~\bibnamefont
  {Chikazumi}},\ }\href@noop {} {\emph {\bibinfo {title} {Physics of
  magnetism}}}\ (\bibinfo  {publisher} {J. Wiley and Sons, New York},\ \bibinfo
  {year} {1964})\BibitemShut {NoStop}%
\bibitem [{\citenamefont {Marshak}(1986)}]{Marshak1986}%
  \BibitemOpen
  \bibfield  {author} {\bibinfo {author} {\bibfnamefont {H.}~\bibnamefont
  {Marshak}},\ }in\ \href@noop {} {\emph {\bibinfo {booktitle} {Low-Temperature
  Nuclear Orientation}}},\ \bibinfo {editor} {edited by\ \bibinfo {editor}
  {\bibfnamefont {N.}~\bibnamefont {Stone}}\ and\ \bibinfo {editor}
  {\bibfnamefont {H.}~\bibnamefont {Postma}}}\ (\bibinfo  {publisher}
  {North-Holland, Amsterdam},\ \bibinfo {year} {1986})\ Chap.~\bibinfo
  {chapter} {16}, p.\ \bibinfo {pages} {769}\BibitemShut {NoStop}%
\bibitem [{\citenamefont {Funk}\ \emph {et~al.}(1999)\citenamefont {Funk},
  \citenamefont {Beck}, \citenamefont {Brewer}, \citenamefont {Bobek},\ and\
  \citenamefont {Klein}}]{Funk1999}%
  \BibitemOpen
  \bibfield  {author} {\bibinfo {author} {\bibfnamefont {T.}~\bibnamefont
  {Funk}}, \bibinfo {author} {\bibfnamefont {E.}~\bibnamefont {Beck}}, \bibinfo
  {author} {\bibfnamefont {W.}~\bibnamefont {Brewer}}, \bibinfo {author}
  {\bibfnamefont {C.}~\bibnamefont {Bobek}}, \ and\ \bibinfo {author}
  {\bibfnamefont {E.}~\bibnamefont {Klein}},\ }\href {\doibase
  10.1016/S0304-8853(99)00287-5} {\bibfield  {journal} {\bibinfo  {journal}
  {Journal of Magnetism and Magnetic Materials}\ }\textbf {\bibinfo {volume}
  {195}},\ \bibinfo {pages} {406} (\bibinfo {year} {1999})}\BibitemShut
  {NoStop}%
\bibitem [{\citenamefont {Shaw}\ and\ \citenamefont {Stone}(1989)}]{Shaw1989}%
  \BibitemOpen
  \bibfield  {author} {\bibinfo {author} {\bibfnamefont {T.~L.}\ \bibnamefont
  {Shaw}}\ and\ \bibinfo {author} {\bibfnamefont {N.~J.}\ \bibnamefont
  {Stone}},\ }\href {\doibase 10.1016/0092-640X(89)90011-9} {\bibfield
  {journal} {\bibinfo  {journal} {Atomic Data and Nuclear Data Tables}\
  }\textbf {\bibinfo {volume} {42}},\ \bibinfo {pages} {339 } (\bibinfo {year}
  {1989})}\BibitemShut {NoStop}%
\bibitem [{\citenamefont {Golovko}\ \emph {et~al.}(2006)\citenamefont
  {Golovko}, \citenamefont {Kraev}, \citenamefont {Phalet}, \citenamefont
  {Severijns}, \citenamefont {V\'{e}nos}, \citenamefont {Z\'{a}kouck\'{y}},
  \citenamefont {Herzog}, \citenamefont {Tramm}, \citenamefont {K\"{o}ster},
  \citenamefont {Srnka}, \citenamefont {Honusek}, \citenamefont {Delaur\'{e}},
  \citenamefont {Beck}, \citenamefont {Kozlov},\ and\ \citenamefont
  {Lindroth}}]{Golovko2006}%
  \BibitemOpen
  \bibfield  {author} {\bibinfo {author} {\bibfnamefont {V.~V.}\ \bibnamefont
  {Golovko}}, \bibinfo {author} {\bibfnamefont {I.~S.}\ \bibnamefont {Kraev}},
  \bibinfo {author} {\bibfnamefont {T.}~\bibnamefont {Phalet}}, \bibinfo
  {author} {\bibfnamefont {N.}~\bibnamefont {Severijns}}, \bibinfo {author}
  {\bibfnamefont {D.}~\bibnamefont {V\'{e}nos}}, \bibinfo {author}
  {\bibfnamefont {D.}~\bibnamefont {Z\'{a}kouck\'{y}}}, \bibinfo {author}
  {\bibfnamefont {P.}~\bibnamefont {Herzog}}, \bibinfo {author} {\bibfnamefont
  {C.}~\bibnamefont {Tramm}}, \bibinfo {author} {\bibfnamefont
  {U.}~\bibnamefont {K\"{o}ster}}, \bibinfo {author} {\bibfnamefont
  {D.}~\bibnamefont {Srnka}}, \bibinfo {author} {\bibfnamefont
  {M.}~\bibnamefont {Honusek}}, \bibinfo {author} {\bibfnamefont
  {B.}~\bibnamefont {Delaur\'{e}}}, \bibinfo {author} {\bibfnamefont
  {M.}~\bibnamefont {Beck}}, \bibinfo {author} {\bibfnamefont {V.~Y.}\
  \bibnamefont {Kozlov}}, \ and\ \bibinfo {author} {\bibfnamefont
  {A.}~\bibnamefont {Lindroth}},\ }\href {\doibase 10.1103/PhysRevC.74.044313}
  {\bibfield  {journal} {\bibinfo  {journal} {Physical Review C}\ }\textbf
  {\bibinfo {volume} {74}},\ \bibinfo {eid} {044313} (\bibinfo {year}
  {2006})}\BibitemShut {NoStop}%
\bibitem [{\citenamefont {Kontani}\ \emph {et~al.}(1972)\citenamefont
  {Kontani}, \citenamefont {Hioki},\ and\ \citenamefont
  {Masuda}}]{Kontani1972}%
  \BibitemOpen
  \bibfield  {author} {\bibinfo {author} {\bibfnamefont {M.}~\bibnamefont
  {Kontani}}, \bibinfo {author} {\bibfnamefont {T.}~\bibnamefont {Hioki}}, \
  and\ \bibinfo {author} {\bibfnamefont {Y.}~\bibnamefont {Masuda}},\ }\href
  {\doibase 10.1143/JPSJ.32.416} {\bibfield  {journal} {\bibinfo  {journal}
  {Journal of the Physical Society of Japan}\ }\textbf {\bibinfo {volume}
  {32}},\ \bibinfo {pages} {416} (\bibinfo {year} {1972})}\BibitemShut
  {NoStop}%
\bibitem [{\citenamefont {Klein}(1986)}]{Klein1986}%
  \BibitemOpen
  \bibfield  {author} {\bibinfo {author} {\bibfnamefont {E.}~\bibnamefont
  {Klein}},\ }in\ \href@noop {} {\emph {\bibinfo {booktitle} {Low-Temperature
  Nuclear Orientation}}},\ \bibinfo {editor} {edited by\ \bibinfo {editor}
  {\bibfnamefont {N.}~\bibnamefont {Stone}}\ and\ \bibinfo {editor}
  {\bibfnamefont {H.}~\bibnamefont {Postma}}}\ (\bibinfo  {publisher}
  {North-Holland, Amsterdam},\ \bibinfo {year} {1986})\ Chap.~\bibinfo
  {chapter} {19}, p.\ \bibinfo {pages} {579}\BibitemShut {NoStop}%
\bibitem [{\citenamefont {van Walle}\ \emph {et~al.}(1986)\citenamefont {van
  Walle}, \citenamefont {Vandeplassche}, \citenamefont {Wouters}, \citenamefont
  {Severijns},\ and\ \citenamefont {Vanneste}}]{Walle1986}%
  \BibitemOpen
  \bibfield  {author} {\bibinfo {author} {\bibfnamefont {E.}~\bibnamefont {van
  Walle}}, \bibinfo {author} {\bibfnamefont {D.}~\bibnamefont {Vandeplassche}},
  \bibinfo {author} {\bibfnamefont {J.}~\bibnamefont {Wouters}}, \bibinfo
  {author} {\bibfnamefont {N.}~\bibnamefont {Severijns}}, \ and\ \bibinfo
  {author} {\bibfnamefont {L.}~\bibnamefont {Vanneste}},\ }\href {\doibase
  10.1103/PhysRevB.34.2014} {\bibfield  {journal} {\bibinfo  {journal}
  {Physical Review B}\ }\textbf {\bibinfo {volume} {34}},\ \bibinfo {pages}
  {2014} (\bibinfo {year} {1986})}\BibitemShut {NoStop}%
\bibitem [{\citenamefont {Herzog}(1985)}]{Herzog1985}%
  \BibitemOpen
  \bibfield  {author} {\bibinfo {author} {\bibfnamefont {P.}~\bibnamefont
  {Herzog}},\ }\href {\doibase 10.1007/BF02063986} {\bibfield  {journal}
  {\bibinfo  {journal} {Hyperfine Interactions}\ }\textbf {\bibinfo {volume}
  {22}},\ \bibinfo {pages} {151} (\bibinfo {year} {1985})}\BibitemShut
  {NoStop}%
\bibitem [{\citenamefont {Herzog}\ \emph {et~al.}(1987)\citenamefont {Herzog},
  \citenamefont {D\"{a}mmrich}, \citenamefont {Freitag}, \citenamefont
  {Prillwitz}, \citenamefont {Prinz},\ and\ \citenamefont
  {Schl\"{o}sser}}]{Herzog1987}%
  \BibitemOpen
  \bibfield  {author} {\bibinfo {author} {\bibfnamefont {P.}~\bibnamefont
  {Herzog}}, \bibinfo {author} {\bibfnamefont {U.}~\bibnamefont
  {D\"{a}mmrich}}, \bibinfo {author} {\bibfnamefont {K.}~\bibnamefont
  {Freitag}}, \bibinfo {author} {\bibfnamefont {B.}~\bibnamefont {Prillwitz}},
  \bibinfo {author} {\bibfnamefont {J.}~\bibnamefont {Prinz}}, \ and\ \bibinfo
  {author} {\bibfnamefont {K.}~\bibnamefont {Schl\"{o}sser}},\ }\href@noop {}
  {\bibfield  {journal} {\bibinfo  {journal} {Nuclear Instruments and Methods
  in Physics Research, Section B: Beam Interactions with Materials and Atoms}\
  }\textbf {\bibinfo {volume} {26}},\ \bibinfo {pages} {471 } (\bibinfo {year}
  {1987})}\BibitemShut {NoStop}%
\bibitem [{\citenamefont {D\"ammrich}\ and\ \citenamefont
  {Herzog}(1988)}]{Dammrich1988}%
  \BibitemOpen
  \bibfield  {author} {\bibinfo {author} {\bibfnamefont {U.}~\bibnamefont
  {D\"ammrich}}\ and\ \bibinfo {author} {\bibfnamefont {P.}~\bibnamefont
  {Herzog}},\ }\href {\doibase 10.1007/BF02398298} {\bibfield  {journal}
  {\bibinfo  {journal} {Hyperfine Interactions}\ }\textbf {\bibinfo {volume}
  {43}},\ \bibinfo {pages} {167} (\bibinfo {year} {1988})}\BibitemShut
  {NoStop}%
\bibitem [{\citenamefont {K\"{o}ster}\ \emph
  {et~al.}(2000{\natexlab{b}})\citenamefont {K\"{o}ster}, \citenamefont
  {Catherall}, \citenamefont {Fedoseyev}, \citenamefont {Franchoo},
  \citenamefont {Georg}, \citenamefont {Huyse}, \citenamefont {Kruglov},
  \citenamefont {Lettry}, \citenamefont {Mishin}, \citenamefont {Oinonen},
  \citenamefont {Ravn}, \citenamefont {Seliverstov}, \citenamefont {Simon},
  \citenamefont {Van~Duppen}, \citenamefont {Van~Roosbroeck},\ and\
  \citenamefont {Weissman}}]{Koster2000a}%
  \BibitemOpen
  \bibfield  {author} {\bibinfo {author} {\bibfnamefont {U.}~\bibnamefont
  {K\"{o}ster}}, \bibinfo {author} {\bibfnamefont {R.}~\bibnamefont
  {Catherall}}, \bibinfo {author} {\bibfnamefont {V.}~\bibnamefont
  {Fedoseyev}}, \bibinfo {author} {\bibfnamefont {S.}~\bibnamefont {Franchoo}},
  \bibinfo {author} {\bibfnamefont {U.}~\bibnamefont {Georg}}, \bibinfo
  {author} {\bibfnamefont {M.}~\bibnamefont {Huyse}}, \bibinfo {author}
  {\bibfnamefont {K.}~\bibnamefont {Kruglov}}, \bibinfo {author} {\bibfnamefont
  {J.}~\bibnamefont {Lettry}}, \bibinfo {author} {\bibfnamefont
  {V.}~\bibnamefont {Mishin}}, \bibinfo {author} {\bibfnamefont
  {M.}~\bibnamefont {Oinonen}}, \bibinfo {author} {\bibfnamefont
  {H.}~\bibnamefont {Ravn}}, \bibinfo {author} {\bibfnamefont {M.}~\bibnamefont
  {Seliverstov}}, \bibinfo {author} {\bibfnamefont {H.}~\bibnamefont {Simon}},
  \bibinfo {author} {\bibfnamefont {P.}~\bibnamefont {Van~Duppen}}, \bibinfo
  {author} {\bibfnamefont {J.}~\bibnamefont {Van~Roosbroeck}}, \ and\ \bibinfo
  {author} {\bibfnamefont {L.}~\bibnamefont {Weissman}},\ }\href {\doibase
  10.1023/A:1012661532704} {\bibfield  {journal} {\bibinfo  {journal}
  {Hyperfine Interactions}\ }\textbf {\bibinfo {volume} {127}},\ \bibinfo
  {pages} {417} (\bibinfo {year} {2000}{\natexlab{b}})}\BibitemShut {NoStop}%
\bibitem [{\citenamefont {Agostinelli}\ \emph {et~al.}(2003)\citenamefont
  {Agostinelli}, \citenamefont {Allison}, \citenamefont {Amako}, \citenamefont
  {Apostolakis}, \citenamefont {Araujo}, \citenamefont {Arce}, \citenamefont
  {Asai}, \citenamefont {Axen}, \citenamefont {Banerjee}, \citenamefont
  {Barrand}, \citenamefont {Behner}, \citenamefont {Bellagamba}, \citenamefont
  {Boudreau}, \citenamefont {Broglia}, \citenamefont {Brunengo}, \citenamefont
  {Burkhardt}, \citenamefont {Chauvie}, \citenamefont {Chuma}, \citenamefont
  {Chytracek}, \citenamefont {Cooperman}, \citenamefont {Cosmo}, \citenamefont
  {Degtyarenko}, \citenamefont {Dell'Acqua}, \citenamefont {Depaola},
  \citenamefont {Dietrich}, \citenamefont {Enami}, \citenamefont {Feliciello},
  \citenamefont {Ferguson}, \citenamefont {Fesefeldt}, \citenamefont {Folger},
  \citenamefont {Foppiano}, \citenamefont {Forti}, \citenamefont {Garelli},
  \citenamefont {Giani}, \citenamefont {Giannitrapani}, \citenamefont {Gibin},
  \citenamefont {Cadenas}, \citenamefont {Gonz\'{a}lez}, \citenamefont {Abril},
  \citenamefont {Greeniaus}, \citenamefont {Greiner}, \citenamefont {Grichine},
  \citenamefont {Grossheim}, \citenamefont {Guatelli}, \citenamefont
  {Gumplinger}, \citenamefont {Hamatsu}, \citenamefont {Hashimoto},
  \citenamefont {Hasui}, \citenamefont {Heikkinen}, \citenamefont {Howard},
  \citenamefont {Ivanchenko}, \citenamefont {Johnson}, \citenamefont {Jones},
  \citenamefont {Kallenbach}, \citenamefont {Kanaya}, \citenamefont {Kawabata},
  \citenamefont {Kawabata}, \citenamefont {Kawaguti}, \citenamefont {Kelner},
  \citenamefont {Kent}, \citenamefont {Kimura}, \citenamefont {Kodama},
  \citenamefont {Kokoulin}, \citenamefont {Kossov}, \citenamefont {Kurashige},
  \citenamefont {Lamanna}, \citenamefont {Lamp\'{e}n}, \citenamefont {Lara},
  \citenamefont {Lefebure}, \citenamefont {Lei}, \citenamefont {Liendl},
  \citenamefont {Lockman}, \citenamefont {Longo}, \citenamefont {Magni},
  \citenamefont {Maire}, \citenamefont {Medernach}, \citenamefont {Minamimoto},
  \citenamefont {de~Freitas}, \citenamefont {Morita}, \citenamefont {Murakami},
  \citenamefont {Nagamatu}, \citenamefont {Nartallo}, \citenamefont {Nieminen},
  \citenamefont {Nishimura}, \citenamefont {Ohtsubo}, \citenamefont {Okamura},
  \citenamefont {O'Neale}, \citenamefont {Oohata}, \citenamefont {Paech},
  \citenamefont {Perl}, \citenamefont {Pfeiffer}, \citenamefont {Pia},
  \citenamefont {Ranjard}, \citenamefont {Rybin}, \citenamefont {Sadilov},
  \citenamefont {Salvo}, \citenamefont {Santin}, \citenamefont {Sasaki},
  \citenamefont {Savvas}, \citenamefont {Sawada}, \citenamefont {Scherer},
  \citenamefont {Sei}, \citenamefont {Sirotenko}, \citenamefont {Smith},
  \citenamefont {Starkov}, \citenamefont {Stoecker}, \citenamefont {Sulkimo},
  \citenamefont {Takahata}, \citenamefont {Tanaka}, \citenamefont {Tcherniaev},
  \citenamefont {Tehrani}, \citenamefont {Tropeano}, \citenamefont {Truscott},
  \citenamefont {Uno}, \citenamefont {Urban}, \citenamefont {Urban},
  \citenamefont {Verderi}, \citenamefont {Walkden}, \citenamefont {Wander},
  \citenamefont {Weber}, \citenamefont {Wellisch}, \citenamefont {Wenaus},
  \citenamefont {Williams}, \citenamefont {Wright}, \citenamefont {Yamada},
  \citenamefont {Yoshida},\ and\ \citenamefont {Zschiesche}}]{Agostinelli2003}%
  \BibitemOpen
  \bibfield  {author} {\bibinfo {author} {\bibfnamefont {S.}~\bibnamefont
  {Agostinelli}}, \bibinfo {author} {\bibfnamefont {J.}~\bibnamefont
  {Allison}}, \bibinfo {author} {\bibfnamefont {K.}~\bibnamefont {Amako}},
  \bibinfo {author} {\bibfnamefont {J.}~\bibnamefont {Apostolakis}}, \bibinfo
  {author} {\bibfnamefont {H.}~\bibnamefont {Araujo}}, \bibinfo {author}
  {\bibfnamefont {P.}~\bibnamefont {Arce}}, \bibinfo {author} {\bibfnamefont
  {M.}~\bibnamefont {Asai}}, \bibinfo {author} {\bibfnamefont {D.}~\bibnamefont
  {Axen}}, \bibinfo {author} {\bibfnamefont {S.}~\bibnamefont {Banerjee}},
  \bibinfo {author} {\bibfnamefont {G.}~\bibnamefont {Barrand}}, \bibinfo
  {author} {\bibfnamefont {F.}~\bibnamefont {Behner}}, \bibinfo {author}
  {\bibfnamefont {L.}~\bibnamefont {Bellagamba}}, \bibinfo {author}
  {\bibfnamefont {J.}~\bibnamefont {Boudreau}}, \bibinfo {author}
  {\bibfnamefont {L.}~\bibnamefont {Broglia}}, \bibinfo {author} {\bibfnamefont
  {A.}~\bibnamefont {Brunengo}}, \bibinfo {author} {\bibfnamefont
  {H.}~\bibnamefont {Burkhardt}}, \bibinfo {author} {\bibfnamefont
  {S.}~\bibnamefont {Chauvie}}, \bibinfo {author} {\bibfnamefont
  {J.}~\bibnamefont {Chuma}}, \bibinfo {author} {\bibfnamefont
  {R.}~\bibnamefont {Chytracek}}, \bibinfo {author} {\bibfnamefont
  {G.}~\bibnamefont {Cooperman}}, \bibinfo {author} {\bibfnamefont
  {G.}~\bibnamefont {Cosmo}}, \bibinfo {author} {\bibfnamefont
  {P.}~\bibnamefont {Degtyarenko}}, \bibinfo {author} {\bibfnamefont
  {A.}~\bibnamefont {Dell'Acqua}}, \bibinfo {author} {\bibfnamefont
  {G.}~\bibnamefont {Depaola}}, \bibinfo {author} {\bibfnamefont
  {D.}~\bibnamefont {Dietrich}}, \bibinfo {author} {\bibfnamefont
  {R.}~\bibnamefont {Enami}}, \bibinfo {author} {\bibfnamefont
  {A.}~\bibnamefont {Feliciello}}, \bibinfo {author} {\bibfnamefont
  {C.}~\bibnamefont {Ferguson}}, \bibinfo {author} {\bibfnamefont
  {H.}~\bibnamefont {Fesefeldt}}, \bibinfo {author} {\bibfnamefont
  {G.}~\bibnamefont {Folger}}, \bibinfo {author} {\bibfnamefont
  {F.}~\bibnamefont {Foppiano}}, \bibinfo {author} {\bibfnamefont
  {A.}~\bibnamefont {Forti}}, \bibinfo {author} {\bibfnamefont
  {S.}~\bibnamefont {Garelli}}, \bibinfo {author} {\bibfnamefont
  {S.}~\bibnamefont {Giani}}, \bibinfo {author} {\bibfnamefont
  {R.}~\bibnamefont {Giannitrapani}}, \bibinfo {author} {\bibfnamefont
  {D.}~\bibnamefont {Gibin}}, \bibinfo {author} {\bibfnamefont {J.~J.~G.}\
  \bibnamefont {Cadenas}}, \bibinfo {author} {\bibfnamefont {I.}~\bibnamefont
  {Gonz\'{a}lez}}, \bibinfo {author} {\bibfnamefont {G.~G.}\ \bibnamefont
  {Abril}}, \bibinfo {author} {\bibfnamefont {G.}~\bibnamefont {Greeniaus}},
  \bibinfo {author} {\bibfnamefont {W.}~\bibnamefont {Greiner}}, \bibinfo
  {author} {\bibfnamefont {V.}~\bibnamefont {Grichine}}, \bibinfo {author}
  {\bibfnamefont {A.}~\bibnamefont {Grossheim}}, \bibinfo {author}
  {\bibfnamefont {S.}~\bibnamefont {Guatelli}}, \bibinfo {author}
  {\bibfnamefont {P.}~\bibnamefont {Gumplinger}}, \bibinfo {author}
  {\bibfnamefont {R.}~\bibnamefont {Hamatsu}}, \bibinfo {author} {\bibfnamefont
  {K.}~\bibnamefont {Hashimoto}}, \bibinfo {author} {\bibfnamefont
  {H.}~\bibnamefont {Hasui}}, \bibinfo {author} {\bibfnamefont
  {A.}~\bibnamefont {Heikkinen}}, \bibinfo {author} {\bibfnamefont
  {A.}~\bibnamefont {Howard}}, \bibinfo {author} {\bibfnamefont
  {V.}~\bibnamefont {Ivanchenko}}, \bibinfo {author} {\bibfnamefont
  {A.}~\bibnamefont {Johnson}}, \bibinfo {author} {\bibfnamefont {F.~W.}\
  \bibnamefont {Jones}}, \bibinfo {author} {\bibfnamefont {J.}~\bibnamefont
  {Kallenbach}}, \bibinfo {author} {\bibfnamefont {N.}~\bibnamefont {Kanaya}},
  \bibinfo {author} {\bibfnamefont {M.}~\bibnamefont {Kawabata}}, \bibinfo
  {author} {\bibfnamefont {Y.}~\bibnamefont {Kawabata}}, \bibinfo {author}
  {\bibfnamefont {M.}~\bibnamefont {Kawaguti}}, \bibinfo {author}
  {\bibfnamefont {S.}~\bibnamefont {Kelner}}, \bibinfo {author} {\bibfnamefont
  {P.}~\bibnamefont {Kent}}, \bibinfo {author} {\bibfnamefont {A.}~\bibnamefont
  {Kimura}}, \bibinfo {author} {\bibfnamefont {T.}~\bibnamefont {Kodama}},
  \bibinfo {author} {\bibfnamefont {R.}~\bibnamefont {Kokoulin}}, \bibinfo
  {author} {\bibfnamefont {M.}~\bibnamefont {Kossov}}, \bibinfo {author}
  {\bibfnamefont {H.}~\bibnamefont {Kurashige}}, \bibinfo {author}
  {\bibfnamefont {E.}~\bibnamefont {Lamanna}}, \bibinfo {author} {\bibfnamefont
  {T.}~\bibnamefont {Lamp\'{e}n}}, \bibinfo {author} {\bibfnamefont
  {V.}~\bibnamefont {Lara}}, \bibinfo {author} {\bibfnamefont {V.}~\bibnamefont
  {Lefebure}}, \bibinfo {author} {\bibfnamefont {F.}~\bibnamefont {Lei}},
  \bibinfo {author} {\bibfnamefont {M.}~\bibnamefont {Liendl}}, \bibinfo
  {author} {\bibfnamefont {W.}~\bibnamefont {Lockman}}, \bibinfo {author}
  {\bibfnamefont {F.}~\bibnamefont {Longo}}, \bibinfo {author} {\bibfnamefont
  {S.}~\bibnamefont {Magni}}, \bibinfo {author} {\bibfnamefont
  {M.}~\bibnamefont {Maire}}, \bibinfo {author} {\bibfnamefont
  {E.}~\bibnamefont {Medernach}}, \bibinfo {author} {\bibfnamefont
  {K.}~\bibnamefont {Minamimoto}}, \bibinfo {author} {\bibfnamefont {P.~M.}\
  \bibnamefont {de~Freitas}}, \bibinfo {author} {\bibfnamefont
  {Y.}~\bibnamefont {Morita}}, \bibinfo {author} {\bibfnamefont
  {K.}~\bibnamefont {Murakami}}, \bibinfo {author} {\bibfnamefont
  {M.}~\bibnamefont {Nagamatu}}, \bibinfo {author} {\bibfnamefont
  {R.}~\bibnamefont {Nartallo}}, \bibinfo {author} {\bibfnamefont
  {P.}~\bibnamefont {Nieminen}}, \bibinfo {author} {\bibfnamefont
  {T.}~\bibnamefont {Nishimura}}, \bibinfo {author} {\bibfnamefont
  {K.}~\bibnamefont {Ohtsubo}}, \bibinfo {author} {\bibfnamefont
  {M.}~\bibnamefont {Okamura}}, \bibinfo {author} {\bibfnamefont
  {S.}~\bibnamefont {O'Neale}}, \bibinfo {author} {\bibfnamefont
  {Y.}~\bibnamefont {Oohata}}, \bibinfo {author} {\bibfnamefont
  {K.}~\bibnamefont {Paech}}, \bibinfo {author} {\bibfnamefont
  {J.}~\bibnamefont {Perl}}, \bibinfo {author} {\bibfnamefont {A.}~\bibnamefont
  {Pfeiffer}}, \bibinfo {author} {\bibfnamefont {M.~G.}\ \bibnamefont {Pia}},
  \bibinfo {author} {\bibfnamefont {F.}~\bibnamefont {Ranjard}}, \bibinfo
  {author} {\bibfnamefont {A.}~\bibnamefont {Rybin}}, \bibinfo {author}
  {\bibfnamefont {S.}~\bibnamefont {Sadilov}}, \bibinfo {author} {\bibfnamefont
  {E.~D.}\ \bibnamefont {Salvo}}, \bibinfo {author} {\bibfnamefont
  {G.}~\bibnamefont {Santin}}, \bibinfo {author} {\bibfnamefont
  {T.}~\bibnamefont {Sasaki}}, \bibinfo {author} {\bibfnamefont
  {N.}~\bibnamefont {Savvas}}, \bibinfo {author} {\bibfnamefont
  {Y.}~\bibnamefont {Sawada}}, \bibinfo {author} {\bibfnamefont
  {S.}~\bibnamefont {Scherer}}, \bibinfo {author} {\bibfnamefont
  {S.}~\bibnamefont {Sei}}, \bibinfo {author} {\bibfnamefont {V.}~\bibnamefont
  {Sirotenko}}, \bibinfo {author} {\bibfnamefont {D.}~\bibnamefont {Smith}},
  \bibinfo {author} {\bibfnamefont {N.}~\bibnamefont {Starkov}}, \bibinfo
  {author} {\bibfnamefont {H.}~\bibnamefont {Stoecker}}, \bibinfo {author}
  {\bibfnamefont {J.}~\bibnamefont {Sulkimo}}, \bibinfo {author} {\bibfnamefont
  {M.}~\bibnamefont {Takahata}}, \bibinfo {author} {\bibfnamefont
  {S.}~\bibnamefont {Tanaka}}, \bibinfo {author} {\bibfnamefont
  {E.}~\bibnamefont {Tcherniaev}}, \bibinfo {author} {\bibfnamefont {E.~S.}\
  \bibnamefont {Tehrani}}, \bibinfo {author} {\bibfnamefont {M.}~\bibnamefont
  {Tropeano}}, \bibinfo {author} {\bibfnamefont {P.}~\bibnamefont {Truscott}},
  \bibinfo {author} {\bibfnamefont {H.}~\bibnamefont {Uno}}, \bibinfo {author}
  {\bibfnamefont {L.}~\bibnamefont {Urban}}, \bibinfo {author} {\bibfnamefont
  {P.}~\bibnamefont {Urban}}, \bibinfo {author} {\bibfnamefont
  {M.}~\bibnamefont {Verderi}}, \bibinfo {author} {\bibfnamefont
  {A.}~\bibnamefont {Walkden}}, \bibinfo {author} {\bibfnamefont
  {W.}~\bibnamefont {Wander}}, \bibinfo {author} {\bibfnamefont
  {H.}~\bibnamefont {Weber}}, \bibinfo {author} {\bibfnamefont {J.~P.}\
  \bibnamefont {Wellisch}}, \bibinfo {author} {\bibfnamefont {T.}~\bibnamefont
  {Wenaus}}, \bibinfo {author} {\bibfnamefont {D.~C.}\ \bibnamefont
  {Williams}}, \bibinfo {author} {\bibfnamefont {D.}~\bibnamefont {Wright}},
  \bibinfo {author} {\bibfnamefont {T.}~\bibnamefont {Yamada}}, \bibinfo
  {author} {\bibfnamefont {H.}~\bibnamefont {Yoshida}}, \ and\ \bibinfo
  {author} {\bibfnamefont {D.}~\bibnamefont {Zschiesche}},\ }\href {\doibase
  10.1016/S0168-9002(03)01368-8} {\bibfield  {journal} {\bibinfo  {journal}
  {Nuclear Instruments and Methods in Physics Research, Section A:
  Accelerators, Spectrometers, Detectors, and Associated Equipment}\ }\textbf
  {\bibinfo {volume} {506}},\ \bibinfo {pages} {250 } (\bibinfo {year}
  {2003})}\BibitemShut {NoStop}%
\bibitem [{\citenamefont {Wauters}\ \emph
  {et~al.}(2009{\natexlab{b}})\citenamefont {Wauters}, \citenamefont {Kraev},
  \citenamefont {Z\'{a}kouck\'{y}}, \citenamefont {Beck}, \citenamefont
  {Golovko}, \citenamefont {Kozlov}, \citenamefont {Phalet}, \citenamefont
  {Tandecki}, \citenamefont {Traykov}, \citenamefont {Gorp},\ and\
  \citenamefont {Severijns}}]{Wauters2009b}%
  \BibitemOpen
  \bibfield  {author} {\bibinfo {author} {\bibfnamefont {F.}~\bibnamefont
  {Wauters}}, \bibinfo {author} {\bibfnamefont {I.}~\bibnamefont {Kraev}},
  \bibinfo {author} {\bibfnamefont {D.}~\bibnamefont {Z\'{a}kouck\'{y}}},
  \bibinfo {author} {\bibfnamefont {M.}~\bibnamefont {Beck}}, \bibinfo {author}
  {\bibfnamefont {V.}~\bibnamefont {Golovko}}, \bibinfo {author} {\bibfnamefont
  {V.}~\bibnamefont {Kozlov}}, \bibinfo {author} {\bibfnamefont
  {T.}~\bibnamefont {Phalet}}, \bibinfo {author} {\bibfnamefont
  {M.}~\bibnamefont {Tandecki}}, \bibinfo {author} {\bibfnamefont
  {E.}~\bibnamefont {Traykov}}, \bibinfo {author} {\bibfnamefont {S.~V.}\
  \bibnamefont {Gorp}}, \ and\ \bibinfo {author} {\bibfnamefont
  {N.}~\bibnamefont {Severijns}},\ }\href {\doibase 10.1016/j.nima.2009.08.026}
  {\bibfield  {journal} {\bibinfo  {journal} {Nuclear Instruments and Methods
  in Physics Research, Section A: Accelerators, Spectrometers, Detectors, and
  Associated Equipment}\ }\textbf {\bibinfo {volume} {609}},\ \bibinfo {pages}
  {156 } (\bibinfo {year} {2009}{\natexlab{b}})}\BibitemShut {NoStop}%
\bibitem [{\citenamefont {Wauters}(2009)}]{Wauters2009thesis}%
  \BibitemOpen
  \bibfield  {author} {\bibinfo {author} {\bibfnamefont {F.}~\bibnamefont
  {Wauters}},\ }\emph {\bibinfo {title} {Search for tensor type weak currents
  by measuring the $\beta$-asymmetry parameter in nuclear decays}},\ \href@noop
  {} {Ph.D. thesis},\ \bibinfo  {school} {KU Leuven} (\bibinfo {year}
  {2009})\BibitemShut {NoStop}%
\bibitem [{\citenamefont {V\'enos}\ \emph {et~al.}(2003)\citenamefont
  {V\'enos}, \citenamefont {Z\'akouck\'y},\ and\ \citenamefont
  {Severijns}}]{Venos2003}%
  \BibitemOpen
  \bibfield  {author} {\bibinfo {author} {\bibfnamefont {D.}~\bibnamefont
  {V\'enos}}, \bibinfo {author} {\bibfnamefont {D.}~\bibnamefont
  {Z\'akouck\'y}}, \ and\ \bibinfo {author} {\bibfnamefont {N.}~\bibnamefont
  {Severijns}},\ }\href {\doibase 10.1016/S0092-640X(02)00016-5} {\bibfield
  {journal} {\bibinfo  {journal} {Atomic Data and Nuclear Data Tables}\
  }\textbf {\bibinfo {volume} {83}},\ \bibinfo {pages} {1} (\bibinfo {year}
  {2003})}\BibitemShut {NoStop}%
\bibitem [{\citenamefont {Holstein}(1974)}]{Holstein1974}%
  \BibitemOpen
  \bibfield  {author} {\bibinfo {author} {\bibfnamefont {B.~R.}\ \bibnamefont
  {Holstein}},\ }\href {\doibase 10.1103/RevModPhys.46.789} {\bibfield
  {journal} {\bibinfo  {journal} {Reviews of Modern Physics}\ }\textbf
  {\bibinfo {volume} {46}},\ \bibinfo {pages} {789} (\bibinfo {year}
  {1974})}\BibitemShut {NoStop}%
\bibitem [{\citenamefont {Holstein}(1976)}]{Holstein1976}%
  \BibitemOpen
  \bibfield  {author} {\bibinfo {author} {\bibfnamefont {B.~R.}\ \bibnamefont
  {Holstein}},\ }\href {\doibase 10.1103/RevModPhys.48.673} {\bibfield
  {journal} {\bibinfo  {journal} {Reviews of Modern Physics}\ }\textbf
  {\bibinfo {volume} {48}},\ \bibinfo {pages} {673} (\bibinfo {year}
  {1976})}\BibitemShut {NoStop}%
\bibitem [{\citenamefont {Yokoo}\ \emph {et~al.}(1973)\citenamefont {Yokoo},
  \citenamefont {Suzuki},\ and\ \citenamefont {Morita}}]{Yokoo1973}%
  \BibitemOpen
  \bibfield  {author} {\bibinfo {author} {\bibfnamefont {Y.}~\bibnamefont
  {Yokoo}}, \bibinfo {author} {\bibfnamefont {S.}~\bibnamefont {Suzuki}}, \
  and\ \bibinfo {author} {\bibfnamefont {M.}~\bibnamefont {Morita}},\ }\href
  {\doibase 10.1143/PTP.50.1894} {\bibfield  {journal} {\bibinfo  {journal}
  {Progress of Theoretical Physics}\ }\textbf {\bibinfo {volume} {50}},\
  \bibinfo {pages} {1894} (\bibinfo {year} {1973})}\BibitemShut {NoStop}%
\bibitem [{\citenamefont {de~Leebeeck}()}]{DeLeebeeck2014}%
  \BibitemOpen
  \bibfield  {author} {\bibinfo {author} {\bibfnamefont {V.}~\bibnamefont
  {de~Leebeeck}},\ }\href@noop {} {}\bibinfo {note} {\emph{in
  preparation}}\BibitemShut {NoStop}%
\bibitem [{\citenamefont {Calaprice}\ and\ \citenamefont
  {Holstein}(1976)}]{Calaprice1976}%
  \BibitemOpen
  \bibfield  {author} {\bibinfo {author} {\bibfnamefont {F.}~\bibnamefont
  {Calaprice}}\ and\ \bibinfo {author} {\bibfnamefont {B.}~\bibnamefont
  {Holstein}},\ }\href {\doibase 10.1016/0375-9474(76)90593-5} {\bibfield
  {journal} {\bibinfo  {journal} {Nuclear Physiscs A}\ }\textbf {\bibinfo
  {volume} {273}},\ \bibinfo {pages} {301 } (\bibinfo {year}
  {1976})}\BibitemShut {NoStop}%
\bibitem [{\citenamefont {Hardy}\ and\ \citenamefont
  {Towner}(2009)}]{Hardy2009}%
  \BibitemOpen
  \bibfield  {author} {\bibinfo {author} {\bibfnamefont {J.~C.}\ \bibnamefont
  {Hardy}}\ and\ \bibinfo {author} {\bibfnamefont {I.~S.}\ \bibnamefont
  {Towner}},\ }\href {\doibase 10.1103/PhysRevC.79.055502} {\bibfield
  {journal} {\bibinfo  {journal} {Physical Review C}\ }\textbf {\bibinfo
  {volume} {79}},\ \bibinfo {eid} {055502} (\bibinfo {year}
  {2009})}\BibitemShut {NoStop}%
\bibitem [{\citenamefont {Towner}(1987)}]{Towner1987}%
  \BibitemOpen
  \bibfield  {author} {\bibinfo {author} {\bibfnamefont {I.}~\bibnamefont
  {Towner}},\ }\href@noop {} {\bibfield  {journal} {\bibinfo  {journal}
  {Physics Reports}\ }\textbf {\bibinfo {volume} {155}},\ \bibinfo {pages}
  {263} (\bibinfo {year} {1987})}\BibitemShut {NoStop}%
\bibitem [{\citenamefont {Martinez-Pinedo}\ \emph {et~al.}(1997)\citenamefont
  {Martinez-Pinedo}, \citenamefont {Zuker}, \citenamefont {Poves},\ and\
  \citenamefont {Caurier}}]{Martinez1997}%
  \BibitemOpen
  \bibfield  {author} {\bibinfo {author} {\bibfnamefont {G.}~\bibnamefont
  {Martinez-Pinedo}}, \bibinfo {author} {\bibfnamefont {A.~P.}\ \bibnamefont
  {Zuker}}, \bibinfo {author} {\bibfnamefont {A.}~\bibnamefont {Poves}}, \ and\
  \bibinfo {author} {\bibfnamefont {E.}~\bibnamefont {Caurier}},\ }\href
  {\doibase 10.1103/PhysRevC.55.187} {\bibfield  {journal} {\bibinfo  {journal}
  {Physical Review C}\ }\textbf {\bibinfo {volume} {55}},\ \bibinfo {pages}
  {187} (\bibinfo {year} {1997})}\BibitemShut {NoStop}%
\bibitem [{\citenamefont {Siiskonen}\ \emph {et~al.}(2001)\citenamefont
  {Siiskonen}, \citenamefont {Hjorth-Jensen},\ and\ \citenamefont
  {Suhonen}}]{Siiskonen2001}%
  \BibitemOpen
  \bibfield  {author} {\bibinfo {author} {\bibfnamefont {T.}~\bibnamefont
  {Siiskonen}}, \bibinfo {author} {\bibfnamefont {M.}~\bibnamefont
  {Hjorth-Jensen}}, \ and\ \bibinfo {author} {\bibfnamefont {J.}~\bibnamefont
  {Suhonen}},\ }\href {\doibase 10.1103/PhysRevC.63.055501} {\bibfield
  {journal} {\bibinfo  {journal} {Physical Review C}\ }\textbf {\bibinfo
  {volume} {63}},\ \bibinfo {pages} {055501} (\bibinfo {year}
  {2001})}\BibitemShut {NoStop}%
\bibitem [{\citenamefont {Honma}\ \emph {et~al.}(2005)\citenamefont {Honma},
  \citenamefont {Otsuka}, \citenamefont {Brown},\ and\ \citenamefont
  {Mizusaki}}]{Honma2005}%
  \BibitemOpen
  \bibfield  {author} {\bibinfo {author} {\bibfnamefont {M.}~\bibnamefont
  {Honma}}, \bibinfo {author} {\bibfnamefont {T.}~\bibnamefont {Otsuka}},
  \bibinfo {author} {\bibfnamefont {B.}~\bibnamefont {Brown}}, \ and\ \bibinfo
  {author} {\bibfnamefont {T.}~\bibnamefont {Mizusaki}},\ }\href {\doibase
  {10.1140/epjad/i2005-06-032-2}} {\bibfield  {journal} {\bibinfo  {journal}
  {The European Physical Journal A}\ }\textbf {\bibinfo {volume} {25}},\
  \bibinfo {pages} {499} (\bibinfo {year} {2005})},\ \bibinfo {note} {{4th
  International Conference on Exotic Nuclei and Atomic Masses, Pine Mt, GA, SEP
  12-16, 2004}}\BibitemShut {NoStop}%
\bibitem [{\citenamefont {Johnson}\ \emph {et~al.}(1963)\citenamefont
  {Johnson}, \citenamefont {Pleasonton},\ and\ \citenamefont
  {Carlson}}]{Johnson1963}%
  \BibitemOpen
  \bibfield  {author} {\bibinfo {author} {\bibfnamefont {C.}~\bibnamefont
  {Johnson}}, \bibinfo {author} {\bibfnamefont {F.}~\bibnamefont {Pleasonton}},
  \ and\ \bibinfo {author} {\bibfnamefont {T.}~\bibnamefont {Carlson}},\ }\href
  {\doibase 10.1103/PhysRev.132.1149} {\bibfield  {journal} {\bibinfo
  {journal} {Physical Review}\ }\textbf {\bibinfo {volume} {132}},\ \bibinfo
  {pages} {1149} (\bibinfo {year} {1963})}\BibitemShut {NoStop}%
\bibitem [{\citenamefont {Gl\"uck}(1998)}]{Gluck1998}%
  \BibitemOpen
  \bibfield  {author} {\bibinfo {author} {\bibfnamefont {F.}~\bibnamefont
  {Gl\"uck}},\ }\href {\doibase 10.1016/S0375-9474(97)00643-X} {\bibfield
  {journal} {\bibinfo  {journal} {Nuclear Physics A}\ }\textbf {\bibinfo
  {volume} {628}},\ \bibinfo {pages} {493} (\bibinfo {year}
  {1998})}\BibitemShut {NoStop}%
\bibitem [{\citenamefont {Carnoy}\ \emph {et~al.}(1991)\citenamefont {Carnoy},
  \citenamefont {Deutsch}, \citenamefont {Girard},\ and\ \citenamefont
  {Prieels}}]{Carnoy1991}%
  \BibitemOpen
  \bibfield  {author} {\bibinfo {author} {\bibfnamefont {A.~S.}\ \bibnamefont
  {Carnoy}}, \bibinfo {author} {\bibfnamefont {J.}~\bibnamefont {Deutsch}},
  \bibinfo {author} {\bibfnamefont {T.~A.}\ \bibnamefont {Girard}}, \ and\
  \bibinfo {author} {\bibfnamefont {R.}~\bibnamefont {Prieels}},\ }\href
  {\doibase 10.1103/PhysRevC.43.2825} {\bibfield  {journal} {\bibinfo
  {journal} {Physical Review C}\ }\textbf {\bibinfo {volume} {43}},\ \bibinfo
  {pages} {2825} (\bibinfo {year} {1991})}\BibitemShut {NoStop}%
\bibitem [{\citenamefont {Bhattacharya}\ \emph {et~al.}(2012)\citenamefont
  {Bhattacharya}, \citenamefont {Cirigliano}, \citenamefont {Cohen},
  \citenamefont {Filipuzzi}, \citenamefont {Gonz\'{a}lez-Alonso}, \citenamefont
  {Graesser}, \citenamefont {Gupta},\ and\ \citenamefont
  {Lin}}]{Bhattacharya2012}%
  \BibitemOpen
  \bibfield  {author} {\bibinfo {author} {\bibfnamefont {T.}~\bibnamefont
  {Bhattacharya}}, \bibinfo {author} {\bibfnamefont {V.}~\bibnamefont
  {Cirigliano}}, \bibinfo {author} {\bibfnamefont {S.~D.}\ \bibnamefont
  {Cohen}}, \bibinfo {author} {\bibfnamefont {A.}~\bibnamefont {Filipuzzi}},
  \bibinfo {author} {\bibfnamefont {M.}~\bibnamefont {Gonz\'{a}lez-Alonso}},
  \bibinfo {author} {\bibfnamefont {M.~L.}\ \bibnamefont {Graesser}}, \bibinfo
  {author} {\bibfnamefont {R.}~\bibnamefont {Gupta}}, \ and\ \bibinfo {author}
  {\bibfnamefont {H.-W.}\ \bibnamefont {Lin}},\ }\href {\doibase
  10.1103/PhysRevD.85.054512} {\bibfield  {journal} {\bibinfo  {journal}
  {Physical Review D}\ }\textbf {\bibinfo {volume} {85}},\ \bibinfo {pages}
  {054512} (\bibinfo {year} {2012})}\BibitemShut {NoStop}%
\bibitem [{\citenamefont {Naviliat-Cuncic}\ and\ \citenamefont
  {Gonz\'{a}lez-Alonso}(2013)}]{Naviliat2013a}%
  \BibitemOpen
  \bibfield  {author} {\bibinfo {author} {\bibfnamefont {O.}~\bibnamefont
  {Naviliat-Cuncic}}\ and\ \bibinfo {author} {\bibfnamefont {M.}~\bibnamefont
  {Gonz\'{a}lez-Alonso}},\ }\href {\doibase 10.1002/andp.201300072} {\bibfield
  {journal} {\bibinfo  {journal} {Annalen der Physik}\ }\textbf {\bibinfo
  {volume} {20}},\ \bibinfo {pages} {1} (\bibinfo {year} {2013})}\BibitemShut
  {NoStop}%
\bibitem [{\citenamefont {Lojek}\ \emph {et~al.}(2009)\citenamefont {Lojek},
  \citenamefont {Bodek},\ and\ \citenamefont {Kuzniak}}]{Lojek2009}%
  \BibitemOpen
  \bibfield  {author} {\bibinfo {author} {\bibfnamefont {K.}~\bibnamefont
  {Lojek}}, \bibinfo {author} {\bibfnamefont {K.}~\bibnamefont {Bodek}}, \ and\
  \bibinfo {author} {\bibfnamefont {M.}~\bibnamefont {Kuzniak}},\ }\href
  {\doibase 10.1016/j.nima.2009.07.075} {\bibfield  {journal} {\bibinfo
  {journal} {Nuclear Instruments and Methods in Physics Research, Section A:
  Accelerators, Spectrometers, Detectors, and Associated Equipment}\ }\textbf
  {\bibinfo {volume} {611}},\ \bibinfo {pages} {284 } (\bibinfo {year}
  {2009})}\BibitemShut {NoStop}%
\bibitem [{\citenamefont {Ban}\ \emph {et~al.}(2009)\citenamefont {Ban},
  \citenamefont {Bialek}, \citenamefont {Bodek}, \citenamefont {Bozek},
  \citenamefont {Gorel}, \citenamefont {Kirch}, \citenamefont {Kistryn},
  \citenamefont {Kozela}, \citenamefont {Kuzniak},\ and\ \citenamefont
  {Naviliat-Cuncic}}]{Ban2009a}%
  \BibitemOpen
  \bibfield  {author} {\bibinfo {author} {\bibfnamefont {G.}~\bibnamefont
  {Ban}}, \bibinfo {author} {\bibfnamefont {A.}~\bibnamefont {Bialek}},
  \bibinfo {author} {\bibfnamefont {K.}~\bibnamefont {Bodek}}, \bibinfo
  {author} {\bibfnamefont {J.}~\bibnamefont {Bozek}}, \bibinfo {author}
  {\bibfnamefont {P.}~\bibnamefont {Gorel}}, \bibinfo {author} {\bibfnamefont
  {K.}~\bibnamefont {Kirch}}, \bibinfo {author} {\bibfnamefont
  {S.}~\bibnamefont {Kistryn}}, \bibinfo {author} {\bibfnamefont
  {A.}~\bibnamefont {Kozela}}, \bibinfo {author} {\bibfnamefont
  {M.}~\bibnamefont {Kuzniak}}, \ and\ \bibinfo {author} {\bibfnamefont
  {O.}~\bibnamefont {Naviliat-Cuncic}},\ }\href {\doibase
  10.1016/j.nima.2009.07.094} {\bibfield  {journal} {\bibinfo  {journal}
  {Nuclear Instruments and Methods in Physics Research Section A: Accelerators,
  Spectrometers, Detectors and Associated Equipment}\ }\textbf {\bibinfo
  {volume} {611}},\ \bibinfo {pages} {198} (\bibinfo {year}
  {2009})}\BibitemShut {NoStop}%
\bibitem [{\citenamefont {Severijns}\ and\ \citenamefont
  {Naviliat-Cuncic}(2013)}]{Severijns2013}%
  \BibitemOpen
  \bibfield  {author} {\bibinfo {author} {\bibfnamefont {N.}~\bibnamefont
  {Severijns}}\ and\ \bibinfo {author} {\bibfnamefont {O.}~\bibnamefont
  {Naviliat-Cuncic}},\ }\href {\doibase 10.1088/0031-8949/2013/T152/014018}
  {\bibfield  {journal} {\bibinfo  {journal} {Physica Scripta}\ }\textbf
  {\bibinfo {volume} {T152}},\ \bibinfo {pages} {014018} (\bibinfo {year}
  {2013})}\BibitemShut {NoStop}%
\end{thebibliography}
\end{document}